\documentclass[preprint,aps,12pt,showpacs,nofootinbib,tightenlines]{revtex4}
\usepackage{mathrsfs}
\usepackage{amsmath}
\usepackage{amssymb}
\usepackage{epsfig}
\usepackage{epstopdf}
\usepackage{graphicx}
\usepackage{subfigure}
\usepackage{booktabs}
\usepackage{float}
\usepackage{amsmath}
\usepackage{color}
\usepackage{multirow}
\usepackage{bm}

\def\be{\begin{eqnarray}}
\def\en{\end{eqnarray}}
\def\non{\nonumber\\}

\begin{document}
\title{Systematic analysis of $D_{(s)}$ meson semi-leptonic decays in the covariant light-front quark model}
\author{Hao Yang$^1$, Shao-Qin Guo$^1$ and Zhi-Qing Zhang$^1$
\footnote{zhangzhiqing@haut.edu.cn} } 
\affiliation{\it \small $^1$  School of Physics and advanced energy, Henan University
of Technology, Zhengzhou, Henan 450052, China } 
\date{\today}
\begin{abstract}
The weak decays of the $D_{(s)}$ meson provide a pivotal platform to advance our understanding of the Standard Model (SM) and to explore New Physics (NP).  In recent years, experiments have collected a significant amount of data on the $D_{(s)}$ meson decays, particularly from BESIII, which provides substantial support for theoretical research.
In this work, we systematically investigate the semi-leptonic decays of $D_{(s)}$ meson to pseudoscalar (P), scalar (S), vector (V), and axial-vector (A) mesons within the framework of the covariant light-front quark model (CLFQM). We calculate the form factors of the transitions $D_{(s)}\to P,S,V,A$ and the branching ratios of the corresponding semi-leptonic decays, then compare them with experimental data and results from other theoretical models.
The form factor values of the transitions $D_{(s)}\to P, V$ obtained from the CLFQM are consistent with those of other theoretical models and available data in most cases. However, significant discrepancies are found in some specific $D_{(s)}\to S,A$ transitions, such as $D\to a_0(980), a_0(1450)$ and $D_{(s)} \to K_{1B}$, compared to other theoretical calculations. The predicted branching ratios for the semi-leptonic decays $D\to P(V)\ell\nu_\ell$ with $\ell=e,\mu$ agree well with experimental data and other theoretical results in most decay channels, validating the reliability of the model. However, for some $D\to S(V)\ell\nu_\ell$ decays, tension exists among different theoretical predictions and experimental results. Further clarification of such differences is necessary.
Our study provides important insights into the internal structures for some scalar and axial-vector mesons and serves as a theoretical reference for future experiments.
\end{abstract}

 \vspace{1cm}

\maketitle

\section{Introduction}\label{intro}
As we know, the Cabibbo-Kobayashi-Maskawa (CKM) matrix exhibits rigorous unitarity relation in the SM, and any deviation from such relation is believed to be a signal of NP. The semi-leptonic decays of $D_{(s)}$ meson play a pivotal role in the extraction of the CKM matrix elements and in the search for NP. In order to precisely extract the CKM matrix elements, one should obtain the form factors for the relevant decay channels. The CLFQM based on relativistic quantum mechanics and quantum field theory, which has shown strong predictive ability in describing the radiation and semi-leptonic decays. The core of CLFQM lies in its description of the internal structure and dynamics of mesons. It regards mesons as bound states composed of quarks and antiquarks, and uses light-front formalism to deal with relativistic effects and dynamics, which has significant advantages. In early research, Jaus \cite{AJaus,BJaus,CJaus} put forward the CLFQM to handle the electroweak transitions, then Cheng \textit{et al.} \cite{Cheng0310} systematically studied the decay constants and form factors of s-wave and p-wave mesons by using this approach. It was subsequently extended in the $D_{(s)}\to P,S,V,A$ transitions form factors with $P,S,V,A$ being a pseudoscalar (P), scalar (S), vector (V), and axial-vector (A) mesons in Ref. \cite{CLF1103}.
These works provide a solid theoretical foundation for using the CLFQM to calculate the semi-leptonic decays of $D_{(s)}$ meson.

This article aims to systematically analyze the branching ratios of the semi-leptonic  $D_{(s)}\to (P,S,V,A)\ell\nu_\ell$ decays based on the CLFQM. The results are compared with existing experimental data from BESIII, CLEO, Belle, and BABAR collaborations, as well as with other theoretical predictions, such as those obtained using the Light-Cone Sum Rules (LCSR), the lattice QCD (LQCD), the covariant confined quark model (CCQM), the constituent quark model (CQM), the QCD Sum Rules (QCDSRs), the relativistic quark model (RQM), the chiral unitarity approach ($\chi$UT), SU(3) flavor symmetry and so on.
On the experimental side, many semi-leptonic decays of  $D_{(s)}$ meson have been observed by experiments. Especially, some scalar and axial-vector mesons were studied in such kind of decay by BESIII recently. For example, the light scalar meson $a_0(980)$ was studied in the decay $D^0\to a_0(980)^-e^+\nu_e$ in Ref. \cite{BESIII2411}, where the branching ratio and the transition form factor were measured as $Br(D^0\to a_0(980)^-e^+\nu_e, a_0(980)^-\to \eta\pi^-)=(8.6\pm1.7\pm0.5)\times10^{-5}$ and  $F^{Da_0}_0=0.559\pm0.056\pm0.013$, respectively. These data will serve as important inputs for understanding the internal structure of $a_0(980)$ and the hadronization information of the transition $D\to a_0(980)$. Similar, the light scalar meson $f_0(980)$ was studied in the decay $D^+_s\to f_0(980)e^+\nu_e$ in Ref. \cite{BESIII2303}, where the branching ratio and the transition form factor were determined as $Br(D^+_s\to f_0(980)e^+\nu_e, f_0(980)\to \pi^+\pi^-)=(1.72\pm0.13\pm0.10)\times10^{-3}$ and  $F^{D_sf_0}_0=0.518\pm0.018\pm0.036$, respectively. These results are important to probe the quark component in $f_0(980)$ and understand the non-perturbative dynamics of charmed meson decays. Furthermore, the semi-leptonic decays $D\to K_1(1270)\ell\nu_\ell$ were also observed by BESIII \cite{BESIII250302,BESIII250203}, where the branching ratios are measured to be
\be
Br(D^0\to K_1(1270)^-e^+\nu_e)&=&(1.02\pm0.09)\times10^{-3}, \\
Br(D^+\to \bar K_1(1270)^0e^+\nu_e)&=&(2.27\pm0.15)\times10^{-3},\\
Br(D^0\to K_1(1270)^-\mu^+\nu_\mu)&=&(0.78^{+0.19}_{-0.21})\times10^{-3},\\
Br(D^+\to \bar K_1(1270)^0\mu^+\nu_\mu)&=&(2.36^{+0.55}_{-0.59})\times10^{-3}.
\en
Compared with theoretical predictions, these measurements can provide crucial information on the $K_1(1270)-K_1(1400)$ mixing angle. At present, $\theta_{K_1}=33^\circ$ or $58^\circ$ is supported and $\theta_{K_1}=-(34\pm13)^\circ$ is disfavored.
On the theoretical side, the difference between different theoretical predictions for the $D_{(s)}\to P, V$ transition form factors is slight, while for the $D_{(s)}\to S, A$ transition form factors, the situation differs completely. Obviously, it is intrinsically related to the uncertain internal structures of scalar and axial-vector mesons. For example, the transition form factor $D\to a_0(980)$ predicted by some theoretical approaches, such as QCDSR, CCQM, generally ranges from 0.53 to 0.55 with small deviations from experimental measurement $F^{Da_0}_0=0.559$, which is much smaller than 0.85 and 1.75 given by LCSR. The form factor of the transition
$D\to K_1(1270)$ is determined from those of the transitions $D\to K_{1A}$ and $D\to K_{1B}$ through the mixing angle $\theta_{K_1}$. In order to calculate the branching ratio of the semi-leptonic decay $D\to K_1(1270)\ell\nu_\ell$, the form factors $V^{DK_{1A,B}}, A^{DK_{1A,B}}_{0,1,2}$ are need to be determined simultaneously. So significant discrepancies among theoretical predictions are expected and the uncertainty from the mixing angle increases the difficulty of theoretical calculations. Given the problems existing in the
$D_{(s)}$ transitions to some hadronic final states and the ongoing precise experimental measurements, the systematic analysis of the semi-leptonic decays  $D_{(s)}\to (P,S,V,A)\ell\nu_\ell$ is necessary, which is helpful to understand the non-perturbative dynamics in charm meson decays and probe the internal properties of the final-state hadrons.

The paper is organized as follows, in Sec. II, we elaborate on the framework of the CLFQM and its application in calculating the branching ratios of semi-leptonic decays. By computing the vector and aixal-vector current matrix elements, the analytical expressions for the hadronic transition form factors can be obtained. In Sec. III, we present the numerical results of the $D_{(s)}\to P,S,V,A$ transition form factors and their $q^2$ dependence. Then using these transition form factors, we calculate the branching ratios of semi-leptonic $D_{(s)}$ meson decays. In addition, detailed data analysis and discussion, including a comparison with other theoretical calculations and experimental measurements, are carried out. The conclusions are presented in the final part. Detailed results on the transition form factors, including fitting parameters, are tabulated in Appendix A. Some specific rules in the process of performing $p^-$ integration as well as the expressions for the form factors are collected in Appendix B.

\section{FORMALISM AND DECAY RATES}\label{form1}
In calculations, the Bauer-Stech-Wirbel (BSW) form factors \cite{Wirbel} for the transitions $D_{(s)}\to P,S,V, A$ are used and defined as follows,
\begin{footnotesize}
\begin{eqnarray}
\left\langle P\left(P^{\prime \prime}\right)\left| V_{\mu}\right|D_{(s)}\left(P^{\prime}\right)\right\rangle &=&\left(P_{\mu}-\frac{m_{D_{(s)}}^{2}-m_{P}^{2}}{q^{2}} q_{\mu}\right) F_{1}^{D_{(s)} P}\left(q^{2}\right)+\frac{m_{D_{(s)}}^{2}-m_{P}^{2}}{q^{2}} q_{\mu} F_{0}^{D_{(s)} P}\left(q^{2}\right), \label{DtoP}\\
\left\langle S\left(P^{\prime \prime}\right)\left| A_{\mu}\right|D_{(s)}\left(P^{\prime}\right)\right\rangle &=&\left(P_{\mu}-\frac{m_{D_{(s)}}^{2}-m_{S}^{2}}{q^{2}} q_{\mu}\right) F_{1}^{D_{(s)} S}\left(q^{2}\right)+\frac{m_{D_{(s)}}^{2}-m_{S}^{2}}{q^{2}} q_{\mu} F_{0}^{D_{(s)} S}\left(q^{2}\right),\\
\left\langle V\left(P^{\prime \prime}, \varepsilon^{\mu *}\right)\left| V_{\mu}\right|D_{(s)}\left(P^{\prime}\right)\right\rangle &=&-\frac{1}{m_{D_{(s)}}+m_{V}} \epsilon_{\mu \nu \alpha \beta} \varepsilon^{* \nu} P^{\alpha} q^{\beta} V^{D_{(s)} V}\left(q^{2}\right),\\
\left\langle V\left(P^{\prime \prime}, \varepsilon^{\mu *}\right)\left| A_{\mu}\right|D_{(s)}\left(P^{\prime}\right)\right\rangle &=& i\left\{\left(m_{D_{(s)}}+m_{V}\right) \varepsilon_{\mu}^{*} A_{1}^{D_{(s)} V}\left(q^{2}\right)-\frac{\varepsilon^{*} \cdot P}{m_{D_{(s)}}+m_{V}} P_{\mu} A_{2}^{D_{(s)} V}\left(q^{2}\right)\right. \nonumber\\
&& \left.-2 m_{V} \frac{\varepsilon^{*} \cdot P}{q^{2}} q_{\mu}\left[A_{3}^{D_{(s)} V}\left(q^{2}\right)-A_{0}^{D_{(s)} V}\left(q^{2}\right)\right]\right\},\\
\left\langle A\left(P^{\prime \prime}, \varepsilon^{\mu *}\right)\left| V_{\mu}\right|D_{(s)}\left(P^{\prime}\right)\right\rangle &=& -i\left\{\left(m_{D_{(s)}}-m_{A}\right) \varepsilon_{\mu}^{*} V_{1}^{D_{(s)} A}\left(q^{2}\right)-\frac{\varepsilon^{*} \cdot P}{m_{D_{(s)}}-m_{A}} P_{\mu} V_{2}^{D_{(s)} A}\left(q^{2}\right)\right.\nonumber\\
&& \left.-2 m_{A} \frac{\varepsilon^{*} \cdot P}{q^{2}} q_{\mu}\left[V_{3}^{D_{(s)} A}\left(q^{2}\right)-V_{0}^{D_{(s)} A}\left(q^{2}\right)\right]\right\},\\
\left\langle A\left(P^{\prime \prime}, \varepsilon^{\mu *}\right)\left| A_{\mu}\right|D_{(s)}\left(P^{\prime}\right)\right\rangle &=&-\frac{1}{m_{D_{(s)}}-m_{A}} \epsilon_{\mu \nu \alpha \beta} \varepsilon^{* \nu} P^{\alpha} q^{\beta} A^{D_{(s)} A}\left(q^{2}\right),\label{DtoA}
\end{eqnarray}
\end{footnotesize}
where $P=P^{\prime}+P^{\prime\prime}, q=P^{\prime}-P^{\prime\prime}$ and the convention $\varepsilon_{0123}=1$ is adopted. In the above formulas, the following relations hold,
\be
F_{1}^{D_{(s)} P}(0)&=&F_{0}^{D_{(s)} P}(0), \;\;A_{3}^{D_{(s)} V}(0)=A_{0}^{D_{(s)} V}(0),\\
A_{3}^{D_{(s)} V}\left(q^{2}\right)&=&\frac{M^{\prime}+M^{\prime \prime}}{2 M^{\prime \prime}} A_{1}^{D_{(s)} V}\left(q^{2}\right)-\frac{M^{\prime}-M^{\prime \prime}}{2 M^{\prime \prime}} A_{2}^{D_{(s)} V}\left(q^{2}\right).
\en
\begin{figure}[htbp]
 	\centering \subfigure{
 		\begin{minipage}{5cm}
 			\centering
 			\includegraphics[width=5cm]{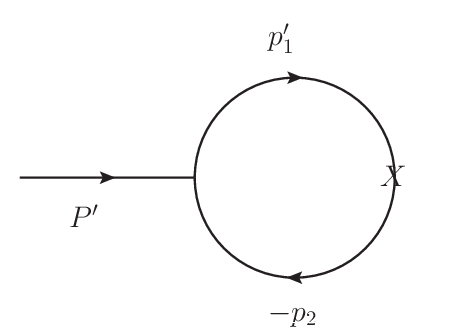}
 	\end{minipage}}
 	\subfigure{
 		\begin{minipage}{6cm}
 			\centering
 			\includegraphics[width=6cm]{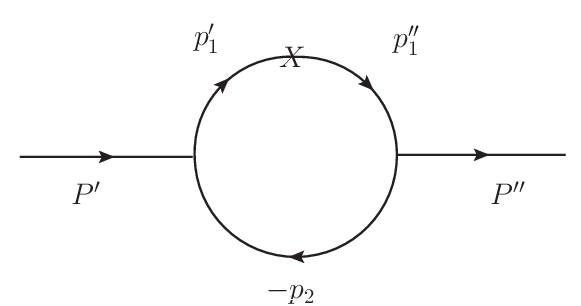}
 	\end{minipage}}
 	\caption{Feynman diagrams for $D_{(s)}$ decay (left) and transition (right) amplitudes, where $P^{\prime(\prime\prime)}$ is the
 		incoming (outgoing) meson momentum, X denotes the vector or axial-vector transition vertex.}
 	\label{feyn}
 \end{figure}

The lowest order form factor can be obtained by calculating the Feynman diagram shown in Figure \ref{feyn}. In the covariant quark mdoel, we adopt the light-front coordiate form for a momentum $p$, that is $p=(p^-,p^+,p_\perp)$ with $p^\pm=p^0\pm p_z, p^2=p^+p^--p^2_\perp$. The incoming (outgoing) meson has the mass $M^\prime (M^{\prime\prime})$ with the momentum $P^\prime=p_1^\prime+p_2 (P^{\prime\prime}=p_1^{\prime\prime}+p_2)$, where $p_{1}^{\prime(\prime\prime)}$ and $p_{2}$ are the momenta of the quark and anti-quark inside the incoming (outgoing) meson with the mass $m_{1}^{\prime(\prime\prime)}$and $m_{2}$, respectively. These momenta can be expressed in terms of the internal variables $(x_{i},p{'}_{\perp})$ as
\be
p_{1,2}^{\prime+}=x_{1,2} P^{\prime+}, \quad p_{1,2 \perp}^{\prime}=x_{1,2} P_{\perp}^{\prime} \pm p_{\perp}^{\prime},
\en
with $x_{1}+x_{2}=1$. To obtain the form factors of the transitions $D_{(s)}\to M$ with $M=P, S,V, A$, we need to consider the matrix elements
\be
\left\langle M(P^\prime\prime)|V_\mu-A_\mu|D_{(s)}(P^\prime)\right\rangle =\mathcal{B}^{D_{(s)}M}_\mu.
\en
For the $D_{(s)}\to P$ transition amplitude, we have
\be
\mathcal{B}_{\mu}^{D_{(s)} P}=-i^{3} \frac{N_{c}}{(2 \pi)^{4}} \int d^{4} p_{1}^{\prime} \frac{H_{D_{(s)}}^{\prime} H_{P}^{\prime \prime}}{N_{1}^{\prime} N_{1}^{\prime \prime} N_{2}} S_{\mu}^{D_{(s)} P},
\en
where $N_{1}^{\prime(\prime \prime)}=p_{1}^{\prime(\prime \prime) 2}-m_{1}^{\prime (\prime\prime) 2}, N_{2}=p_{2}^{2}-m_{2}^{2}$ arise from the quark propagators, and the trace $S_{\mu}^{D_{(s)} P}$ can be obtained by using the Lorentz contraction,
\be
S_{\mu}^{D_{(s)} P}=\operatorname{Tr}\left[\gamma_{5}\left(\not p_{1}^{\prime \prime}+m_{1}^{\prime \prime}\right) \gamma_{\mu}\left(\not p_{1}^{\prime}+m_{1}^{\prime}\right) \gamma_{5}\left(-\not p_{2}+m_{2}\right)\right].
\en

It is similar for the $D_{(s)} \to S, V, A$ transition amplitudes, which are listed as following

\be
\mathcal{B}_{\mu}^{D_{(s)} S}=-i^{3} \frac{N_{c}}{(2 \pi)^{4}} \int d^{4} p_{1}^{\prime} \frac{H_{D_{(s)}}^{\prime} H_{S}^{\prime \prime}}{N_{1}^{\prime} N_{1}^{\prime \prime} N_{2}} S_{\mu}^{D_{(s)} S},
\en
where
\be
S_{ \mu}^{D_{(s)} S} & =&\operatorname{Tr}\left[(-i)\left(\not p_{1}^{\prime \prime}+m_{1}^{\prime \prime}\right) \gamma_{\mu} \gamma_{5}\left(\not p_{1}^{\prime}+m_{1}^{\prime}\right) \gamma_{5}\left(-\not p_{2}+m_{2}\right)\right] \nonumber\\
& =&-i S_{V \mu}^{D_{(s)} P}\left(m_{1}^{\prime \prime} \rightarrow-m_{1}^{\prime \prime}\right) .
\en

\be
\mathcal{B}_{\mu}^{D_{(s)} V}=-i^{3} \frac{N_{c}}{(2 \pi)^{4}} \int d^{4} p_{1}^{\prime} \frac{H_{D_{(s)}}^{\prime}\left(i H_{V}^{\prime \prime}\right)}{N_{1}^{\prime} N_{1}^{\prime \prime} N_{2}} S_{\mu \nu}^{D_{(s)} V} \varepsilon^{\prime \prime * \nu},
\en
where
\begin{small}
\begin{eqnarray}
S_{\mu \nu}^{D_{(s)} V} & =&\left(S_{V}^{D_{(s)} V}-S_{A}^{D_{(s)} V}\right)_{\mu \nu} \nonumber\\
& =&\operatorname{Tr}\left[\left(\gamma_{\nu}-\frac{1}{W_{V}^{\prime \prime}}\left(p_{1}^{\prime \prime}-p_{2}\right)_{\nu}\right)\left(\not p_{1}^{\prime \prime}+m_{1}^{\prime \prime}\right)\left(\gamma_{\mu}-\gamma_{\mu} \gamma_{5}\right)\left(\not p_{1}^{\prime}+m_{1}^{\prime}\right) \gamma_{5}\left(-\not p_{2}+m_{2}\right)\right].
\end{eqnarray}
\end{small}
There are two types axial-vector mesons with $J^{PC}=1^{++}$ and $1^{+-}$, respectively, which are denoted as $^3A$ and $^1A$, and the corresponding transition amplitudes are expressed as
\be
\mathcal{B}_{\mu}^{D_{(s)} \;^{3}A} &=&-i^{3} \frac{N_{c}}{(2 \pi)^{4}} \int d^{4} p_{1}^{\prime} \frac{H_{D_{(s)}}^{\prime}\left(-i H_{^3A}^{\prime \prime}\right)}{N_{1}^{\prime} N_{1}^{\prime \prime} N_{2}} S_{\mu \nu}^{D_{(s)} \;^{3}A} \varepsilon^{\prime \prime * \nu}, \\
\mathcal{B}_{\mu}^{D_{(s)}\; ^{1}A} &=&-i^{3} \frac{N_{c}}{(2 \pi)^{4}} \int d^{4} p_{1}^{\prime} \frac{H_{D_{(s)}}^{\prime}\left(-i H_{^1A}^{\prime \prime}\right)}{N_{1}^{\prime} N_{1}^{\prime \prime} N_{2}} S_{\mu \nu}^{D_{(s)}\; ^{1}A} \varepsilon^{\prime \prime * \nu},
\en
where
\begin{small}
\begin{eqnarray}
S_{\mu \nu}^{D_{(s)} \;^3A} &=&\left(S_{V}^{D_{(s)}\; ^3A}-S_{A}^{D_{(s)} \;^{3}A}\right)_{\mu \nu} \nonumber\\
&=&\operatorname{Tr}\left[\left(\gamma_{\nu}-\frac{1}{W_{^3A}^{\prime \prime}}\left(p_{1}^{\prime \prime}-p_{2}\right)_{\nu}\right) \gamma_{5}\left(\not p_{1}^{\prime \prime}+m_{1}^{\prime \prime}\right)\left(\gamma_{\mu}-\gamma_{\mu} \gamma_{5}\right)\left(\not p_{1}^{\prime}+m_{1}^{\prime}\right) \gamma_{5}\left(-\not p_{2}+m_{2}\right)\right] \nonumber\\
&=&\operatorname{Tr}\left[\left(\gamma_{\nu}-\frac{1}{W_{^3A}^{\prime \prime}}\left(p_{1}^{\prime \prime}-p_{2}\right)_{\nu}\right)\left(\not p_{1}^{\prime \prime}-m_{1}^{\prime \prime}\right)\left(\gamma_{\mu} \gamma_{5}-\gamma_{\mu}\right)\left(\not p_{1}^{\prime}+m_{1}^{\prime}\right) \gamma_{5}\left(-\not p_{2}+m_{2}\right)\right], \\
S_{\mu \nu}^{D_{(s)}\; ^{1}A} &=&\left(S_{V}^{D_{(s)} \;^{1}A}-S_{A}^{D_{(s)}\; ^{1}A}\right)_{\mu \nu} \nonumber\\
&=&\operatorname{Tr}\left[\left(-\frac{1}{W_{^1A}^{\prime \prime}}\left(p_{1}^{\prime \prime}-p_{2}\right)_{\nu}\right) \gamma_{5}\left(\not p_{1}^{\prime \prime}+m_{1}^{\prime \prime}\right)\left(\gamma_{\mu}-\gamma_{\mu} \gamma_{5}\right)\left(\not p_{1}^{\prime}+m_{1}^{\prime}\right) \gamma_{5}\left(-\not p_{2}+m_{2}\right)\right] \nonumber\\
&=&\operatorname{Tr}\left[\left(-\frac{1}{W_{^1A}^{\prime \prime}}\left(p_{1}^{\prime \prime}-p_{2}\right)_{\nu}\right)\left(\not p_{1}^{\prime \prime}-m_{1}^{\prime \prime}\right)\left(\gamma_{\mu} \gamma_{5}-\gamma_{\mu}\right)\left(\not p_{1}^{\prime}+m_{1}^{\prime}\right) \gamma_{5}\left(-\not p_{2}+m_{2}\right)\right] .
\end{eqnarray}
\end{small}

The specific expressions for the traces $S_{\mu}^{D_{(s)} P}$, $S_{\mu}^{D_{(s)} S}$, $S_{\mu}^{D_{(s)} V}$, $S_{\mu \nu}^{D_{(s)} \;^3A}$ and $S_{\mu \nu}^{D_{(s)} \;^1A}$ are listed in Appendix B. In practice, we use the light-front decomposition of the Feynman loop momentum and integrate out the minus component through the contour method.
If the covariant vertex functions are not singular when performing integration, the transition amplitudes will
pick up the singularities in the anti-quark propagators. The integration then leads to
\be
N_{1}^{\prime(\prime \prime)} & \to& \hat{N}_{1}^{\prime(\prime\prime)}=x_{1}\left(M^{\prime(\prime \prime) 2}-M_{0}^{\prime(\prime \prime )2}\right), \\
H_{M}^{\prime(\prime \prime)} & \to& h_{M}^{\prime(\prime \prime)}, \\
W_{M}^{\prime \prime} & \to& w_{M}^{\prime \prime}, \\
\int \frac{d^{4} p_{1}^{\prime}}{N_{1}^{\prime} N_{1}^{\prime \prime} N_{2}} H_{D_{(s)}}^{\prime} H_{M}^{\prime \prime} S^{D_{(s)} M} & \to& -i \pi \int \frac{d x_{2} d^{2} p_{\perp}^{\prime}}{x_{2} \hat{N}_{1}^{\prime} \hat{N}_{1}^{\prime \prime}} h_{D_{(s)}}^{\prime} h_{M}^{\prime \prime} \hat{S}^{D_{(s)} M},
\en
where the subscript $M$ represents a $P, S, V$ or $A$ meson. The explicit forms of $h^{\prime\prime}_M$ and $w^{\prime\prime}_M$ are given by \cite{Cheng0310} \footnote{Here we only give the vertex functions for the final sate mesons, that
for the inital meson $D_{(s)}$ is similar with $h^{\prime\prime}_P$.}
\be
h_{P}^{\prime\prime}&=&h_{V}^{\prime\prime}=\left(M^{\prime\prime 2}-M_{0}^{\prime\prime 2}\right) \sqrt{\frac{x_{1} x_{2}}{N_{c}}} \frac{1}{\sqrt{2} \widetilde{M}_{0}^{\prime\prime}} \varphi, \\
h_{S}^{\prime\prime}&=&\sqrt{\frac{2}{3}} h_{^3A}^{\prime\prime}=\left(M^{\prime\prime 2}-M_{0}^{\prime\prime 2}\right) \sqrt{\frac{x_{1} x_{2}}{N_{c}}} \frac{1}{\sqrt{2} \widetilde{M}_{0}^{\prime}} \frac{\widetilde{M_{0}^{\prime 2}}}{2 \sqrt{3} M_{0}^{\prime}} \varphi_{p},
\en
\be
h_{^1{A}}^{\prime\prime}&=&\left(M^{\prime\prime 2}-M_{0}^{\prime\prime 2}\right) \sqrt{\frac{x_{1} x_{2}}{N_{c}}} \frac{1}{\sqrt{2} \widetilde{M}_{0}^{\prime\prime}} \varphi_{p}, \\
w_{V}^{\prime\prime}&=& M_{0}^{\prime\prime}+m_{1}^{\prime\prime}+m_{2}, \quad w_{^3A}^{\prime\prime}=\frac{\widetilde{M_{0}^{\prime\prime 2}}}{m_{1}^{\prime\prime}-m_{2}}, \quad w_{^1A}^{\prime\prime}=2,
\en
where $M^{\prime\prime}$ and $M^{\prime\prime}_0$ refer to the mass and the kinetic invariant mass of each final state meson, respectively, the latter can be expressed as the energies
 $e^{(\prime\prime)}_i (i=1,2)$ of the constituent quark and anti-quark with masses (momentum fractions) being $m^{\prime\prime}_1(x_1)$ and $m_2(x_2)$, respectively. The related definitions are given as follows
\be
M_{0}^{\prime\prime 2} &=&\left(e_{1}^{\prime\prime}+e_{2}\right)^{2}=\frac{p_{\perp}^{\prime\prime 2}+m_{1}^{\prime\prime 2}}{x_{1}}
+\frac{p_{\perp}^{2}+m_{2}^{2}}{x_{2}}, \quad \widetilde{M}_{0}^{\prime\prime}=\sqrt{M_{0}^{\prime\prime 2}-\left(m_{1}^{\prime\prime}-m_{2}\right)^{2}},\\
e_{i}^{(\prime\prime)} &=&\sqrt{m_{i}^{(\prime\prime) 2}+p_{\perp}^{\prime\prime 2}+p_{z}^{\prime\prime 2}}\;\;(i=1,2), \quad \quad x_{1}+x_{2}=1,
\en
where $p_{\perp}^{\prime \prime}=p_{\perp}^{\prime}-x_{2} q_{\perp}$,  $p_{z}^{\prime\prime}
=\frac{x_{2} M_{0}^{\prime\prime}}{2}-\frac{m_{2}^{2}+p_{\perp}^{\prime\prime 2}}{2 x_{2} M_{0}^{\prime\prime}}$. The $\varphi$ and $\varphi_p$ are the light-front momentum distribution amplitudes for the $s$-wave and $p$-wave mesons. In the present work, we use the phenomenological Gaussian-type wave function \cite{Gaussian1988}
\be
\varphi &=& \varphi\left(x_{2}, p_{\perp}^{\prime\prime}\right)=4\left(\frac{\pi}{\beta^{ 2}}\right)^{\frac{3}{4}} \sqrt{\frac{d p_{z}^{\prime\prime}}{d x_{2}}} \exp \left(-\frac{p_{z}^{\prime\prime 2}+p_{\perp}^{\prime\prime 2}}{2 \beta^{2}}\right), \\
\varphi_{p} &=& \varphi_{p}\left(x_{2}, p_{\perp}^{\prime\prime}\right)=\sqrt{\frac{2}{\beta^{ 2}}} \varphi, \quad \frac{d p_{z}^{\prime\prime}}{d x_{2}}=\frac{e_{1}^{\prime\prime} e_{2}}{x_{1} x_{2} M_{0}^{\prime\prime}}
\en
where $\beta$ is a phenomenological parameter and can be fixed by fitting the corresponding decay constant (the left Feynman diagram in Figure \ref{feyn}). The specific expressions for the $D_{(s)}\to P, S, V, A$ transition form factors obtained by matching the matrix elements $\left\langle M\left(P^{\prime \prime}\right)\left| V_{\mu}-A_\mu\right|D_{(s)}\left(P^{\prime}\right)\right\rangle$ and the corresponding BSW form factors defined in Eqs.(\ref{DtoP})-(\ref{DtoA}) are listed in Appendix B.

Equipped with these form factors, one can calculate the branching ratios of the considered semi-leptonic decays.
The differential decay rate for the decays $D_{(s)} \rightarrow P(S) \ell \nu_{\ell}$ reads
\begin{footnotesize}
\begin{eqnarray}
\frac{d \Gamma}{d q^{2}}&=&(\frac{q^2-m_l^2}{q^2})^2 \frac{\sqrt{\lambda (m_{D_{(s)}}^{2}, m_{P(S)}^{2}, q^{2})} G_{F}^{2}\left|V_{c q}\right|^{2}}{384 m_{D_{(s)}}^{3} \pi^{3}} \Bigg\{(2+\frac{m_l^2}{q^2})\lambda (m_{D_{(s)}}^{2}, m_{P(S)}^{2}, q^{2})F_{1}^2(q^{2}) \non
&&+3\frac{m_l^2}{q^2} (m_{D_{(s)}}^{2}-m_{P(S)}^{2})^2F_{0}^2(q^{2}) \Bigg\},
\label{eq:wid}
\end{eqnarray}
\end{footnotesize}
where $V_{cq}$ with $q=s$ or $d$ is the CKM matrix element, $\lambda(m^{2}_{D_{(s)}},m^{2}_{P(S)},q^{2})=(m^{2}_{D_{(s)}}+m^{2}_{P(S)}-q^{2})^{2}-4m^{2}_{D_{(s)}}m^{2}_{P(S)}$ and $m_{\ell}$ is the mass of the lepton $\ell$.

The differential decay width of the decays $D_{(s)} \rightarrow V \ell \nu_{\ell}$ is listed as
\be
\frac{d \Gamma}{d q^{2}}=\frac{d \Gamma_{L}}{d q^{2}}+\frac{d \Gamma_{+}}{d q^{2}}+\frac{d \Gamma_{-}}{d q^{2}},
\en
with
\begin{footnotesize}
\begin{eqnarray}
 \frac{d\Gamma_L}{dq^2}&=&(\frac{q^2-m_l^2}{q^2})^2\frac{ {\sqrt{\lambda(m_{D_{(s)}}^2,m_{V}^2,q^2)}} G_F^2 |V_{cq}|^2} {384m_{D_{(s)}}^3\pi^3}
 \times \frac{1}{q^2} \Bigg\{ 3 m_l^2 \lambda(m_{D_{(s)}}^2,m_{V}^2,q^2) A_0^2(q^2)  \non
 &&+\frac{m_l^2+2q^2}{4m^2_{V}}  \left|
 (m_{D_{(s)}}^2-m_{V}^2-q^2)(m_{D_{(s)}}+m_{V})A_1(q^2)-\frac{\lambda(m_{D_{(s)}}^2,m_{V}^2,q^2)}{m_{D_{(s)}}+m_{V}}A_2(q^2)\right|^2
 \Bigg\},\label{eq:AL}\;\;\;\;\;\;\\
\frac{d\Gamma_\pm}{dq^2}&=&(\frac{q^2-m_l^2}{q^2})^2\frac{ {\sqrt{\lambda(m_{D_{(s)}}^2,m_{V}^2,q^2)}} G_F^2 |V_{cq}|^2} {384m_{D_{(s)}}^3\pi^3}
  \nonumber\\
 &&\;\;\times \Bigg\{ (m_l^2+2q^2) \lambda(m_{D_{(s)}}^2,m_{V}^2,q^2)\left[\frac{V(q^2)}{m_{D_{(s)}}+m_{V}}\mp
 \frac{(m_{D_{(s)}}+m_{V})A_1(q^2)}{\sqrt{\lambda(m_{D_{(s)}}^2,m_{V}^2,q^2)}}\right]^2
 \Bigg\}.\label{eq:AL+}
\end{eqnarray}
\end{footnotesize}
Although the masses of electron and muon are small, we retain them in our calculations to systematically examine the mass effects.

The decay width for the decays $D_{(s)} \rightarrow A \ell \nu_{\ell}$ can be derived from Eqs. (\ref{eq:AL}) and (\ref{eq:AL+}) through the subsequent substitutions:
\be
\left\{V\left(q^{2}\right), A_{0}\left(q^{2}\right), A_{1}\left(q^{2}\right), A_{2}\left(q^{2}\right)\right\} &\longrightarrow & \left\{A\left(q^{2}\right), V_{0}\left(q^{2}\right), V_{1}\left(q^{2}\right), V_{2}\left(q^{2}\right)\right\} \non
m_{V} & \longrightarrow & m_{A}, \non
m_{D_{(s)}} \pm m_{V} & \longrightarrow & m_{D_{(s)}} \mp m_{A}.
\en

\section{Numerical results and discussions} \label{numer}
Within this segment, our focus is on the branching ratios of the semi-leptonic decays $D_{(s)}\to (P,V,S,A)\ell\nu_\ell$, before which the $D_{(s)}\to P, V, S, A$ transition form factors will be calculated and discussed carefully.

\subsection{Transition Form Factors}
\begin{table}[H]
\caption{The values of the input parameters\cite{PDG2024,yang,Momeni2004,BESIII2110,HFLAV,BESIII1804,Zuo0604,Momeni1903,Cheng0310,CLF1103}.}
\label{table1}
\begin{tabular*}{16.5cm}{@{\extracolsep{\fill}}l|ccccc}
  \hline\hline

&$m_{c}=1.4$&$m_{s}=0.37$&$m_{u,d}=0.25$&$m_e=0.000511$    \\[1ex]
&$m_{\mu}=0.106$&$m_{D^{0}}=1.865$&$m_{D^{\pm}}=1.87$&$m_{D_{s}^\pm}=1.968$\\[1ex]
&$m_{K^\pm}=0.494$&$m_{K^{0}}=0.498$&$m_{\pi^\pm}=0.140$&$m_{\pi^0}=0.135$\\[1ex]
&$m_{\eta}=0.548$&$m_{\eta^{'}}=0.958$&$m_{\eta_q}=0.741$&$m_{\eta_s}=0.802$\\[1ex]
&$m_{K^{*}(892)}=0.89$&$m_{\rho}=0.775$&$m_{\omega}=0.783$&$m_{\phi}=1.019$\\[1ex]
\textbf{Masses(\text{GeV})}
&$m_{f_{0q}}=1.474$&$m_{f_{0s}}=1.5$&$m_{a_0(980)}=0.98$&$m_{f_0(980)}=0.99$\\[1ex]
&$m_{K_1(1270)}=1.253$&$m_{K_1(1400)}=1.403$&$m_{K_{1A}}=1.31$&$m_{K_{1B}}=1.34$\\[1ex]
&$m_{h_{1}(1170)}=1.166$&$m_{h_{1}(1380)}=1.423$&$m_{h_{1q}}=1.242$&$m_{h_{1s}}=1.314$\\[1ex]
&$m_{f_{1}(1285)}=1.282$&$m_{f_{1}(1420)}=1.428$&$m_{f_{1q}}=1.283$&$m_{f_{1s}}=1.425$\\[1ex]
&$m_{a_1(1260)}=1.209$&$m_{b_1(1235)}=1.23$&$m_{a_0(1450)}=1.439$&$m_{K^*_0(1430)}=1.425$\\[1ex]
&$m_{f_{0}(1500)}=1.522$&$m_{f_{0}(1710)}=1.733$&$m_{f_0(1370)}=1.2-1.5$\\[1ex]
\hline
\end{tabular*}
\begin{tabular*}{16.5cm}{@{\extracolsep{\fill}}l|ccccc}
  \hline
\multirow{2}{*}{\textbf{CKM}}&$V_{cd}=0.221\pm0.004$&$V_{us}=0.2243\pm0.0008$\\[1ex]
&$V_{cs}=0.975\pm0.006$& $V_{ud}=0.97373\pm0.00031$ \\[1ex]
\hline
\end{tabular*}
\begin{tabular*}{16.5cm}{@{\extracolsep{\fill}}l|cccccc}
\hline
 &$f_{D}=0.205^{+0.005}_{-0.005}$&$f_{D_{s}}=0.251^{+0.004}_{-0.004}$  &$f_{\pi}=0.132$ &$f_{K}=0.16$\\[1ex]
\textbf{ Decay constants} &$f_{\eta_{q}}=0.141^{+0.007}_{-0.007}$  &$f_{\eta_{s}}=0.177^{+0.008}_{-0.008}$&$f_{K^{*}}=0.217^{+0.005}_{-0.005}$&$f_{\rho}=0.209^{+0.002}_{-0.002}$\\[1ex]
\;\;\;\;\;\;\textbf{(\text{GeV})} &$f_{\omega}=0.195^{+0.003}_{-0.003} $&$f_{\phi}=0.229^{+0.003}_{-0.003}$&$f_{K_{1A}}=0.250^{+0.038}_{-0.038}$&$f_{K_{1B}}=0.190^{+0.029}_{-0.029}$  \\[1ex]
&$f_{a_1(1260)}=0.238^{+0.036}_{-0.036}$\\[1ex]
\hline\hline
\end{tabular*}
\begin{tabular*}{16.5cm}{@{\extracolsep{\fill}}l|ccccc}
&$\beta_{D}=0.465^{+0.012}_{-0.012}$&$\beta_{D_{s}}=0.545^{+0.013}_{-0.013}$&$\beta_{\pi}=0.328^{+0.002}_{-0.004}$\\[1ex]
&$\beta_{K}=0.394^{+0.003}_{-0.003}$&$\beta_{\eta_{q}}=0.374^{+0.038}_{-0.045}$&$\beta_{\eta_{s}}=0.404^{+0.030}_{-0.032}$\\[1ex]
\textbf{Shape parameters(\text{GeV})}
&$\beta_{K^*}=0.279^{+0.004}_{-0.004}$&$\beta_{\rho}=0.260^{+0.001}_{-0.002}$&$\beta_{\omega}=0.252^{+0.002}_{-0.002}$\\[1ex]
&$\beta_{\phi}=0.322^{+0.002}_{-0.002}$&$\beta_{K_{1A}}=0.246^{+0.017}_{-0.018}$ &$\beta_{K_{1B}}=1.783^{+0.266}_{-0.265}$\\[1ex]
&$\beta_{a_1(1260)}=0.229^{+0.016}_{-0.017}$&\\[1ex]
\hline\hline
\end{tabular*}
\begin{tabular*}{16.5cm}{@{\extracolsep{\fill}}l|ccc}
\multirow{2}{*}{\textbf{Full widths}(\textbf{GeV})}&$\Gamma_{D^0}=(1.583\pm0.004)\times10^{-12}$&$\Gamma_{D^\pm}=(6.288\pm0.03)\times10^{-13}$\\[1ex]
&$\Gamma_{D_s^\pm}=(1.296\pm0.006)\times10^{-12}$\\[1ex]
\hline\hline
\end{tabular*}
\end{table}
In Table \ref{table1} we present all the input parameters required for this study, including the constituent quark masses, the meson masses, the Cabibbo-Kobayashi-Maskawa (CKM) matrix elements, the shape parameters $\beta$ and the full decay widths of $D_{(s)}$ mesons. Other $\beta$ values for the remaining mesons can be found in Table II of Ref. \cite{CLF1103}.

All the calculations are carried out within the $q^+ = 0$ reference frame, where the form factors can only be obtained at space-like momentum transfers $q^2 = - q^2_\bot \leq 0$. Those parameterized form factors are extrapolated from the space-like region to the time-like region by using following expression \cite{Cheng0310},
\be
F\left(q^{2}\right)=\frac{F(0)}{\left(1-q^{2} / m^{2}\right)\left[1-a\left(q^{2} / m^{2}\right)+b\left(q^{2} / m^{2}\right)^{2}\right]},\label{para}
\en
where $F(q^{2})$ denotes different form factors and $m$ represents the initial meson mass. The values of $a$ and $b$ can be obtained by performing a 3-parameter
fit to the form factors in the range -10GeV$^{2} \leq q^{2} \leq 0$. The $F(q^{2}_{max})$ is defined as the value of the form factor evaluated at the maximum value of $q^2$. The calculated values of $F(q^{2}_{max})$, $a$, and $b$ are listed in Appendix A. 
In the following, we will give the $D\to P,V,S$ and $A$ transtion form factors at maximum recoil ($q^2=0$) in Tables \ref{tablePPX}$\sim$\ref{tablePAXX}. Other theoretcial reults and data are also listed for comparsion. 
The errors mainly come from the shape parameters $\beta$ of innial and final state mesons, which are can be fixed by using the corresponding decay constants. At present, the decay constants of the pseudoscalar and vector mesons have been quite preisely determined from the data of their purely leptonic or two-photon decays, so the uncertainty caused by these prarameters are small and less than $5\%$. For the scalar and axial-vector mesons, due to the absence of experimental data, their decay constants are calculated under the heavy quark symmetry and SU(3) flavor constraiant, and uncertainties of the corresponding transition form factors can reach around $15\%$. The values of decay constants $f_{K_{1A}}$ and $f_{K_{1B}}$ were taken from the QCDSR calcualtions \cite{yang}, which would induced larger errors for the $D_{(s)}\to K_{1A}$ and $D_{(s)}\to K_{1B}$ transtion form factors. 
\subsubsection{The $D_{(s)}\to P$ transition form factors }
 For the $D_{(s)}\to \pi(K)$ transtions, the difference from different theoretical predictions and experimental measurements are small and less than $10\%$.
 For the $D_{(s)}\to \eta^{(\prime)}$ transtions, our predictions are well consistent with most other theoretical results except for individual calculations. 
It is noticed that our prediction for the form factor of the transition $D\to \eta$ is slightly larger than that of $D\to \eta^\prime$. This relation is consistent with the LCSR calculations, $F_0^{D\eta^{(\prime)}}=0.552\pm0.051(0.458\pm0.105)$ \cite{LCSROffen} and $F_0^{D \eta^{(\prime)}}=0.429^{+0.165}_{-0.141}(0.292^{+0.113}_{-0.104})$ \cite{LCSRDuplancic}. This behavior might be explained by the mixing mechanism between the flavor eigenstates $\eta_q=(u \bar{u}+d \bar{d}) / \sqrt{2}$ and $\eta_s=s\bar s$,
\begin{equation}
	\left(
	\begin{array}{c}
		\eta \\
		\eta^{'}
	\end{array}
	\right)
	=
	\left(
	\begin{array}{cc}
		\cos\theta & -\sin\theta \\
		\sin\theta & \cos\theta
	\end{array}
	\right)
	\left(
	\begin{array}{c}
		\eta_q \\
		\eta_s
	\end{array}
	\right),
\end{equation}
where the mixing angle $\theta$ has been well determined as $\theta=39.3^{\circ}\pm 1.0^{\circ}$.
 It is contrary for the case of the transitions $D_s\to \eta^{(\prime)}$, that is $F_0^{D_s\to \eta}<F_0^{D_s\to \eta^\prime}$, which is supported by BESIII experiment \cite{2019etap} and many theoretical models, including this work and  LCSR \cite{LCSROffen,LCSRDuplancic}, CCQM \cite{CCQM1904} and RQM \cite{Kang1911}. However, the LQCD calculation \cite{LQCD1406} gives an opposite tendency. Furthermore, the LQCD calcualtions for the $D_s\to \eta(D_s\to \eta^\prime)$ transtion form factors are larger (smaller) than all the avalable theoretical and experimental values. It deserves further investigation to clarify such disagreement. Precise experimental measurements of these form factors are helpful to probe the $\eta-\eta^\prime$ mixing angle and further shed light on the long-standing question on the gluonic component in these two mesons.

\begin{table}[H]
\caption{The $D_{(s)}\to K,\pi,\eta^{(\prime)}$ transtion form factors at $q^{2}=0$, together with other theoretical results and data.}
\begin{center}
\scalebox{1.2}{
\begin{tabular}{c|cccc}
\hline\hline
&$D\to K$&$D\to \pi$&$D_s\to K$&$-$\\
\hline
This work&$0.789^{+0.002}_{-0.001}$&$0.679^{+0.003}_{-0.004}$&$0.714^{+0.004}_{-0.002}$&$-$\\
\hline
CLFQM\cite{Cheng0310}&$0.78$&$0.67$&$-$&$-$\\
CLFQM\cite{CLF1103}&$0.79$&$0.66$&$0.66$&$-$\\
CCQM\cite{CCQM1904}&$0.77$&$0.63$&$0.60$&$-$\\
RQM\cite{Kang1911}&$0.716$&$0.640$&$0.674$&$-$\\
BESIII\cite{BESIII150807}&$0.737$&$0.637$&$-$&$-$\\
BABAR\cite{BABAR0704}&$0.727$&$-$&$-$&$-$\\
\hline
&$D\to \eta$&$D\to \eta^{\prime}$&$D_s\to \eta$&$D_s\to \eta^{\prime}$\\
\hline
This work&$0.558^{+0.019}_{-0.025}$&$0.456^{+0.015}_{-0.021}$&$0.490^{+0.014}_{-0.017}$&$0.599^{+0.017}_{-0.021}$\\
\hline
CLFQM\cite{CLF1103}&$0.55$&$0.45$&$0.48$&$0.59$\\
LCSR\cite{LCSROffen}&$0.552$&$0.458$&$0.432$&$0.520$\\
LCSR\cite{LCSRDuplancic}&$0.429$&$0.292$&$0.495$&$0.557$\\
CCQM\cite{CCQM1904}&$0.36$&$0.36$&$0.49$&$0.59$\\
RQM\cite{Kang1911}&$0.547$&$0.538$&$0.443$&$0.559$\\
LQCD\cite{LQCD1406}&$-$&$-$&$0.564$&$0.437$\\
BESIII\cite{AblikimBES}&$-$&$-$&$0.4576$&$0.490$\\
\hline\hline
\end{tabular}}\label{tablePPX}
\end{center}
\end{table}
\subsubsection{The $D_{(s)}\to V$ transition form factors  }
For most form factors of the transitions $D_{(s)}\to V$, the predictions among different theories are consistent with each other, which are listed in Tables \ref{tablePVX} and \ref{tablePVXX}. It is noticed that the definitions for the $D_{(s)}\to V$ transition form factors used in the CCQM approach \cite{CCQM1904} is not same with ours. For ease of comparison, the CCQM form factors have been converted to the BSW ones as shown in Tables III and IV. The overwhelming majority of theoretical predictions show that
usually $V(0)$ is the largest and $A_2(0)$ is the smallest among these $D_{(s)}\to V$ transition form factors.

\begin{table}[H]
\caption{The $D\to K^*,\rho,\omega$ transition form factors at $q^{2}=0$, together with other theoretical results.}
\begin{center}
\scalebox{1.15}{
\begin{tabular}{c|c|cccc}
\hline\hline
Transition&Reference&$V(0)$&$A_0(0)$&$A_1(0)$&$A_2(0)$\\
\hline
$D \to K^*$&This work&$0.947^{+0.010}_{-0.011}$&$0.635^{+0.002}_{-0.003}$&$0.656^{+0.004}_{-0.004}$&$0.574^{+0.000}_{-0.001}$\\
$$&CLFQM\cite{Cheng0310}&$0.94$&$0.69$&$0.65$&$0.57$\\
$$&CLFQM\cite{CLF1103}&$0.98$&$0.78$&$0.72$&$0.60$\\
$$&CCQM\cite{CCQM1904}&$0.90$&$0.77$&$0.74$&$0.68$\\
$$&RQM\cite{Kang1911}&$0.927$&$0.655$&$0.608$&$0.520$\\
$$&MS\cite{MS0008}&$1.03$&$0.76$&$0.66$&$0.49$\\
$$&QSR\cite{QSR9305}&$1.1$&$0.4$&$0.5$&$0.6$\\
$$&BSW\cite{BSW}&$1.23$&$0.73$&$0.88$&$1.15$\\
\hline
$D \to \rho$&This work&$0.845^{+0.009}_{-0.009}$&$0.546^{+0.002}_{-0.002}$&$0.571^{+0.003}_{-0.003}$&$0.485^{+0.001}_{-0.001}$\\
$$&CLFQM\cite{Cheng0310}&$0.86$&$0.64$&$0.58$&$0.48$\\
$$&CLFQM\cite{CLF1103}&$0.88$&$0.69$&$0.60$&$0.47$\\
$$&CCQM\cite{CCQM1904}&$0.76$&$0.64$&$0.61$&$0.57$\\
$$&RQM\cite{Kang1911}&$0.979$&$0.712$&$0.682$&$0.640$\\
$$&MS\cite{MS0008}&$0.90$&$0.66$&$0.59$&$0.49$\\
$$&QSR\cite{QSR9305}&$1.00$&$0.6$&$0.5$&$0.4$\\
$$&BSW\cite{BSW}&$1.23$&$0.67$&$0.78$&$0.92$\\
\hline
$D \to \omega$&This work&$0.832^{+0.009}_{-0.009}$&$0.539^{+0.002}_{-0.002}$&$0.559^{+0.003}_{-0.003}$&$0.491^{+0.000}_{-0.001}$\\
$$&CLFQM\cite{CLF1103}&$0.85$&$0.64$&$0.58$&$0.49$\\
$$&CCQM\cite{CCQM1904}&$0.72$&$0.60$&$0.58$&$0.55$\\
$$&RQM\cite{Kang1911}&$0.871$&$0.647$&$0.674$&$0.713$\\
\hline
\end{tabular}}\label{tablePVX}
\end{center}
\end{table}

\begin{table}[H]
\caption{The $D_s\to K^*, \phi$ transition form factors at $q^{2}=0$, together with other theoretical results.}
\begin{center}
\scalebox{1.15}{
\begin{tabular}{c|c|cccc}
\hline\hline
Transition&Reference&$V(0)$&$A_0(0)$&$A_1(0)$&$A_2(0)$\\
\hline
$D_s \to K^*$&This work&$0.860^{+0.010}_{-0.010}$&$0.531^{+0.002}_{-0.002}$&$0.549^{+0.003}_{-0.003}$&$0.485^{+0.001}_{-0.002}$\\
$$&CLFQM\cite{CLF1103}&$0.87$&$0.61$&$0.56$&$0.46$\\
$$&CCQM\cite{CCQM1904}&$0.80$&$0.58$&$0.58$&$0.57$\\
$$&RQM\cite{Kang1911}&$0.959$&$0.629$&$0.596$&$0.540$\\
\hline
$D_s \to \phi$&This work&$1.009^{+0.011}_{-0.011}$&$0.634^{+0.002}_{-0.002}$&$0.656^{+0.003}_{-0.004}$&$0.564^{+0.002}_{-0.002}$\\
$$&CLFQM\cite{CLF1103}&$0.98$&$0.72$&$0.69$&$0.59$\\
$$&CCQM\cite{CCQM1904}&$0.91$&$0.68$&$0.68$&$0.67$\\
$$&RQM\cite{Kang1911}&$0.999$&$0.713$&$0.643$&$0.492$\\
\hline
\end{tabular}}\label{tablePVXX}
\end{center}
\end{table}

In addition, we also consider the ratios of the form factors at maximum recoil $(q^2=0)$, which are defined as
\be
r_V=\frac{V(0)}{A_1(0)},\;\;\;\; r_2=\frac{A_2(0)}{A_1(0)}.
\en
Our results show good agreement with the experimental data given by BESIII. For example, we obtain the ratios $r_V=1.44$ and $r_2=0.86$ for the transition $D \to K^*$, which are consistent with the measurement values $r_V=1.46$ and $r_2=0.67$ \cite{BESIII181111}. The ratios for the transition $D \to \rho$ are $r_V=1.49$ and $r_2=0.86$, which are comparable with the data $r_V=1.70$ and $r_2=0.85$ \cite{BESIII180906}. As for the form factor ratios $r_V=1.56$ and $r_2=0.89$ for the transition $D_s \to K^*$, they agree with the BESIII measurements $r_V=1.67$ and $r_2=0.77$\cite{BESIII181102} again.

\subsubsection{The $D_{(s)}\to S$ transition form factors  }
There are two typical schemes for the classification to the scalar mesons with masses near and below $1.5$ GeV. The nonet mesons below 1 GeV, including $f_0(500), f_0(980), K^*(800)$ and $a_0(980)$, are usually viewed as the lowest lying $q\bar q$ states, while the nonet ones near 1.5 GeV, including $f_0(1370), f_0(1500)/f_0(1700), K^*_0(1430)$ and $a_0(1450)$, are suggested as the first excited states. Such a description is called Scheme I, and Scheme II is expressed as: The nonet mesons near 1.5 GeV are treated as $q\bar q$ ground sates, while the nonet mesons below 1 GeV are exotic states beyond the quark model such as four-quark bound states.
\begin{table}[H]
\caption{The $D_{(s)}\to a_0(980), f_0(980), a_0(1450),K^*_0(1430)$ transtion form factors, together with other theoretical results.}
\begin{center}
\scalebox{1.15}{
\begin{tabular}{c|c|c|c}
\hline\hline
&$D \to a_{0}(980)$&$D \to f_{0}(980)$&$D_s \to f_{0}(980)$\\
\hline
This work&$0.515^{+0.042}_{-0.032}$&$0.283^{+0.023}_{-0.017}$&$0.431^{+0.007}_{-0.007}$\\
\hline
LCSR\cite{Huang2102}&$0.85$&$-$&$-$\\
QCDSR\cite{Hong2409}&$0.53$&$-$&$-$\\
CCQM\cite{Soni2020}&$0.55$&$0.45$&$0.39$\\
LCSR\cite{yangmaozhi2017,wangwei1701,Colangelo}&1.75&0.32&0.30\\
LFQM\cite{kehongwei2009}&$-$&$0.22$&$0.43$\\
DR\cite{Bennich2009}&$-$&$0.22$&$0.46$\\
\hline
&$D \to a_0(1450)$&$D \to K^*_0(1430)$&$D_s \to K^*_0(1430)$\\
\hline
This work&$0.515^{+0.017}_{-0.015}$&$0.498^{+0.022}_{-0.015}$&$0.537^{+0.027}_{-0.020}$\\
\hline
CLFQM\cite{Cheng0310}&$-$&$0.48$&$-$\\
CLFQM\cite{CLF1103}&$0.51$&$0.47$&$0.55$\\
LCSR\cite{Huang2102,Yang2409,Huang2211}&0.94&0.60&0.65\\
QCDSR\cite{Hong2409,Yang0509}&0.28&0.57&0.51\\
\hline\hline
\end{tabular}}\label{tablePS}
\end{center}
\end{table}

\begin{table}[H]
\caption{The $D_{(s)}\to f_0(1370), f_0(1500)$ transition form factors.}
\begin{center}
\scalebox{1.15}{
\begin{tabular}{c|c|c|c|c}
\hline\hline
&$D \to f_{0}(1370)$&$D_s \to f_{0}(1370)$&$D \to f_{0}(1500)$&$D_s \to f_{0}(1500)$\\
\hline
This work&$0.399$&$0.269$&$0.283$&$0.441$\\
\hline\hline
\end{tabular}}\label{tablePSX}
\end{center}
\end{table}

QCD predicts the existence of glueballs and the lowest scalar glueball may be located in scalar mesons $f_0(1370), f_0(1500)$ and $f_0(1710)$.The analysis in Ref. \cite{Cheng1503} indicates that $f_0(1710)$ should have a large glueball component and $f_0(1500)$ is mainly a flavor octet, and the mixing matrix is given as
\begin{equation}
	\left(
	\begin{array}{c}
f_{0}(1370) \\
f_{0}(1500) \\
f_{0}(1710)
	\end{array}
	\right)
	=
	\left(
	\begin{array}{ccc}
0.78\pm 0.02 & 0.52\pm0.03 & -0.36\pm0.01 \\
-0.55\pm0.03 & 0.84\pm0.02 & 0.03\pm0.02 \\
0.31\pm0.01 & 0.17\pm0.01 & 0.934\pm0.004
	\end{array}
	\right)
	\left(
	\begin{array}{c}
f_{0 q} \\
f_{0 s} \\
G
	\end{array}
	\right),
\label{eq:fo}
\end{equation}
where G denotes a glueball, $f_{0 q}=(u \bar{u}+d \bar{d}) / \sqrt{2}$ and $f_{0 s}=s\bar{s}$ are pure flavor states. Certainly, some people with different viewpoint consider that $f_0(1500)$ is composed primarily of a glueball and $f_0(1710)$ is dominated by $s\bar s$ content \cite{lee}. Copmared with well established $f_0(1500)$ and $f_0(1710)$, the existence of the $f_0(1370)$ is still debated, whose mass and width have not been well determined \cite{PDG2024}.
 Below 1 GeV, two isospin-singlet scalar mesons $f_0(500)$ and $f_0(980)$ consistent of two flavor states and the mixing formula is written as
\begin{equation}
	\left(
	\begin{array}{c}
f_{0}(500) \\
f_{0}(980) \\
	\end{array}
	\right)
	=
	\left(
	\begin{array}{cc}
\cos\theta & -\sin\theta \\
\sin\theta & \cos\theta
	\end{array}
	\right)
	\left(
	\begin{array}{c}
f_{0 q} \\
f_{0 s}
	\end{array}
	\right),
\label{eq:fo1}
\end{equation}
where  $f_{0q}=(u \bar{u}+d \bar{d}) / \sqrt{2}$ and $f_{0s}=s\bar s$, the mixing angle $\theta$ is in the range $25^{\circ}<\theta<40^{\circ}$.

the difference from different theoretical predictions for the $D_{(s)}\to S$ transitions are more significant. For example, the values of the $D\to a_0(980,1450)$ transition form factors given by the LCSR are about $1.7\sim3.4$ times of our predictions, while the QCDSR result for the $D\to a_0(1450)$ transtion form factor is almost half of our prediction. 
For the form factors of the transitions $D_{(s)}\to K^*_0(1430)$ shown in Table \ref{tablePS}, our predictions are comparable with the results given by the LCSR \cite{Yang2409, Huang2211} and QCDSR \cite{Yang0509}. Obviously, our results are consistent with the previous CLFQM calculations \cite{Cheng0310,CLF1103}. The mixing formula Eq. (\ref{eq:fo}) is used in the calculations of the form factors of the transitions $D_{(s)}\to f_0(1370,1500)$, which are listed in Table \ref{tablePSX}. For the light scalar mesons with masses below 1 GeV, we calculate the corresponding form factors by treating them in the two-quark picture. Our prediction for the form factor of the transition $D\to a_0(980)$ is much smaller than the LCSR calculations \cite{Huang2102,yangmaozhi2017}, while it is consistent the QCDSR \cite{Hong2409} and CCQM results \cite{Soni2020}. One can find that the $D\to a_0(980)$ transition form factor cannot be as large as 1.75 \cite{yangmaozhi2017}, which would correspond to the branching ratios of the decays $D^0\to a^-_0(980)e\nu_e$ and $D^+\to a^0_0(980)e^+\nu_e$  being about $2\sim4$ times larger than the data \cite{BESIII2411,BESIII1803}.  It is noticed that $f_0(980)$ meson is considered as the admixture of $q\bar q$ and $s\bar s$ states characterized by a mixing angle $\theta$ defined in Eq. (\ref{eq:fo1}), which is taken as $34^\circ$ in this work. Regarding the form factors of the transitions $D_{(s)}\to f_0(980)$, the disparities among different theoretical approaches are not significant. Certainly, some of the discrepancies are induced by the uncertainties from the mixing angle.

\subsubsection{The $D_{(s)}\to A$ transition form factors}
In the quark model, axial-vector mesons are considered as the orbitally excited $q\bar q$ states, which are divided into two different types of nonnets $1 ^3P_1$ and $1 ^1P_1$ with $J^{PC}=1^{++}$ and $1^{--}$, respectively. The $1^{++}$ nonet  is composed of $a_1(1260), f_1(1285), f_1(1420)$ and $K_{1A}$, while the $1^{+-}$ one contains $b_1(1235), h_1(1170), h_1(1380)$ and $K_{1B}$.  The neutal $a_1(1260)$ and $b_1(1235)$ cannot have a mixing because of the opposite C-parities. While $K_{1A}$ and $K_{1B}$ can mix with each other to form the physical mass eigenstates $K_1(1270)$ and $K_1(1400)$ due to the mass difference of strange and light quarks,
\be
K_{1}(1270)&=&K_{1 A} \sin \theta_{K_{1}}+K_{1 B} \cos \theta_{K_{1}},  \non
K_{1}(1400)&=&K_{1 A} \cos \theta_{K_{1}}-K_{1 B} \sin \theta_{K_{1}},
\en
where the mixing angle $\theta_{K_{1}}$ serves as a fundamental parameter revealing the internal structure of axial-vertor mesons and is taken as $\theta_{K_{1}}=33^{\circ}$ or $58^\circ$. There also exists mixing between the mesons in the same nonet, that is $f_1(1285)$ and $f_1(1420)$, $h_1(1170)$ and $h_1(1380)$. The mixing scheme of these mesons can be described by their quark states
\be
f_{1}(1285)&=&f_{1 q} \sin \alpha_{f_1}+f_{1 s} \cos \alpha_{f_1}, \non
f_{1}(1420)&=&f_{1 q} \cos \alpha_{f_1}-f_{1 s} \sin \alpha_{f_1}
\en
with $\alpha_{f_1} = 69.7^{\circ}$,  $f_{1q}=(u \bar{u}+d \bar{d}) / \sqrt{2}$ and $f_{1s}=s\bar s$ \cite{Kang1707}, and
\be
h_{1}(1170)=h_{1 q} \sin \alpha_{h_1}+h_{1 s} \cos \alpha_{h_1}, \non
h_{1}(1380)=h_{1 q} \cos \alpha_{h_1}-h_{1 s} \sin \alpha_{h_1},
\en
with $\alpha_{h_1} = 86.7^{\circ}$,  $h_{1q}=(u \bar{u}+d \bar{d}) / \sqrt{2}$ and $h_{1s}=s\bar s$ \cite{Kang1707}. It is noticed that such mixing angle $\alpha$ in the flavor basis is related to the singlet-octet mixing angle $\theta$ by the relation $\alpha=\theta+54.7^\circ$. So $\alpha$ denotes the deviation from the idea mixing with $\theta=35.3^\circ$. Furthermore, the mixing
angle $\alpha$ can be related the masses of $K_{1A}$ and $K_{1B}$ states through the Gell-Mann Okubo relations \cite{axialmix}, which in turn depends on the $K_{1A}-K_{1B}$ mixing angle $\theta_{K_1}$. So calculating the form factors of $D_{(s)}$ to these aixal-vector mesons and the branching ratios of corresponding semi-leptonic decays are helpful to understand their inner structures.

In Table \ref{tablePAX}, we give the values of the $D\to a_1(1260),b_1(1235),h_{1q},f_{1q}$ and $D_s\to h_{1s},f_{1s}$ transition form factors. Other theoretical results are also listed for comparison. Obviously, our predictions are consistent with the previous CLFQM results \cite{Cheng0310,CLF1103}.
 However, significant discrepancies emerge for some form factors of the transitions $D \to a_1(1260)$ and $D \to b_1(1235)$ compared to the results from other approaches, such as LCSR \cite{Momeni1903,Huang2102}, QCDSR \cite{Zuo2016,wuxinggang2107}. Even the same models also yield significant different reults. The primary reasons lie in the differences in the input parameters and the varing orders of LCDAs twist included.
 Further theoretical and experimental investigations are required to clarify these divergences.

 It is noticed that in the calculations of the $D_{(s)}\to K_{1A}, K_{1B}$ transition form factors, the shape parameters for $K_{1A}$ and $K_{1B}$ are determined from their respective decay constants, while it is different in the previous CLFQM work \cite{CLF1103}, where both their shape parameters are estimated from the decay constant of $K_{1A}$. One can find that our new results for the $D_{(s)}\to K_{1A}$ transition form factors are still consistent with the previous CLFQM calculations \cite{CLF1103}, while the values for the $D_{(s)}\to K_{1B}$ transition form factors are very different, which are shown in Table \ref{tablePAXX}. Certainly, our results for the $D_{(s)}\to K_{1B}$ transition form factors are comparable to the magnitude values given by the three-point QCDSRs except for the form factor $A(0)$ \cite{Khosravi0812}. The striking divergence requires further investigation from both theoretical and experimental perspectives.
\begin{table}[H]
\caption{Numerical values of the $D\to a_1(1260),b_1(1235),h_{1q},f_{1q}$ and $D_s\to h_{1s},f_{1s}$ transition form factors, together with other theoretical results for comparison.}
\begin{center}
\scalebox{1.15}{
\begin{tabular}{c|c|cccc}
\hline\hline
Transition&Reference&$A(0)$&$V_0(0)$&$V_1(0)$&$V_2(0)$\\
\hline
$D \to a_1(1260)$&This work&$0.159^{+0.012}_{-0.014}$&$0.307^{+0.006}_{-0.007}$&$1.349^{+0.028}_{-0.043}$&$0.048^{+0.004}_{-0.004}$\\
$$&CLFQM\cite{Cheng0310}&$0.20$&$0.31$&$1.54$&$0.06$\\
$$&CLFQM\cite{CLF1103}&$0.19$&$0.32$&$1.51$&$0.05$\\
$$&LCSR\cite{Momeni1903}&$0.04$&$0.10$&$0.26$&$-0.02$\\
$$&LCSR\cite{Huang2102}&$0.34$&$0.24$&$2.63$&$0.34$\\
$$&QCDSR\cite{Zuo2016}&$0.314$&$-0.114$&$0.039$&$0.112$\\
$$&QCDSR\cite{wuxinggang2107}&$0.13$&$0.217$&$1.898$&$0.228$\\
\hline
$D \to b_1(1235)$&This work&$0.120^{+0.001}_{-0.002}$&$0.498^{+0.020}_{-0.021}$&$1.401^{+0.029}_{-0.037}$&$-0.104^{+0.012}_{-0.011}$\\
$$&CLFQM\cite{Cheng0310}&$0.11$&$0.49$&$1.37$&$-0.10$\\
$$&CLFQM\cite{CLF1103}&$0.12$&$0.50$&$1.39$&$-0.10$\\
$$&LCSR\cite{Momeni1903}&$-0.28$&$-0.23$&$-0.16$&$0.15$\\
$$&LCSR\cite{Huang2102}&$-0.24$&$-0.16$&$-1.78$&$-0.24$\\
\hline
$D \to h_{1q}$&This work&$0.118^{+0.001}_{-0.002}$&$0.492^{+0.019}_{-0.022}$&$1.431^{+0.030}_{-0.037}$&$-0.103^{+0.012}_{-0.011}$\\
$$&CLFQM\cite{CLF1103}&$0.11$&$0.49$&$1.42$&$-0.10$\\
\hline
$D_s \to h_{1s}$&This work&$0.115^{+0.001}_{-0.001}$&$0.551^{+0.012}_{-0.012}$&$1.511^{+0.017}_{-0.019}$&$-0.139^{+0.009}_{-0.009}$\\
$$&CLFQM\cite{CLF1103}&$0.10$&$0.57$&$1.43$&$-0.17$\\
\hline
$D \to f_{1q}$&This work&$0.176^{+0.006}_{-0.007}$&$0.339^{+0.002}_{-0.002}$&$1.736^{+0.023}_{-0.024}$&$0.046^{+0.003}_{-0.003}$\\
$$&CLFQM\cite{CLF1103}&$0.18$&$0.34$&$1.75$&$0.05$\\
\hline
$D_s \to f_{1s}$&This work&$0.162^{+0.005}_{-0.005}$&$0.292^{+0.003}_{-0.003}$&$1.710^{+0.029}_{-0.029}$&$0.028^{+0.002}_{-0.002}$\\
$$&CLFQM\cite{CLF1103}&$0.17$&$0.22$&$1.47$&$0.03$\\
\hline
\end{tabular}}\label{tablePAX}
\end{center}
\end{table}

\begin{table}[H]
\caption{Numerical values of the $D_{(s)}\to K_{1A},K_{1B}$ transition form factors, together with other theoretical results for comparison.}
\begin{center}
\scalebox{1.0}{
\begin{tabular}{c|c|cccc}
\hline\hline
Transition&Reference&$A(0)$&$V_0(0)$&$V_1(0)$&$V_2(0)$\\
\hline
$D \to K_{1A}$&This work&$0.133^{+0.012}_{-0.012}$&$0.311^{+0.004}_{-0.005}$&$1.566^{+0.036}_{-0.040}$&$0.019^{+0.003}_{-0.003}$\\
$$&CLFQM\cite{Cheng0310}&$0.98$&$0.34$&$2.02$&$0.03$\\
$$&CLFQM\cite{CLF1103}&$0.15$&$0.28$&$1.60$&$0.01$\\
$$&LCSR\cite{Momeni1903}&$0.06$&$0.11$&$0.32$&$-0.03$\\
$$&3PSR\cite{Khosravi0812}&$0.11$&$0.04$&$0.02$&$-0.01$\\
\hline
$D \to K_{1B}$&This work&$0.010^{+0.005}_{-0.003}$&$0.091^{+0.037}_{-0.025}$&$0.239^{+0.102}_{-0.066}$&$-0.036^{+0.010}_{-0.014}$\\
$$&CLFQM\cite{Cheng0310}&$0.10$&$0.44$&$1.53$&$-0.09$\\
$$&CLFQM\cite{CLF1103}&$0.10$&$0.48$&$1.58$&$-0.13$\\
$$&LCSR\cite{Momeni1903}&$-0.47$&$-0.42$&$-0.26$&$0.29$\\
$$&3PSR\cite{Khosravi0812}&$-0.75$&$-0.13$&$-0.16$&$0.08$\\
\hline
$D_s \to K_{1A}$&This work&$0.155^{+0.011}_{-0.012}$&$0.293^{+0.005}_{-0.006}$&$1.435^{+0.035}_{-0.041}$&$0.053^{+0.003}_{-0.003}$\\
$$&CLFQM\cite{CLF1103}&$0.19$&$0.29$&$1.68$&$0.07$\\
$$&LCSR\cite{Momeni1903}&$0.05$&$0.10$&$0.28$&$-0.01$\\
$$&3PSR\cite{Khosravi0812}&$0.16$&$0.03$&$0.05$&$-0.02$\\
\hline
$D_s \to K_{1B}$&This work&$0.017^{+0.007}_{-0.004}$&$0.148^{+0.057}_{-0.039}$&$0.339^{+0.135}_{-0.091}$&$-0.055^{+0.014}_{-0.020}$\\
$$&CLFQM\cite{CLF1103}&$0.10$&$0.51$&$1.50$&$-0.12$\\
$$&LCSR\cite{Momeni1903}&$-0.40$&$-0.41$&$-0.22$&$0.24$\\
$$&3PSR\cite{Khosravi0812}&$-0.84$&$-0.26$&$-0.30$&$0.14$\\
\hline
\end{tabular}}\label{tablePAXX}
\end{center}
\end{table}

To characterize the evolution of these $D_{(s)} \to P,V,S,A$ transition form factors, we plot their $q^{2}$-dependence, which is shown in Appendix A.

\subsection{Semi-leptonic decays}
Based on the form factors and helicity amplitudes provided in the preceding section, we calculate the branching ratios of the semi-leptonic decays $D_{(s)}\to (P, V, S, A)\ell\nu_{\ell}$ and list the results in Tables \ref{vtoeta} $\sim$ \ref{tableAA}, where the first uncertainty arises from the decay width of $D_{(s)}$ meosn, the second error is from the CKM matrix element, and the last one comes from the shape parameters of the initial and final state mesons. Usually, the first two errors are smaller, the sum is no more than $5\%$. The main uncertainties are stem from the shape parameters $\beta$. Espeically for the decays $D_{(s)}\to S(A)\ell\nu_\ell$, such uncertainties are in the rang of $(10\sim 20)\%$, and can reach up to $60\%$ at most. The branching ratios of the decays $D_{(s)}\to P(V)\ell\nu_\ell$ are more sensitive to different parameterization forms compared with those of the decays $D_{(s)}\to S(A)\ell\nu_\ell$. 
In the calculations of the branching ratios of the semileptonic decays $D_{(s)}\to P(V)\ell\nu_\ell$, we will adopt the three parameterization schemes above for comparison. 
The difference for those of the decays $D_{(s)}\to P(V)\ell\nu_\ell$ among these parametrizations can reach about $20\%$, even larger for individual decays.

\subsubsection{$D_{(s)}\to P\ell\nu_\ell$ decays}
\begin{table}[H]
	\caption{The branching ratios of the semi-leptonic decays $D^{+}\to \eta^{(')}\ell^+\nu_{\ell}$,$D^{+}_s\to \eta^{(')}\ell^+\nu_{\ell}$,  together with other results for comparison. }
	\begin{center}
		\scalebox{0.8}{
			\begin{tabular}{|c|c|c|c|c|}
				\hline\hline
				$$&$10^{-3}\times \mathcal{B}r(D^{+}\to \eta e^+\nu_e)$&$10^{-3}\times \mathcal{B}r(D^{+}\to \eta\mu^+\nu_{\mu})$&$10^{-4}\times \mathcal{B}r(D^{+}\to \eta^{\prime} e^+\nu_e)$&$10^{-4}\times \mathcal{B}r(D^{+}\to \eta^{\prime} \mu^+\nu_{\mu})$\\
				\hline
			This work &$1.06^{+0.01+0.04+0.07}_{-0.01-0.04-0.09}$&$0.98^{+0.00+0.04+0.07}_{-0.00-0.04-0.09}$&$1.69^{+0.01+0.04+0.12}_{-0.01-0.04-0.15}$&$1.48^{+0.01+0.05+0.10}_{-0.01-0.05-0.13}$\\
				\hline
				LCSR\cite{Zuo0604}&$0.86$&$0.84$&$-$&$-$\\
				CCQM\cite{CCQM1904}&$0.94$&$0.91$&$2.00$&$1.90$\\
				RQM\cite{Kang1911}&$1.24$&$1.21$&$2.25$&$2.11$\\
				CLFQM\cite{Kang1707}&$1.20$&$1.20$&$1.80$&$1.70$\\
				LCSR\cite{LCSR1508}&$1.42$&$-$&$1.52$&$-$\\
				PDG\cite{PDG2024}&$1.11$&$1.04$&$2.0$&$-$\\
				BESIII\cite{BESIII2506,BESIII2410}&0.975&0.908&1.79&1.92\\
				CLEO\cite{CLEO1011}&$1.14$&$-$&$2.16$&$-$\\
				\hline
				$$&$10^{-2}\times \mathcal{B}r(D^{+}_s\to \eta e^+\nu_e)$&$10^{-2}\times \mathcal{B}r(D^{+}_s\to \eta\mu^+\nu_{\mu})$&$10^{-2}\times \mathcal{B}r(D^{+}_s\to \eta^{\prime} e^+\nu_e)$&$10^{-2}\times \mathcal{B}r(D^{+}_s\to \eta^{\prime} \mu^+\nu_{\mu})$\\
				\hline
			This work& $2.14^{+0.01+0.03+0.13}_{-0.01-0.03-0.14}$&$2.00^{+0.01+0.02+0.12}_{-0.01-0.02-0.14}$&$0.88^{+0.00+0.01+0.05}_{-0.00-0.01-0.06}$&$0.78^{+0.00+0.01+0.05}_{-0.00-0.01-0.05}$\\
				\hline
				LCSR\cite{Zuo0604}&$1.27$&$1.25$&$-$&$-$\\
				CCQM\cite{CCQM1904}&$2.24$&$2.18$&$0.83$&$0.79$\\
				RQM\cite{Kang1911}&$2.37$&$2.32$&$0.87$&$0.83$\\
				CLFQM\cite{Kang1707}&$2.26$&$2.22$&$0.89$&$0.85$\\
				LCSR\cite{LCSR1508}&$2.40$&$-$&$0.79$&$-$\\
				PDG\cite{PDG2024}&$2.27$&$2.24$&$0.81$&$0.80$\\
				BESIII\cite{BESIII2306,BESIII2307}&2.255&2.235&0.810&0.801\\
				CLEO\cite{CLEO2015}&$2.28$&$-$&$0.68$&$-$\\
				\hline\hline
		\end{tabular}}\label{vtoeta}
	\end{center}
\end{table}
In Table \ref{vtoeta} and Table \ref{vtop}, we list our predictions for the branching ratios of the semi-leptonic decays $D_{(s)} \to P\ell\nu_{\ell}$ with other theoretical predictions and experimental measurements for comparison. For most of these decays, the results from different theory approaches are consistent with each other and can explain the data in many cases. We can understand the big difference between the branching fraction ratios $\frac{\mathcal{B}r(D^+\rightarrow \eta \ell^+\nu_\ell)}{\mathcal{B}r(D^+\rightarrow \eta^{\prime} \ell^+\nu_\ell)}\approx 5.6$ and $\frac{\mathcal{B}r(D^{+}_s\rightarrow \eta \ell^+\nu_\ell)}{\mathcal{B}r(D^{+}_s\rightarrow \eta^{\prime} \ell^+\nu_\ell)}\approx 2.8$ \cite{PDG2024} like this: The decays $D\to \eta\ell\nu_\ell$ have a larger final-state phase space compared with that of the decays $D\to \eta^\prime\ell\nu_\ell$. Furthermore, the form factor of transition $D\to \eta$ is larger than that of $D\to \eta^\prime$, that is $F^{D\eta}_0>F^{D\eta^\prime}_0$. So the branching ratio of the decay $D^+\rightarrow \eta \ell^+\nu_\ell$ is much larger than that of $D^+\rightarrow \eta^\prime \ell^+\nu_\ell$. While it is not the case for the decays $D_s\to \eta^{(\prime)}\ell\nu_\ell$. The discrepancy between the branching ratios of the decays $D_s\to \eta\ell\nu_\ell$ and $D_s\to \eta^{\prime}\ell\nu_\ell$ is narrowed by the form factors, since $F^{D_s\eta}_0<F^{D_s\eta^\prime}_0$.
\begin{table}[H]
\caption{The branching ratios of the semi-leptonic decays $D_{(s)}\to (\pi,K) \ell\nu_{\ell}$, together with other results for comparison. }
\begin{center}
	\scalebox{0.8}{
		\begin{tabular}{|c|c|c|c|c|}
			\hline\hline
			$$&$10^{-3}\times \mathcal{B}r(D^{0}\to \pi^-e^+\nu_e)$&$10^{-3}\times \mathcal{B}r(D^{0}\to \pi^-\mu^+\nu_{\mu})$&$10^{-3}\times \mathcal{B}r(D^{+}\to \pi^0e^+\nu_e)$&$10^{-3}\times \mathcal{B}r(D^{+}\to \pi^0\mu^+\nu_{\mu})$\\
			\hline
			This work &$2.69^{+0.01+0.10+0.03}_{-0.01-0.10-0.05}$&$2.57^{+0.01+0.09+0.03}_{-0.01-0.09-0.04}$&$3.46^{+0.02+0.13+0.03}_{-0.02-0.12-0.06}$&$3.30^{+0.02+0.12+0.03}_{-0.02-0.12-0.06}$\\
			\hline
			LCSR\cite{Zuo0604}&$2.78$&$2.75$&$3.52$&$3.49$\\
			CCQM\cite{CCQM1904}&$2.2$&$2.2$&$2.9$&$2.8$\\
			RQM\cite{Kang1911}&$2.78$&$2.74$&$3.53$&$3.47$\\
			CLFQM\cite{Kang1707}&$-$&$-$&$4.1$&$4.1$\\
			HMXT\cite{FajferP}&$2.7$&$-$&$3.3$&$-$\\
			PDG\cite{PDG2024}&$2.91$&$2.67$&$3.72$&$3.50$\\
			BESIII\cite{BESIII1802}&$-$&$2.72$&$3.50$&$-$\\
			\hline
			$$&$10^{-2}\times \mathcal{B}r(D^{0}\to K^-e^+\nu_e)$&$10^{-2}\times \mathcal{B}r(D^{0}\to K^-\mu^+\nu_{\mu})$&$10^{-2}\times \mathcal{B}r(D^{+}\to \bar{K}^0e^+\nu_e)$&$10^{-2}\times \mathcal{B}r(D^{+}\to \bar{K}^0\mu^+\nu_{\mu})$\\
			\hline
This work &$3.70^{+0.01+0.05+0.01}_{-0.01-0.05-0.02}$&$3.45^{+0.01+0.04+0.01}_{-0.01-0.04-0.02}$&$9.39^{+0.05+0.12+0.03}_{-0.04-0.12-0.05}$&$8.75^{+0.04+0.11+0.02}_{-0.04-0.11-0.04}$\\
			\hline
			LCSR\cite{Zuo0604}&$3.20$&$3.15$&$8.12$&$7.98$\\
			CCQM\cite{CCQM1904}&$3.63$&$3.53$&$9.28$&$9.02$\\
			RQM\cite{Kang1911}&$3.56$&$3.49$&$9.02$&$8.85$\\
			CLFQM\cite{Kang1707}&$-$&$-$&$10.32$&$10.07$\\
			HMXT\cite{FajferP}&$3.4$&$-$&$8.4$&$-$\\
			PDG\cite{PDG2024}&$3.538$&$3.418$&$8.81$&$8.68$\\
			BESIII\cite{BESIII2408}&$3.521$&$3.419$&$8.864$&$8.665$\\
			\hline
			$$&$10^{-3}\times \mathcal{B}r(D_s^{+}\to K^0e^+\nu_e)$&$10^{-3}\times \mathcal{B}r(D_s^{+}\to K^0\mu^+\nu_{\mu})$&$-$&$-$\\
			\hline
       This work&$2.65^{+0.01+0.10+0.03}_{-0.01-0.10-0.03}$&$2.49^{+0.02+0.09+0.03}_{-0.01-0.09-0.02}$&$-$&$-$\\
			\hline
			LCSR\cite{Zuo0604}&$3.90$&$3.83$&$-$&$-$\\
			CCQM\cite{CCQM1904}&$2.0$&$2.0$&$-$&$-$\\
			RQM\cite{Kang1911}&$4.0$&$3.9$&$-$&$-$\\
			CLFQM\cite{Kang1707}&$2.7$&$2.6$&$-$&$-$\\
			LCSR\cite{wuxinggang2405}&$3.379$&$3.351$&$-$&$-$\\
			CQM\cite{CQM}&$3.18$&$3.18$&$-$&$-$\\
			PDG\cite{PDG2024}&$2.88$&$-$&$-$&$-$\\
			CLEO\cite{CLEO1505}&$3.9$&$-$&$-$&$-$\\
			BESIII\cite{BESIII2406}&$2.98$&$-$&$-$&$-$\\
			\hline\hline
	\end{tabular}}\label{vtop}
\end{center}
\end{table}

\subsubsection{$D_{(s)}\to V\ell\nu_\ell$ decays}
In Tables \ref{table7} and  \ref{table6}, the branching ratios of the decays $D_{(s)}\to V\ell\nu_\ell$ are presented alongside other theoretical and experimental results for comparison. Several comments should be addressed:
\begin{itemize}
    \item The branching ratios of the decays $D_s\to K^{*}\ell\nu_\ell$ are consistent with most other theoretical predictions and data. However, those of the decays $D\to K^{*}\ell\nu_\ell$ are slightly larger than most other results including data.
    \item Compared with previous  CLFQM calculations \cite{Kang1707}, the disparity between our predictions and other theoretical results or experimental measurements is reduced.
    \item The branching ratios of the decays $D_{(s)}\to V\ell\nu_\ell$ range from $10^{-3}$ to $10^{-2}$. Among these decays, the channels $D^{+}\to \bar{K}^{*0}\ell^+\nu_{\ell}$ have the largest branching ratios, which is approximately 36 times of $Br(D_s^{+}\to K^{*0}\ell^+\nu_\ell)$. It is mainly because that the former have an enhancement factor $|V_{cs}/V_{cd}|^2\approx20$ compared to the later. Furthermore, the decay width of the $D$ meson $\Gamma_{D}$ is about only one half of $\Gamma_{D_s}$.
\end{itemize}
\begin{table}[H]
	\caption{The branching ratios of the semi-leptonic decays $D^{+}_{s}\to (\phi, K^{*0}) \ell^+\nu_{\ell}$, together with other results for comparison.  }
	\begin{center}
		\scalebox{0.8}{
			\begin{tabular}{|c|c|c|c|c|}
				\hline\hline
				$$&$10^{-2}\times \mathcal{B}r(D_s^{+}\to \phi e^+\nu_e)$&$10^{-2}\times \mathcal{B}r(D_s^{+}\to \phi \mu^+\nu_{\mu})$&$10^{-3}\times \mathcal{B}r(D_s^{+}\to K^{*0}e^+\nu_e)$&$10^{-3}\times \mathcal{B}r(D_s^{+}\to K^{*0}\mu^+\nu_{\mu})$\\
				\hline
			This work &$2.88^{+0.01+0.04+0.04}_{-0.01-0.04-0.04}$&$2.70^{+0.01+0.03+0.04}_{-0.01-0.03-0.04}$&$1.85^{+0.01+0.07+0.03}_{-0.01-0.07-0.03}$&$1.76^{+0.01+0.06+0.03}_{-0.01-0.06-0.02}$\\
				\hline
				LCSR\cite{Zuo0604}&$2.53$&$2.40$&$1.87$&$1.79$\\
				CCQM\cite{CCQM1904}&$3.01$&$2.85$&$1.8$&$1.7$\\
				RQM\cite{Kang1911}&$2.69$&$2.54$&$2.1$&$2.0$\\
				CLFQM\cite{Kang1707}&$3.1$&$2.9$&$1.9$&$1.9$\\
				CQM\cite{CQM}&$2.57$&$2.57$&$1.9$&$1.9$\\
				UChPT\cite{Sekihara}&$2.12$&$1.94$&$2.02$&$1.89$\\
				PDG\cite{PDG2024}&$2.34$&$2.24$&$2.05$&$-$\\
				CLEO\cite{CLEO1505}&$2.14$&$-$&$1.8$&$-$\\
				BESIII\cite{BESIII181102,BESIII230703}&$-$&$2.25$&$2.37$&$-$\\
				BABAR\cite{BABAR0807}&$2.61$&$-$&$-$&$-$\\
				\hline\hline
		\end{tabular}}\label{table7}
	\end{center}
\end{table}

\begin{itemize}
 \item Most the branching ratios of the decays $D\to V \ell\nu_\ell$ measured by BESIII (shown in Table \ref{table6}) are slightly smaller than the present theoretical predictions. Certainly, they are also smaller than the data given by CLEO.
\end{itemize}
\begin{table}[H]
	\caption{The branching ratios of the semi-leptonic decays $D^{0}\to (K^{*-},\rho^-)\ell^+\nu_{\ell}$, $D^{+}\to (\bar{K}^{*0},\rho^0,\omega) \ell^+\nu_{\ell}$, together with other results for comparison. }
	\begin{center}
		\scalebox{0.8}{
			\begin{tabular}{|c|c|c|c|c|}
				\hline\hline
				$$&$10^{-2}\times \mathcal{B}r(D^{0}\to K^{*-}e^+\nu_e)$&$10^{-2}\times \mathcal{B}r(D^{0}\to K^{*-}\mu^+\nu_{\mu})$&$10^{-2}\times \mathcal{B}r(D^{+}\to \bar{K}^{*0}e^+\nu_e)$&$10^{-2}\times \mathcal{B}r(D^{+}\to \bar{K}^{*0}\mu^+\nu_{\mu})$\\
				\hline
		This work &$2.59^{+0.01+0.03+0.03}_{-0.01-0.03-0.04}$&$2.44^{+0.01+0.03+0.03}_{-0.01-0.03-0.04}$&$6.65^{+0.03+0.08+0.10}_{-0.03-0.08-0.10}$&$6.26^{+0.03+0.08+0.43}_{-0.03-0.08-0.22}$\\
				\hline
				LCSR\cite{Zuo0604}&$2.12$&$2.01$&$5.37$&$5.10$\\
				CCQM\cite{CCQM1904}&$2.96$&$2.80$&$7.61$&$7.21$\\
				RQM\cite{Kang1911}&$1.92$&$1.82$&$4.87$&$4.62$\\
				CLFQM\cite{Kang1707}&$-$&$-$&$7.5$&$7.0$\\
				CQM\cite{CQM}&$2.46$&$-$&$6.24$&$$\\
				UChPT\cite{Sekihara}&$2.15$&$1.98$&$5.56$&$5.12$\\
				PDG\cite{PDG2024}&$2.16$&$2.06$&$5.40$&$5.27$\\
				BESIII\cite{BESIII240310}&$-$&$2.062$&$-$&$-$\\
				CLEO\cite{CLEO0506,CLEO1004}&2.16&$-$&$-$&5.27\\
				\hline
				$$&$10^{-3}\times \mathcal{B}r(D^{0}\to \rho^- e^+\nu_e)$&$10^{-3}\times \mathcal{B}r(D^{0}\to \rho^-\mu^+\nu_{\mu})$&$10^{-3}\times \mathcal{B}r(D^{+}\to \rho^0 e^+\nu_e)$&$10^{-3}\times \mathcal{B}r(D^{+}\to \rho^0\mu^+\nu_{\mu})$\\
				\hline
This work &$1.69^{+0.00+0.06+0.02}_{-0.00-0.06-0.02}$&$1.60^{+0.00+0.06+0.02}_{-0.00-0.06-0.02}$&$2.17^{+0.01+0.08+0.03}_{-0.01-0.08-0.03}$&$2.07^{+0.01+0.08+0.03}_{-0.01-0.07-0.03}$\\

				\hline
				LCSR\cite{Zuo0604}&$1.81$&$1.73$&$2.29$&$2.20$\\
				CCQM\cite{CCQM1904}&$1.62$&$1.55$&$2.09$&$2.01$\\
				RQM\cite{Kang1911}&$1.96$&$1.88$&$2.49$&$2.39$\\
				CLFQM\cite{Kang1707}&$-$&$-$&$2.3$&$2.2$\\
				UChPT\cite{Sekihara}&$1.97$&$1.84$&$2.54$&$2.37$\\
				PDG\cite{PDG2024}&$1.46$&$1.35$&$1.87$&$1.64$\\
				BESIII\cite{BESIII2409,BESIII2106,BESIII1809}&1.439&1.35&1.86&$-$\\
				CLEO\cite{CLEO}&$1.77$&$-$&$2.17$&$-$\\
				\hline
				$$&$10^{-3}\times \mathcal{B}r(D^{+}\to \omega e^+\nu_e)$&$10^{-3}\times \mathcal{B}r(D^{+}\to \omega\mu^+\nu_{\mu})$&$-$&$-$\\
				\hline
				This work&$2.02^{+0.01+0.07+0.01}_{-0.01-0.07-0.05}$&$2.00^{+0.01+0.01+0.08}_{-0.01-0.02-0.13}$&$-$&$-$\\
				\hline
				LCSR\cite{Zuo0604}&$1.93$&$1.85$&$-$&$-$\\
				CCQM\cite{CCQM1904}&$1.85$&$1.78$&$-$&$-$\\
				RQM\cite{Kang1911}&$2.17$&$2.08$&$-$&$-$\\
				CLFQM\cite{Kang1707}&$2.1$&$2.0$&$-$&$-$\\
				UChPT\cite{Sekihara}&$2.46$&$2.29$&$-$&$-$\\
				PDG\cite{PDG2024}&$1.69$&$1.77$&$-$&$-$\\
				CLEO\cite{CLEO}&$1.82$&$-$&$-$&$-$\\
				BESIII\cite{BESIII2002}&$-$&$1.77$&$-$&$-$\\
				\hline\hline
		\end{tabular}}\label{table6}
	\end{center}
\end{table}

\subsubsection{$D_{(s)}\to S\ell\nu_\ell$ decays}
\begin{table}[H]
\caption{The branching ratios of the decays $D_{(s)}\to a_0(980)\ell\nu_{\ell}, f_0(980)\ell\nu_{\ell}, f_0(500)\ell\nu_\ell$, together with other results for comparison. For the decays $D_{(s)}\to f_0(980)\ell\nu_{\ell}, f_0(500)\ell\nu_{\ell}$, their values correspond to the mixing angle $\theta=34^\circ$. For simplicity, we replace $a_0(980), f_0(980)$ and $f_0(500)$ with $a_0, f_0$ and $\sigma$, respectively, in this table.  }
\begin{center}
\scalebox{0.8}{
\begin{tabular}{|c|c|c|c|c|}
\hline\hline
$$&$10^{-4}\times \mathcal{B}r(D^{0}\to a_0^- e^+\nu_e)$&$10^{-4}\times \mathcal{B}r(D^{0}\to a_0^- \mu^+\nu_{\mu})$&$10^{-4}\times \mathcal{B}r(D^{+}\to a_0 e^+\nu_e)$&$10^{-4}\times \mathcal{B}r(D^{+}\to a_0\mu^+\nu_{\mu})$\\
\hline
This work&$1.49^{+0.00+0.05+0.13}_{-0.00-0.05-0.06}$&$1.29^{+0.00+0.05+0.11}_{-0.00-0.05-0.05}$&$1.92^{+0.01+0.07+0.24}_{-0.01-0.07-0.29}$&$1.67^{+0.01+0.06+0.22}_{-0.01-0.06-0.25}$\\
\hline
LCSR\cite{Huang2102}&$1.36$&$1.21$&$1.79$&$1.59$\\
CCQM\cite{Soni2020}&$1.68$&$1.63$&$2.18$&$2.12$\\
LCSR\cite{wuxinggang2211}&$1.57$&$1.50$&$1.98$&$1.89$\\
SU(3)\cite{wangrumin2023}&$1.18$&$0.98$&$1.55$&$1.28$\\
BESIII\cite{BESIII2411,BESIII1803}&1.08&$-$&2.08&$-$\\
\hline
$$&$10^{-5}\times \mathcal{B}r(D^{+}\to f_0 e^+\nu_e)$&$10^{-5}\times \mathcal{B}r(D^{+}\to f_0\mu^+\nu_{\mu})$&$10^{-3}\times \mathcal{B}r(D^{+}_s\to f_0 e^+\nu_e)$&$10^{-3}\times \mathcal{B}r(D^{+}_s\to f_0 \mu^+\nu_{\mu})$\\
\hline
This work&$5.73^{+0.03+0.21+0.73}_{-0.03-0.21-0.87}$&$4.95^{+0.02+0.18+0.64}_{-0.02-0.18-0.75}$&$4.00^{+0.02+0.05+0.22}_{-0.02-0.05-0.67}$&$3.52^{+0.02+0.04+0.14}_{-0.02-0.04-0.14}$\\
\hline
CCQM\cite{Soni2020}&$7.78$&$7.87$&$2.1$&$2.1$\\
LCSR\cite{Colangelo}&$-$&$-$&$2$&$-$\\
LFQM\cite{kehongwei2009}&$5.7$&$-$&$1.32$&$-$\\
SU(3)\cite{wangrumin2023}&$3.92$&$3.23$&$2.3$&$1.95$\\
BESIII\cite{BESIII1809,BESIII2303}&$\textless 5.28$&$-$&3.25&$-$\\
CLEO\cite{CLEO0907}&$3.77$&$-$&$-$&$-$\\
\hline
$$&$10^{-4}\times \mathcal{B}r(D^{+}\to \sigma e^+\nu_e)$&$10^{-4}\times \mathcal{B}r(D^{+}\to \sigma\mu^+\nu_{\mu})$&$10^{-3}\times \mathcal{B}r(D^{+}_s\to \sigma e^+\nu_e)$&$10^{-3}\times \mathcal{B}r(D^{+}_s\to \sigma \mu^+\nu_{\mu})$\\
\hline
This work&$5.36^{+0.03+0.20+0.05}_{-0.03-0.19-0.05}$&$4.91^{+0.02+0.18+0.05}_{-0.02-0.18-0.05}$&$6.55^{+0.03+0.08+0.03}_{-0.04-0.08-0.03}$&$6.04^{+0.03+0.07+0.03}_{-0.04-0.07-0.03}$\\
\hline
SU3 \cite{wangrumin2023}&$4.05$&$3.69$&$6.73$&$6.21$\\
\hline
\end{tabular}}\label{tableSs}
\end{center}
\end{table}

For the semi-leptonic $D_{(s)}$ decays to scalar mesons with masses below 1 GeV, we predict their branching ratios, which are listed in Table \ref{tableSs} with other theoretical results and available data for comparison.
$Br(D^0 \to a_0(980)^- e^+\nu_{e}, a_0(980)^-\to\eta\pi^-)$ was determined as $(0.86\pm0.17\pm0.05)\times10^{-4}$ by BESIII \cite{BESIII2411}, and $Br(a_0(980)^-\to\eta\pi^-)=0.80\pm0.38$ can be obtained from $\Gamma_{\gamma\gamma}Br(a_0(980)\to\eta\pi)=0.24\pm0.08$ keV and $\Gamma_{\gamma\gamma}=0.30\pm0.10$ keV \cite{PDG2024}. So the branching raito of the decay $D^0 \to a_0(980)^- e^+\nu_{e}$ can be estimated to be $(1.08\pm0.56)\times10^{-4}$ under the narrow width approximation. It is similar for the charged decay $D^{+}\to a_0(980) e^+\nu_e$. Although the central value of the measured branching ratio for the decay $D^0 \to a_0(980)^- e^+\nu_{e}$ is smaller than all the theoretical predictions, they can still be consistent with each other within errors.  For the decay $D^{+}\to f_0(980) e^+\nu_e$, its branching ratio predicted by several approaches, such as CCQM and LFQM, is much larger than the experimental measurement given by CLEO \cite{CLEO0907} and also larger than the upper limit set by BESIII \cite{BESIII1809}. While taking into account the uncertainties of the mixing angle and using its possible range $25^\circ \textless \theta \textless 40^\circ$, one can find that the value of $Br(D^{+}\to f_0(980) e^+\nu_e)$ falls in the range of $(3.28\sim7.58)\times10^{-5}$.

\begin{table}[H]
\caption{The branching ratios of the decays $D_{(s)}\to (K^*_0(1430),a_0(1450),f_0(1370),f_0(1500)$, $f_0(1710))\ell^+\nu_{\ell}$.
For simplicity, we replace $K^*_0(1430),a_0(1450),f_0(1370),f_0(1500)$ and $f_0(1710)$ with $K^*_0, a_0, f_0, f^\prime_0$ and $f^{\prime\prime}_0$, respectively, in this table.}
\begin{center}
\scalebox{0.8}{
\begin{tabular}{|c|c|c|c|c|}
\hline\hline
$$&$10^{-4}\times \mathcal{B}r(D^{+}\to \bar{K^*_0} e^+\nu_e)$&$10^{-4}\times \mathcal{B}r(D^{+}\to \bar{K^*_0}\mu^+\nu_{\mu})$&$10^{-5}\times \mathcal{B}r(D^{+}_s\to K^*_0 e^+\nu_e)$&$10^{-5}\times \mathcal{B}r(D^{+}_s\to K^*_0 \mu^+\nu_{\mu})$\\
\hline
This work&$3.34^{+0.02+0.04+0.14}_{-0.02-0.04-0.41}$&$2.32^{+0.01+0.03+0.07}_{-0.01-0.03-0.27}$&$2.48^{+0.01+0.09+0.09}_{-0.01-0.09-0.27}$&$1.85^{+0.01+0.07+0.08}_{-0.01-0.07-0.20}$\\
\hline
LCSR\cite{Yang2409,Huang2211}&4.59&3.52&3.6&3.1\\
QCDSR\cite{Yang0509}&$4.6$&$4.6$&$2.4$&$2.4$\\
CLFQM\cite{Kang1707}&$2.9$&$2.2$&$2.7$&$2.2$\\
\hline
$$&$10^{-6}\times \mathcal{B}r(D^{0}\to a_0^- e^+\nu_e)$&$10^{-6}\times \mathcal{B}r(D^{0}\to a_0^- \mu^+\nu_{\mu})$&$10^{-6}\times \mathcal{B}r(D^{+}\to a_0 e^+\nu_e)$&$10^{-6}\times \mathcal{B}r(D^{+}\to a_0\mu^+\nu_{\mu})$\\
\hline
This work&$5.80^{+0.01+0.21+0.46}_{-0.01-0.21-0.23}$&$3.98^{+0.01+0.15+0.30}_{-0.01-0.14-0.19}$&$8.01^{+0.04+0.29+0.31}_{-0.04-0.29-0.68}$&$5.48^{+0.03+0.20+0.21}_{-0.03-0.20-0.46}$\\
\hline
LCSR\cite{Huang2102}&$3.14$&$2.01$&$4.28$&$2.76$\\
CLFQM\cite{Kang1707}&$-$&$-$&$5.4$&$3.8$\\
\hline
&$10^{-5}\times \mathcal{B}r(D^{+}\to f_0 e^+\nu_e)$&$10^{-5}\times \mathcal{B}r(D^{+}\to f_0\mu^+\nu_{\mu})$&$10^{-4}\times \mathcal{B}r(D^{+}_s\to f_0 e^+\nu_e)$&$10^{-4}\times \mathcal{B}r(D^{+}_s\to f_0 \mu^+\nu_{\mu})$\\
\hline
This work&$0.23-3.46$&$0.15-2.79$&$0.61-5.66$&$0.43-4.71$\\
\hline
$$&$10^{-7}\times \mathcal{B}r(D^{+}\to f_0^\prime e^+\nu_e)$&$10^{-7}\times \mathcal{B}r(D^{+}\to f_0^\prime\mu^+\nu_{\mu})$&$10^{-4}\times \mathcal{B}r(D^{+}_s\to f_0^\prime e^+\nu_e)$&$10^{-5}\times \mathcal{B}r(D^{+}_s\to f_0^\prime \mu^+\nu_{\mu})$\\
\hline
This work&$8.58^{+0.04+0.31+0.69}_{-0.04-0.31-0.33}$&$5.21^{+0.02+0.19+0.42}_{-0.02-0.19-0.20}$&$1.27^{+0.01+0.02+0.06}_{-0.01-0.02-0.00}$&$8.73^{+0.04+0.11+0.33}_{-0.04-0.11-0.00}$\\
\hline
CLFQM\cite{Kang1707}&$11$&$7$&$1.5$&$12$\\
\hline
$$&$10^{-9}\times \mathcal{B}r(D^{+}\to f_0^{\prime\prime} e^+\nu_e)$&$10^{-10}\times \mathcal{B}r(D^{+}\to f_0^{\prime\prime}\mu^+\nu_{\mu})$&$10^{-7}\times \mathcal{B}r(D^{+}_s\to f_0^{\prime\prime} e^+\nu_e)$&$10^{-7}\times \mathcal{B}r(D^{+}_s\to f_0^{\prime\prime} \mu^+\nu_{\mu})$\\
\hline
This work&$3.07^{+0.01+0.11+0.28}_{-0.01-0.11-0.12}$&$1.56^{+0.01+0.06+0.16}_{-0.01-0.06-0.06}$&$2.51^{+0.01+0.03+0.14}_{-0.01-0.03-0.00}$&$1.01^{+0.00+0.01+0.06}_{-0.00-0.01-0.00}$\\
\hline
CLFQM\cite{Kang1707}&$4.7$&$2.5$&$3.4$&$1.4$\\
\hline
\end{tabular}}\label{tableS}
\end{center}
\end{table}

For the semi-leptonic $D_{(s)}$ decays to scalar mesons with masses above 1 GeV, we list their branching ratios in Table \ref{tableSs} with other theoretical results and previous CLFQM calculations for comparison.
Compared with the decays $D^{0}\to a_0(1450)^- \ell^+\nu_\ell$ and $D^{+}\to a_0(1450) \ell^+\nu_\ell$, the discrepancy between their branching ratios mainly comes from the initial meson widths, $\Gamma_{D^0}\approx2.5\Gamma_{D^+}$, as well as the factor $1/\sqrt{2}$ from the wave functions of the neutral $a_0(1450)$, that is $Br(D^{+}\to a_0(1450) \ell^+\nu_\ell)$ is about 1.25 times $Br(D^{0}\to a_0(1450)^- \ell^+\nu_\ell)$. Except for the values of the transition form factors at $q^2=0$, the variation trend of the transtion form factor dependence on $q^2$  also plays an important role to affect the predicted branching ratios of the corresponding semileptonic decays. For example, the difference between the QCDSR and CLFQM predictions for the $D\to K^*_0(1430)$ transition form factors is about 10 percent, while the discripancy from the predicted branching ratios of the decays $D^+\to \bar K^*_0(1430)\mu^+\nu_\mu$ can be a fator of two.
Calculating the branching ratios of the decays $D^{+}\to (f_0(1370), f_0(1500), f_0(1710)) \ell^+\nu_\ell$ from quark component contributoins, one can find that there exsits a distinct hierarchical relationship among them,
\be
Br(D^{+}\to f_0(1370) \ell^+\nu_\ell)>>Br(D^{+}\to f_0(1500)\ell^+\nu_\ell)>>Br(D^{+}\to f_0(1710)\ell^+\nu_\ell).
\en
If such a relationship can be confirmed in future experiments, it is helpful to probe their internal structures and verify the corresponding mixing mechanisms. For the branching ratios of the decays $D^+_{(s)} \to f_0(1370)\ell^+\nu_\ell$, we calculated the ranges by taking the mass of $f_0(1370)$ from 1.2 GeV to  1.5 GeV and find that their branching raitos are sensitive to the $f_0(1370)$ meson mass.

\subsubsection{$D_{(s)}\to A\ell\nu_\ell$ decays}

\begin{table}[H]
	\caption{The branching ratios of the semi-leptonic decays $D_{(s)}\to A\ell\nu_{\ell}$ with $A=a_1, b_1, f_1^{(\prime)}, h_1^{(\prime)}$.}
	\begin{center}
		\scalebox{0.7}{
			\begin{tabular}{|c|c|c|c|c|}
				\hline\hline
				$$&$10^{-5}\times \mathcal{B}r(D^{0}\to a_1^- e^+\nu_e)$&$10^{-5}\times \mathcal{B}r(D^{0}\to a_1^-\mu^+\nu_{\mu})$&$10^{-5}\times \mathcal{B}r(D^{+}\to a_1 e^+\nu_e)$&$10^{-5}\times \mathcal{B}r(D^{+}\to a_1 \mu^+\nu_{\mu})$\\
				\hline
				This work&$4.39^{+0.01+0.16+0.28}_{-0.01-0.16-0.25}$&$3.88^{+0.01+0.14+0.23}_{-0.01-0.14-0.22}$&$5.80^{+0.03+0.21+0.37}_{-0.03-0.21-0.32}$&$5.13^{+0.02+0.19+0.32}_{-0.02-0.19-0.30}$\\
				\hline
				LCSR\cite{Momeni1903}&$2.85$&$-$&$3.76$&$-$\\
				LCSR\cite{Huang2102}&$6.90$&$6.27$&$9.38$&$8.52$\\
				QCDSR\cite{Zuo2016}&$1.11$&$-$&$1.47$&$-$\\
				QCDSR\cite{wuxinggang2107}&$5.261$&$4.732$&$6.673$&$6.002$\\
				CLFQM\cite{zhaozhenxing2015}&$4.1$&$3.6$&$-$&$-$\\
				\hline
				$$&$10^{-5}\times \mathcal{B}r(D^{0}\to b_1^- e^+\nu_e)$&$10^{-5}\times \mathcal{B}r(D^{0}\to b_1^-\mu^+\nu_{\mu})$&$10^{-5}\times \mathcal{B}r(D^{+}\to b_1 e^+\nu_e)$&$10^{-5}\times \mathcal{B}r(D^{+}\to b_1 \mu^+\nu_{\mu})$\\
				\hline
				This work&$5.52^{+0.01+0.20+0.36}_{-0.01-0.20-0.35}$&$4.74^{+0.01+0.17+0.31}_{-0.01-0.17-0.29}$&$7.29^{+0.03+0.27+0.47}_{-0.03-0.26-0.46}$&$6.27^{+0.03+0.23+0.41}_{-0.03-0.22-0.38}$\\
				\hline
				LCSR\cite{Momeni1903}&$1.88$&$-$&$2.47$&$-$\\
				LCSR\cite{Huang2102}&$4.85$&$4.40$&$6.58$&$6.00$\\
				CLFQM\cite{Kang1707}&$-$&$-$&$7.4$&$6.4$\\
				\hline
				$$&$10^{-5}\times \mathcal{B}r(D^{+}\to f_1 e^+\nu_e)$&$10^{-5}\times \mathcal{B}r(D^{+}\to f_1\mu^+\nu_{\mu})$&$10^{-4}\times \mathcal{B}r(D^{+}_s\to f_1 e^+\nu_e)$&$10^{-4}\times \mathcal{B}r(D^{+}_s\to f_1 \mu^+\nu_{\mu})$\\
				\hline
				This work&$3.75^{+0.02+0.14+0.09}_{-0.02-0.13-0.08}$&$3.21^{+0.02+0.12+0.07}_{-0.02-0.12-0.08}$&$2.58^{+0.01+0.03+0.10}_{-0.01-0.03-0.09}$&$2.22^{+0.01+0.03+0.09}_{-0.01-0.02-0.07}$\\
				\hline
				CLFQM\cite{Kang1707}&$3.7$&$3.2$&$0.6-3.6$&$0.52-3.06$\\
				QCDSR\cite{Zuo2016}&$1.07$&$-$&$-$&$-$\\
				\hline
				$$&$10^{-7}\times \mathcal{B}r(D^{+}\to f^\prime_1 e^+\nu_e)$&$10^{-7}\times \mathcal{B}r(D^{+}\to f^\prime_1\mu^+\nu_{\mu})$&$10^{-4}\times \mathcal{B}r(D^{+}_s\to f^\prime_1 e^+\nu_e)$&$10^{-4}\times \mathcal{B}r(D^{+}_s\to f^\prime_1 \mu^+\nu_{\mu})$\\
				\hline
				This work&$6.95^{+0.03+0.25+0.15}_{-0.03-0.25-0.19}$&$5.70^{+0.03+0.21+0.11}_{-0.03-0.20-0.16}$&$3.49^{+0.02+0.04+0.16}_{-0.02-0.04-0.13}$&$2.89^{+0.01+0.04+0.12}_{-0.01-0.04-0.09}$\\
				CLFQM\cite{Kang1707}&$2-14$&$2-12$&$2.5$&$2.1$\\
				QCDSR\cite{Zuo2016}&$1.22$&$-$&$-$&$-$\\
				\hline
				$$&$10^{-4}\times \mathcal{B}r(D^{+}\to h_1 e^+\nu_e)$&$10^{-4}\times \mathcal{B}r(D^{+}\to h_1\mu^+\nu_{\mu})$&$10^{-5}\times \mathcal{B}r(D^{+}_s\to h_1 e^+\nu_e)$&$10^{-5}\times \mathcal{B}r(D^{+}_s\to h_1 \mu^+\nu_{\mu})$\\
				\hline
				This work&$1.45^{+0.01+0.05+0.14}_{-0.01-0.05-0.17}$&$1.25^{+0.01+0.05+0.28}_{-0.01-0.05-0.17}$&$2.45^{+0.01+0.03+0.06}_{-0.01-0.03-0.06}$&$2.16^{+0.01+0.03+0.05}_{-0.01-0.03-0.05}$\\
				\hline
				CLFQM\cite{Kang1707}&$1.4$&$$1.22&$0-19.7$&$0-17.4$\\
				\hline
				$$&$10^{-8}\times \mathcal{B}r(D^{+}\to h^\prime_1 e^+\nu_e)$&$10^{-8}\times \mathcal{B}r(D^{+}\to h^\prime_1\mu^+\nu_{\mu})$&$10^{-4}\times \mathcal{B}r(D^{+}_s\to h^\prime_1 e^+\nu_e)$&$10^{-4}\times \mathcal{B}r(D^{+}_s\to h^\prime_1 \mu^+\nu_{\mu})$\\
				\hline
				This work&$2.25^{+0.01+0.08+0.03}_{-0.01-0.08-0.06}$&$2.75^{+0.01+0.10+0.19}_{-0.01-0.10-0.21}$&$5.39^{+0.03+0.07+0.14}_{-0.02-0.07-0.16}$&$6.77^{+0.03+0.08+0.18}_{-0.03-0.08-0.22}$\\
				\hline
				CLFQM\cite{Kang1707}&$0-20$&$0-20$&$6.4$&$5.4$\\
				\hline
		\end{tabular}}\label{tableAAA}
	\end{center}
\end{table}

For the semi-leptonic decays $D_{(s)} \to A \ell\nu_\ell$, we employ the following notations for the final-state mesons: $a_1(1260)$ and $b_1(1235)$ are simplified as $a_1$ and $b_1$,  the lighter mesons $f_1(1285),h_1(1170),K_1(1270)$ and the heavier mesons $f_1(1420), h_1(1380), K_1(1400)$ are denoted as $f_1,h_1,K_1$ and $f_1^\prime, h_1^\prime, K_1^\prime$, respectivly.

We present the branching ratios of the decays $D_{(s)} \to (a_1, b_1, f^{(\prime)}_1, h^{(\prime)}_1)\ell\nu_{\ell}$ in Table \ref{tableAAA}, where the results from other theoretical approaches are also listed for comparison. One can find significant discrepancies exist in the different theoretical results even when the same method was used, such as between
 Refs. \cite{Momeni1903} and \cite{Huang2102},  Refs. \cite{Zuo2016} and \cite{wuxinggang2107}, which need further clarification.  Among the form factors $A(q^2), V_0(q^2), V_1(q^2)$ and $V_2(q^2)$, the branching ratios of the decays $D_{(s)}\to A\ell\nu_\ell$ are most sensitive to the variation of $V_1(q^2)$. If there is a significant difference for the values of $V_1(q^2)$ between two calcualtions, the corresponding branching ratios will also vary noticeably. Certainly, our predictions are consistent with the previous CLFQM calculations except for those of some decays with only a range provided.

Compared with the formulae for the branching ratios between the decays $D^{+}\to h_1 \ell^+\nu_\ell$ and $D^{+}\to h^\prime_1 \ell^+\nu_\ell$, the difference is mainly from the mixing factors $\sin(\alpha_{h_1})$ and $\cos(\alpha_{h_1})$, that is $Br(D^{+}\to h_1 \ell^+\nu_\ell)\propto\sin^2(86.7^\circ)=0.997$ and $Br(D^{+}\to h^\prime_1 \ell^+\nu_\ell)\propto\cos^2(86.7^\circ)=0.03$, so $Br(D^{+}\to h_1 \ell^+\nu_\ell)$ is much larger than $Br(D^{+}\to h^\prime_1 \ell^+\nu_\ell)$. Additionally, the final-state phase space widens the gap between their branching ratios. While it is contrary for the decays $D^{+}_s\to h^{(\prime)}_1 \ell^+\nu_\ell$, that is $Br(D^{+}_s\to h^{\prime}_1 \ell^+\nu_\ell)$ is much larger than $Br(D^{+}_s\to h_1 \ell^+\nu_\ell)$ since the decays $D^{+}_s\to h^{\prime}_1 \ell^+\nu_\ell$ are connected with the large factor $\sin^2(86.7^\circ)$, while the final-state phase space effect partially compensates for the discrepancies induced by the mixing angle. The decays $D^{+}_{(s)}\to f^{(\prime)}_1 \ell^+\nu_\ell$ exhibit a similar pattern, but the difference in their branching ratios is less pronounced than that in those of the decays $D^{+}_{(s)}\to h^{(\prime)}_1 \ell^+\nu_\ell$, primarily due to the mixing angle.

\begin{figure}[H]
	\vspace{0.60cm}
	\centering
	\subfigure[]{\includegraphics[width=0.28\textwidth]{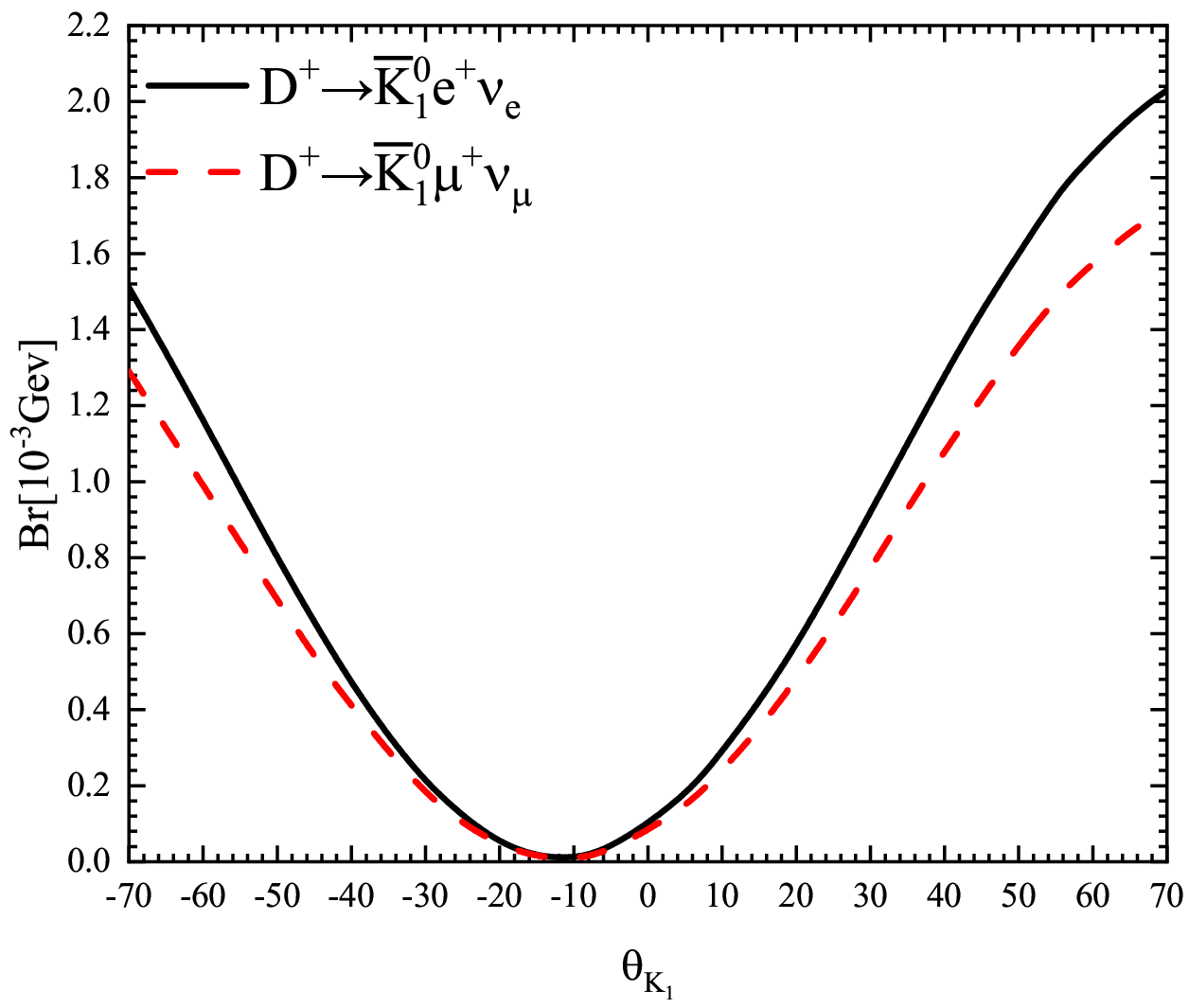}\quad}
	\subfigure[]{\includegraphics[width=0.28\textwidth]{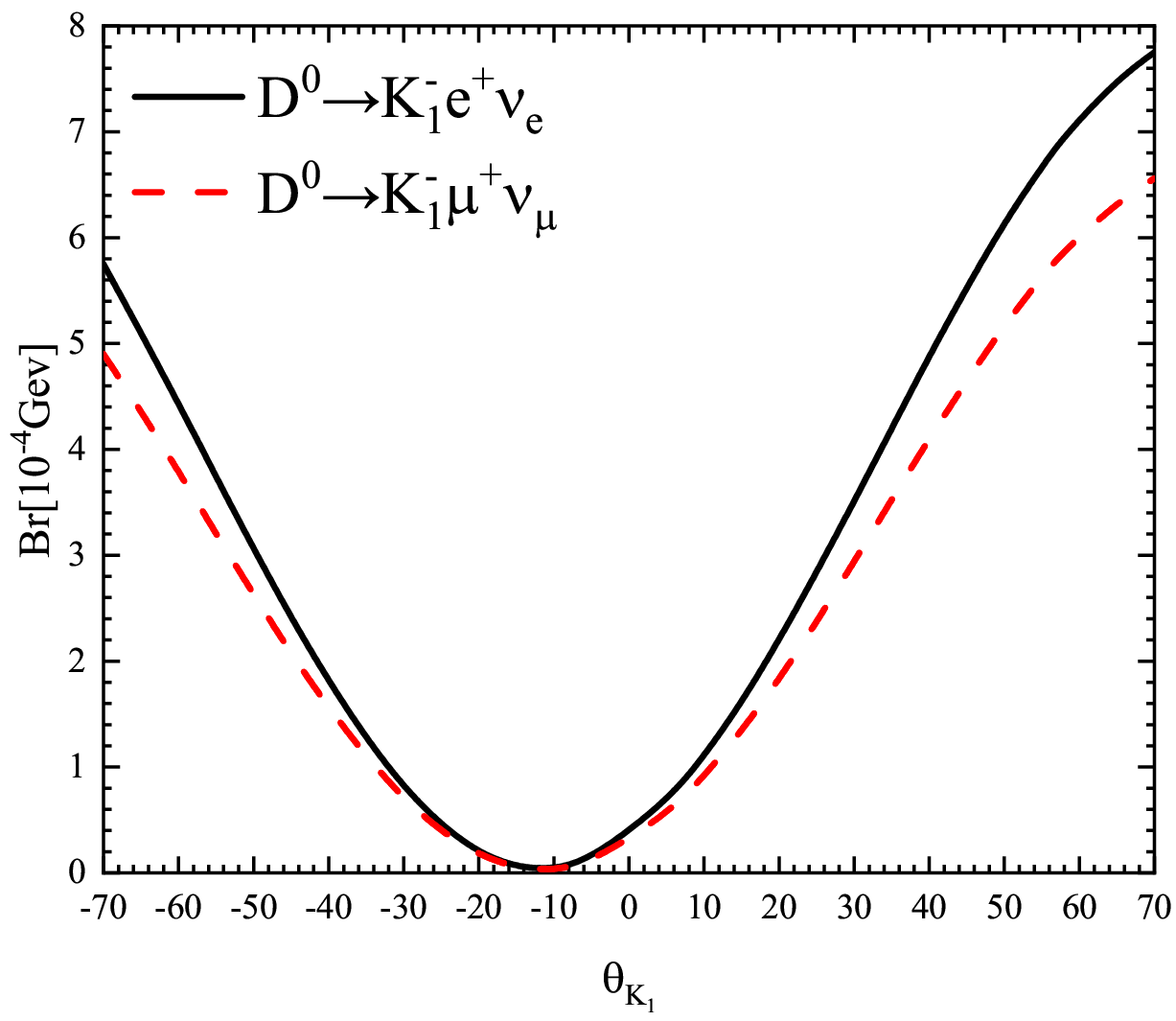}\quad}\\
	\subfigure[]{\includegraphics[width=0.28\textwidth]{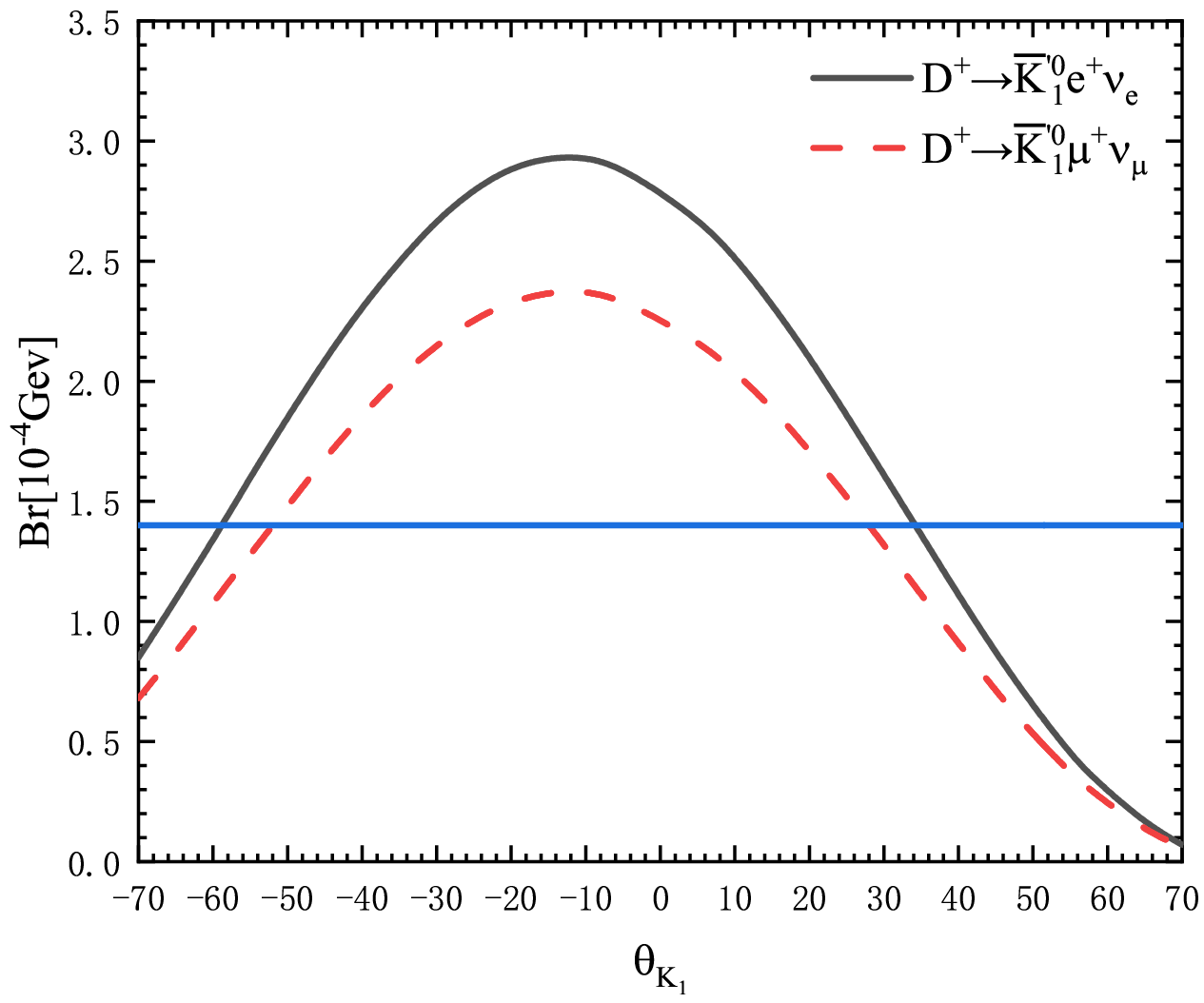}\quad}
	\subfigure[]{\includegraphics[width=0.28\textwidth]{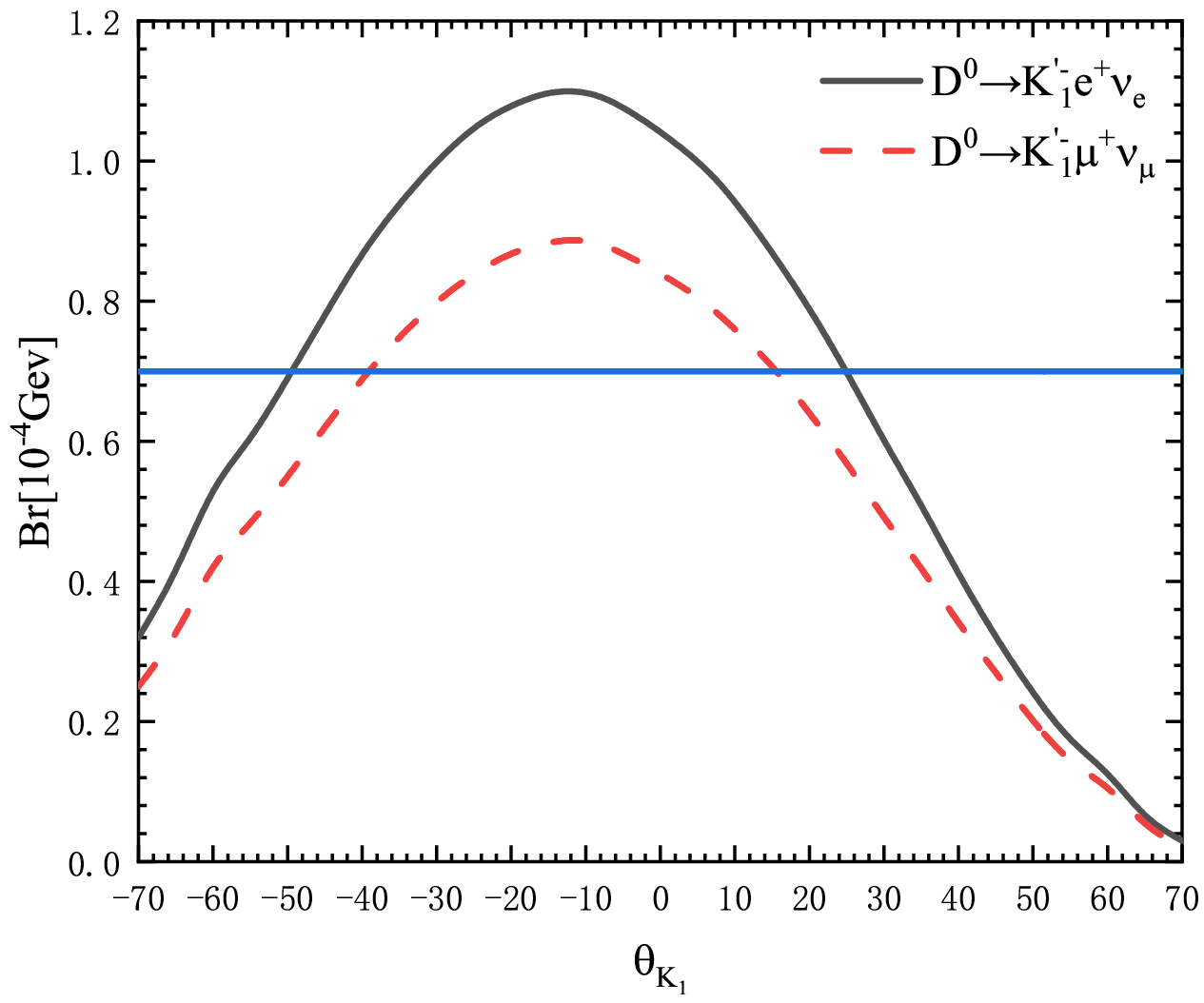}\quad}
	\caption{The branching ratios of the decays $D^+\to \bar K_1(1270)^0\ell^+\nu_\ell$ (a) and $D^0\to K_1(1270)^-\ell^+\nu_\ell$ (b),
		$D^+\to \bar K_1(1400)^0\ell^+\nu_\ell$ (c) and $D^0\to K_1(1400)^-\ell^+\nu_\ell$ (d). The horizontal lines in (c) and (d) represent the upper limits of decays $D^+\to \bar K_1(1400)^0 e^+\nu_e$ and $D^0\to K_1(1400)^-e^+\nu_e$, respectively.}\label{1270}
\end{figure}

In Table \ref{tableAA}, we present the branching ratios of the decays $D_{(s)} \to K^{(\prime)}_1\ell\nu_{\ell}$. Compared to the previous CLFQM results in Ref. \cite{Kang1707}, the discrepancies mainly stem from different choices for the shape parameter of the $K_{1B}$ state as mentioned in the previous section.
Given the present measured value $(2.27\pm0.15)\times10^{-3}$ for the branching raito of the decay $D^{+}\to \bar{K_1^0} e^+\nu_e$ \cite{BESIII250302}, the large mixing angle $58^\circ$ appears to be favored, while if taken the previous data $Br(D^{+}\to \bar{K_1^0} e^+\nu_e)=(1.29^{+0.46}_{-0.42})\times10^{-3}$ \cite{BESIII240319}, the small mixing angle $33^\circ$ is favored, as shown in Figure \ref{1270}(a). Furthermore, in Ref. \cite{BESIII250302}, the upper limits of the branching ratios of the decays $D^+\to \bar K_1^\prime e^+\nu_e$ and $D^0\to K_1^{\prime-} e^+\nu_e$ were set as $1.4\times10^{-4}$ and $0.7\times10^{-4}$, which are shown in Figure \ref{1270}(c) and Figure \ref{1270}(d), respectively. One can find that the branching ratios of the decays
$D^{+}\to \bar K_1^{\prime0} e^+\nu_e$ and $D^{0}\to \bar K_1^{\prime-} e^+\nu_e$ corresponding to $\theta_{K_1}=33^\circ$ are very close to the upper limits. So here we take the mixing angle $\theta_{K_1}=58^\circ$ to calculate the branching ratios of these decays $D\to K_1^{(\prime)}\ell\nu_{\ell}$, which are listed  in Table \ref{tableAA}. It is obvious that the branching ratios of the decays $D_{(s)}\to K_1\ell\nu_{\ell}$ are much larger than those of the corresponding decays $D_{(s)}\to K^\prime_1\ell\nu_{\ell}$.
\begin{table}[H]
	\caption{The branching ratios of the semi-leptonic decays $D_{(s)}\to K_1^{(\prime)} \ell^+\nu_{\ell}$.}
	\begin{center}
		\scalebox{0.7}{
			\begin{tabular}{|c|c|c|c|c|}
				\hline\hline
				&$10^{-3}\times \mathcal{B}r(D^{0}\to K_1^- e^+\nu_e)$&$10^{-3}\times \mathcal{B}r(D^{0}\to K_1^-\mu^+\nu_{\mu})$&$10^{-5}\times \mathcal{B}r(D^{0}\to K_1^{\prime-} e^+\nu_e)$&$10^{-5}\times \mathcal{B}r(D^{0}\to K_1^{\prime-} \mu^+\nu_{\mu})$\\
				\hline
				This work&$0.70^{+0.00+0.01+0.13}_{-0.00-0.01-0.18}$&$0.59^{+0.00+0.01+0.10}_{-0.00-0.01-0.15}$&$1.30^{+0.01+0.02+0.79}_{-0.01-0.02-0.95}$&$1.08^{+0.01+0.02+0.63}_{-0.01-0.02-0.74}$\\
				\hline
				CLFQM\cite{Bian2105}&$1.56$&$1.3$&$3.7$&$2.9$\\
				BESIII\cite{BESIII250302,BESIII250203}&$1.02$&$0.78$&$-$&$-$\\
				\hline
				&$10^{-3}\times \mathcal{B}r(D^{+}\to \bar{K_1^0} e^+\nu_e)$&$10^{-3}\times \mathcal{B}r(D^{+}\to \bar{K_1^0}\mu^+\nu_{\mu})$&$10^{-4}\times \mathcal{B}r(D^{+}_s\to K_1^0 e^+\nu_e)$&$10^{-4}\times \mathcal{B}r(D^{+}_s\to K_1^0 \mu^+\nu_{\mu})$\\
				\hline
				This work&$1.82^{+0.02+0.02+0.34}_{-0.02-0.02-0.48}$&$1.54^{+0.02+0.02+0.29}_{-0.02-0.02-0.38}$&$1.03^{+0.01+0.04+0.22}_{-0.01-0.04-0.24}$&$0.91^{+0.01+0.03+0.20}_{-0.01-0.03-0.21}$\\
				\hline
				CLFQM\cite{Kang1707}&$3.2$&$2.6$&$1.7$&$1.5$\\
				CLFQM\cite{Bian2105}&$3.97$&$3.31$&$-$&$-$\\
				BESIII\cite{BESIII250302,BESIII250203,BESIII2309}&$2.27^a$&$2.36$&$\textless 4.1$&$-$\\
				\hline
				&$10^{-4}\times \mathcal{B}r(D^+ \to \bar K_1^{\prime0} e^+\nu_e)$&$10^{-4}\times \mathcal{B}r(D^{+}\to \bar K_1^{\prime0}\mu^+\nu_{\mu})$&$10^{-6}\times \mathcal{B}r(D^{+}_s\to  K_1^{\prime0} e^+\nu_e)$&$10^{-6}\times \mathcal{B}r(D^{+}_s\to K_1^{\prime0} \mu^+\nu_{\mu})$\\
				\hline
				This work&$0.35^{+0.01+0.00+0.14}_{-0.01-0.00-0.26}$&$0.29^{+0.01+0.00+0.12}_{-0.01-0.00-0.21}$&$1.39^{+0.04+0.05+0.43}_{-0.04-0.05-0.67}$&$1.22^{+0.03+0.05+0.44}_{-0.03-0.05-0.57}$\\
				\hline
				CLFQM\cite{Kang1707}&$0.05-0.20$&$0.04-0.17$&$0.5-1.4$&$0.5-1.2$\\
				CLFQM\cite{Bian2105}&$0.94$&$0.74$&$-$&$-$\\
				\hline
		\end{tabular}}\label{tableAA}
	\end{center}
	{\tiny $^a$  The measured value for the decay $D^+ \to \bar K_1^{0} e^+\nu_e$ given by BESIII was $1.29^{+0.46}_{-0.42}$ \cite{BESIII240319} last year.}
\end{table}

\section{Summary}\label{sum}
The form factors of the transitions $D_{(s)}\to P,S, V, A $ are systematically investigated within the CLFQM framework.
For the form factors of the transitions $D_{(s)}\to P,V$, our predictions are consistent well with those given by other theoretical approaches, such as LCSR, CCQM, QCDSR and RQM. It proves that these calculations are reliable. However, significant discrepancies emerge in some $D_{(s)}\to S,A$ transitions, such as $D \to a_0(980),a_0(1450)$ and $D_{(s)} \to K_{1B}$, which highlight uncertainties in the internal structures of the scalar and axial-vector mesons. Then the branching ratios of the semi-leptonic decays $D_{(s)}\to (P,S, V, A)\ell\nu_\ell$ are obtained by using the corresponding form factors. Obviously, pronounced differences occur in the decays $D_{(s)}\to (S, A)\ell\nu_\ell$ among different theoretical predictions. It is interesting for the decays $D \to a_0(980)e\nu_e$, the predictions from several approaches, such as CLFQM, LCSR, CCQM and SU(3) flavor symmetry, can explain the data given by BESIII within errors under the two-quark picture for $a_0(980)$. Most theoretical results for the branching ratio of the decay $D^+\to f_0(980)e^+\nu_e$ are larger than the data given by CLEO and BESIII, while accounting for the mixing angle uncertainties and using its possible range $25^\circ \textless \theta \textless 40^\circ$, the two-quark picture for $f_0(980)$ can still account for the measurements. More precise experimental data for the decays $D\to K_1(1270)\ell\nu_\ell$ and $D\to K_1(1400)\ell\nu_\ell$ are needed to further determine the mixing angle $\theta_{K_1}$ between $K_1(1270)$ and $K_1(1400)$.
 In conclusion, systematic comparisons between theoretical predictions and experimental data over a wide range of semi-leptonic $D_{(s)}$ decay processes are essential to test SM and search for NP in charm physics.

\section*{Acknowledgment}
This work is partly supported by the National Natural Science
Foundation of China under Grant No. 11347030 and the Natural Science Foundation of Henan
Province under grant no. 232300420116, 252300421302.

\appendix
\section{Numerical results and $q^2$-dependence of the transition form factors}
\begin{table}[H]
\caption{Form factors of the transitions $D_{(s)}\to K,\pi, \eta_q, \eta_s$ obained in the CLFQM are fitted to Eq. (\ref{para}). The uncertainties are from the decay constants of initial and final state mesons.}
\begin{center}
\scalebox{1.15}{
\begin{tabular}{|c|cc|cc|cc|}
\hline\hline
$$&$D\to K$&$$&$D \to \pi$&$$&$D_s\to K$&$$\\
\hline
$$&$F_1$&$F_0$&$F_1$&$F_0$&$F_1$&$F_0$\\
\hline
$F(0)$&$0.789^{+0.002}_{-0.001}$&$0.789^{+0.003}_{-0.002}$&$0.679^{+0.004}_{-0.006}$&$0.679^{+0.005}_{-0.005}$&$0.714^{+0.004}_{-0.004}$&$0.714^{+0.005}_{-0.003}$\\
$F(q^2_{max})$&$1.391^{+0.002}_{-0.002}$&$1.333^{+0.001}_{-0.002}$&$3.229^{+0.059}_{-0.031}$&$3.325^{+0.004}_{-0.007}$&$1.277^{+0.014}_{-0.019}$&$1.286^{+0.017}_{-0.017}$\\
$a$&$0.326^{+0.001}_{-0.001}$&$0.478^{+0.002}_{-0.002}$&$0.331^{+0.003}_{-0.003}$&$0.410^{+0.002}_{-0.001}$&$0.316^{+0.003}_{-0.002}$&$0.385^{-0.003}_{-0.003}$\\
$b$&$0.187^{+0.004}_{-0.004}$&$0.097^{+0.001}_{-0.001}$&$0.243^{+0.008}_{-0.007}$&$0.093^{+0.001}_{-0.001}$&$0.261^{+0.008}_{-0.008}$&$0.109^{+0.001}_{-0.001}$\\
\hline\hline
$$&$D\to \eta_q$&$$&$D_s \to \eta_s$&$$&$$&$$\\
\hline
$$&$F_1$&$F_0$&$F_1$&$F_0$&$$&$$\\
\hline
$F(0)$&$0.721^{+0.021}_{-0.033}$&$0.721^{+0.021}_{-0.033}$&$0.773^{+0.022}_{-0.027}$&$0.773^{+0.022}_{-0.027}$&$$&$$\\
$F(q^2_{max})$&$0.989^{+0.046}_{-0.031}$&$0.968^{+0.032}_{-0.031}$&$1.039^{+0.032}_{-0.042}$&$1.022^{+0.025}_{-0.038}$&$$&$$\\
$a$&$0.323^{+0.002}_{-0.006}$&$0.432^{+0.027}_{-0.029}$&$0.320^{+0.002}_{-0.003}$&$0.422^{+0.018}_{-0.007}$&$$&$$\\
$b$&$0.207^{+0.029}_{-0.026}$&$0.092^{+0.003}_{-0.002}$&$0.242^{+0.18}_{-0.010}$&$0.107^{+0.003}_{-0.002}$&$$&$$\\
\hline\hline
\end{tabular}}\label{tablePPX1}
\end{center}
\end{table}

\begin{table}[H]
\caption{Same as Table \ref{tablePPX1} except for the transitions $D_{(s)}\to K^*,\rho,\omega,\phi$.}
\begin{center}
\scalebox{1.15}{
\begin{tabular}{c|c|cccc}
\hline\hline
Transition&&$V(0)$&$A_0(0)$&$A_1(0)$&$A_2(0)$\\
\hline
$D \to K^*$&$F(0)$&$0.947^{+0.010}_{-0.011}$&$0.635^{+0.002}_{-0.003}$&$0.656^{+0.004}_{-0.004}$&$0.574^{+0.000}_{-0.001}$\\
\hline
$$&$F(q^2_{max})$&$1.174^{+0.010}_{-0.010}$&$0.781^{+0.000}_{-0.013}$&$0.812^{+0.013}_{-0.001}$&$0.708^{+0.001}_{-0.001}$\\
$$&$a$&$0.348^{+0.004}_{-0.004}$&$0.380^{+0.002}_{-0.002}$&$0.410^{+0.003}_{-0.003}$&$0.349^{+0.002}_{-0.002}$\\
$$&$b$&$0.271^{+0.011}_{-0.010}$&$0.099^{+0.001}_{-0.001}$&$0.097^{+0.001}_{-0.001}$&$0.190^{+0.005}_{-0.004}$\\
\hline
$D \to \rho$&$F(0)$&$0.845^{+0.009}_{-0.009}$&$0.546^{+0.002}_{-0.002}$&$0.571^{+0.003}_{-0.003}$&$0.485^{+0.001}_{-0.001}$\\
\hline
$$&$F(q^2_{max})$&$1.099^{+0.003}_{-0.009}$&$0.735^{+0.014}_{-0.000}$&$0.754^{+0.001}_{-0.000}$&$0.638^{+0.001}_{-0.014}$\\
$$&$a$&$0.363^{+0.006}_{-0.007}$&$0.371^{+0.002}_{-0.002}$&$0.404^{+0.003}_{-0.003}$&$0.353^{+0.003}_{-0.003}$\\
$$&$b$&$0.327^{+0.015}_{-0.014}$&$0.102^{+0.002}_{-0.001}$&$0.099^{+0.001}_{-0.001}$&$0.210^{+0.006}_{-0.005}$\\
\hline
$D \to \omega$&$F(0)$&$0.832^{+0.009}_{-0.009}$&$0.539^{+0.002}_{-0.002}$&$0.559^{+0.003}_{-0.003}$&$0.491^{+0.000}_{-0.001}$\\
\hline
$$&$F(q^2_{max})$&$1.077^{+0.009}_{-0.009}$&$0.718^{+0.000}_{-0.000}$&$0.745^{+0.008}_{-0.007}$&$0.646^{+0.001}_{-0.001}$\\
$$&$a$&$0.370^{+0.007}_{-0.008}$&$0.366^{+0.002}_{-0.002}$&$0.370^{+0.033}_{-0.028}$&$0.354^{+0.003}_{-0.003}$\\
$$&$b$&$0.341^{+0.016}_{-0.015}$&$0.104^{+0.002}_{-0.002}$&$0.084^{+0.015}_{-0.017}$&$0.222^{+0.006}_{-0.006}$\\
\hline
$D_s \to \phi$&$F(0)$&$1.009^{+0.011}_{-0.011}$&$0.634^{+0.002}_{-0.002}$&$0.656^{+0.003}_{-0.004}$&$0.564^{+0.002}_{-0.002}$\\
\hline
$$&$F(q^2_{max})$&$1.200^{+0.010}_{-0.010}$&$0.749^{+0.000}_{-0.012}$&$0.780^{+0.012}_{-0.000}$&$0.668^{+0.001}_{-0.012}$\\
$$&$a$&$0.339^{+0.004}_{-0.005}$&$0.386^{+0.002}_{-0.002}$&$0.414^{+0.003}_{-0.003}$&$0.347^{+0.002}_{-0.002}$\\
$$&$b$&$0.324^{+0.014}_{-0.013}$&$0.117^{+0.002}_{-0.002}$&$0.114^{+0.001}_{-0.001}$&$0.219^{+0.005}_{-0.005}$\\
\hline
$D_s \to K^*$&$F(0)$&$0.860^{+0.010}_{-0.010}$&$0.531^{+0.002}_{-0.002}$&$0.549^{+0.003}_{-0.003}$&$0.485^{+0.001}_{-0.002}$\\
\hline
$$&$F(q^2_{max})$&$1.098^{+0.008}_{-0.008}$&$0.698^{+0.001}_{-0.002}$&$0.362^{+0.001}_{-0.002}$&$0.626^{+0.001}_{-0.014}$\\
$$&$a$&$0.375^{+0.009}_{-0.010}$&$0.369^{+0.004}_{-0.008}$&$0.396^{+0.003}_{-0.003}$&$0.354^{+0.004}_{-0.004}$\\
$$&$b$&$0.438^{+0.023}_{-0.021}$&$0.130^{+0.001}_{-0.005}$&$0.121^{+0.002}_{-0.002}$&$0.279^{+0.009}_{-0.009}$\\
\hline
\end{tabular}}\label{tablePVX1}
\end{center}
\end{table}

\begin{table}[H]
\caption{Same as Table \ref{tablePPX1} except for the transitions $D_{(s)}\to a_0(980),a_0(1450),K^*_0(1430),$ $f_{0q},f_{0s}$.}
\begin{center}
\scalebox{1.15}{
\begin{tabular}{|c|cc|cc|}
\hline\hline
&$D \to a_0(980)$& &$D \to a_0(1450)$&\\
\hline
&$F_1$&$F_0$&$F_1$&$F_0$\\
\hline
$F(0)$&$0.515^{+0.042}_{-0.032}$&$0.515^{+0.042}_{-0.031}$&$0.515^{+0.017}_{-0.015}$&$0.515^{+0.018}_{-0.015}$\\
$F(q^2_{max})$&$0.608^{+0.049}_{-0.047}$&$0.584^{+0.043}_{-0.043}$&$0.529^{+0.015}_{-0.021}$&$0.534^{+0.020}_{-0.010}$\\
$a$&$0.326^{+0.006}_{-0.002}$&$0.542^{+0.019}_{-0.021}$&$0.326^{+0.002}_{-0.003}$&$0.536^{+0.027}_{-0.023}$\\
$b$&$0.203^{+0.019}_{-0.022}$&$0.107^{+0.004}_{-0.004}$&$0.203^{+0.011}_{-0.010}$&$0.099^{+0.006}_{-0.007}$\\
\hline\hline
&$D \to K^*_0(1430)$&$$&$D_s \to K^*_0(1430)$& \\
\hline
&$F_1$&$F_0$&$F_1$&$F_0$\\
\hline
$F(0)$&$0.498^{+0.022}_{-0.015}$&$0.498^{+0.022}_{-0.016}$&$0.537^{+0.027}_{-0.020}$&$0.537^{+0.028}_{-0.020}$\\
$F(q^2_{max})$&$0.520^{+0.031}_{-0.015}$&$0.512^{+0.020}_{-0.014}$&$0.570^{+0.032}_{-0.015}$&$0.560^{+0.030}_{-0.022}$\\
$a$&$0.332^{+0.001}_{-0.001}$&$0.605^{+0.022}_{-0.086}$&$0.324^{+0.003}_{-0.006}$&$0.569^{+0.032}_{-0.024}$\\
$b$&$0.179^{+0.010}_{-0.008}$&$0.115^{+0.006}_{-0.006}$&$0.253^{+0.022}_{-0.017}$&$0.122^{+0.006}_{-0.008}$\\
\hline\hline
&$D \to f_{0q}$&&$D_s \to f_{0s}$& \\
\hline
&$F_1$&$F_0$&$F_1$&$F_0$\\
\hline
$F(0)$&$0.514^{+0.017}_{-0.015}$&$0.514^{+0.018}_{-0.015}$&$0.523^{+0.009}_{-0.008}$&$0.523^{+0.008}_{-0.008}$\\
$F(q^2_{max})$&$0.526^{+0.015}_{-0.021}$&$0.533^{+0.022}_{-0.014}$&$0.541^{+0.010}_{-0.010}$&$0.531^{+0.010}_{-0.010}$\\
$a$&$0.326^{+0.002}_{-0.003}$&$0.536^{+0.030}_{-0.025}$&$0.329^{+0.001}_{-0.001}$&$0.650^{+0.011}_{-0.010}$\\
$b$&$0.203^{+0.011}_{-0.010}$&$0.099^{+0.007}_{-0.008}$&$0.218^{+0.001}_{-0.001}$&$0.139^{+0.003}_{-0.003}$\\
\hline\hline
\end{tabular}}\label{tablePSX1}
\end{center}
\end{table}

\begin{table}[H]
\caption{Same as Table \ref{tablePPX1} except for the transitions $D\to a_1(1260),b_1(1235),h_{1q},f_{1q}$ and $D_s\to h_{1s},f_{1s}$.}
\begin{center}
\scalebox{1.15}{
\begin{tabular}{c|c|cccc}
\hline\hline
Transition&&$A(0)$&$V_0(0)$&$V_1(0)$&$V_2(0)$\\
\hline
$D \to a_1(1260)$&$F(0)$&$0.159^{+0.012}_{-0.014}$&$0.307^{+0.006}_{-0.007}$&$1.349^{+0.028}_{-0.043}$&$0.048^{+0.004}_{-0.004}$\\
\hline
$$&$F(q^2_{max})$&$0.175^{+0.015}_{-0.015}$&$0.337^{+0.015}_{-0.000}$&$1.445^{+0.044}_{-0.038}$&$0.053^{+0.011}_{-0.000}$\\
$$&$a$&$0.332^{+0.009}_{-0.014}$&$0.381^{+0.007}_{-0.011}$&$0.530^{+0.011}_{-0.009}$&$0.631^{+0.016}_{-0.017}$\\
$$&$b$&$0.196^{+0.015}_{-0.012}$&$0.276^{+0.021}_{-0.016}$&$0.110^{+0.001}_{-0.001}$&$0.175^{+0.008}_{-0.006}$\\
\hline
$D \to b_1(1235)$&$F(0)$&$0.120^{+0.001}_{-0.002}$&$0.498^{+0.020}_{-0.021}$&$1.401^{+0.029}_{-0.037}$&$-0.104^{+0.012}_{-0.011}$\\
\hline
$$&$F(q^2_{max})$&$0.124^{+0.000}_{-0.000}$&$0.529^{+0.024}_{-0.022}$&$1.508^{+0.033}_{-0.033}$&$-0.098^{+0.015}_{-0.001}$\\
$$&$a$&$0.767^{+0.021}_{-0.022}$&$0.582^{+0.012}_{-0.014}$&$0.426^{+0.005}_{-0.005}$&$1.148^{+0.076}_{-0.087}$\\
$$&$b$&$0.208^{+0.008}_{-0.007}$&$0.119^{+0.000}_{-0.002}$&$0.108^{+0.002}_{-0.002}$&$1.937^{+0.210}_{-0.181}$\\
\hline
$D \to h_{1q}$&$F(0)$&$0.118^{+0.001}_{-0.002}$&$0.492^{+0.019}_{-0.022}$&$1.431^{+0.030}_{-0.037}$&$-0.103^{+0.012}_{-0.011}$\\
\hline
$$&$F(q^2_{max})$&$0.124^{+0.000}_{-0.000}$&$0.517^{+0.022}_{-0.024}$&$1.536^{+0.044}_{-0.033}$&$-0.098^{+0.015}_{-0.011}$\\
$$&$a$&$0.767^{+0.021}_{-0.021}$&$0.584^{+0.011}_{-0.014}$&$0.426^{+0.005}_{-0.005}$&$1.148^{+0.076}_{-0.087}$\\
$$&$b$&$0.208^{+0.008}_{-0.007}$&$0.119^{+0.002}_{-0.002}$&$0.108^{+0.002}_{-0.002}$&$1.937^{+0.210}_{-0.181}$\\
\hline
$D_s \to h_{1s}$&$F(0)$&$0.115^{+0.001}_{-0.001}$&$0.551^{+0.012}_{-0.012}$&$1.511^{+0.017}_{-0.019}$&$-0.139^{+0.009}_{-0.009}$\\
\hline
$$&$F(q^2_{max})$&$0.121^{+0.010}_{-0.000}$&$0.561^{+0.015}_{-0.015}$&$1.557^{+0.023}_{-0.046}$&$-0.137^{+0.010}_{-0.010}$\\
$$&$a$&$0.789^{+0.011}_{-0.011}$&$0.542^{+0.008}_{-0.008}$&$0.441^{+0.004}_{-0.004}$&$0.764^{+0.019}_{-0.020}$\\
$$&$b$&$0.226^{+0.003}_{-0.003}$&$0.119^{+0.000}_{-0.000}$&$0.122^{+0.002}_{-0.001}$&$1.263^{+0.047}_{-0.044}$\\
\hline
$D \to f_{1q}$&$F(0)$&$0.176^{+0.006}_{-0.007}$&$0.339^{+0.002}_{-0.002}$&$1.736^{+0.023}_{-0.024}$&$0.046^{+0.003}_{-0.003}$\\
\hline
$$&$F(q^2_{max})$&$0.193^{+0.015}_{-0.000}$&$0.363^{+0.000}_{-0.000}$&$1.829^{+0.032}_{-0.023}$&$0.052^{+0.011}_{-0.000}$\\
$$&$a$&$0.308^{+0.001}_{-0.002}$&$0.361^{+0.004}_{-0.005}$&$0.550^{+0.007}_{-0.006}$&$0.609^{+0.026}_{-0.013}$\\
$$&$b$&$0.165^{+0.007}_{-0.007}$&$0.247^{+0.013}_{-0.012}$&$0.110^{+0.001}_{-0.001}$&$0.153^{+0.003}_{-0.003}$\\
\hline
$D_s \to f_{1s}$&$F(0)$&$0.162^{+0.005}_{-0.005}$&$0.292^{+0.003}_{-0.003}$&$1.710^{+0.029}_{-0.029}$&$0.028^{+0.002}_{-0.002}$\\
\hline
$$&$F(q^2_{max})$&$0.169^{+0.000}_{-0.011}$&$0.304^{+0.000}_{-0.000}$&$1.763^{+0.033}_{-0.033}$&$0.031^{+0.000}_{-0.000}$\\
$$&$a$&$0.319^{+0.001}_{-0.002}$&$0.389^{+0.007}_{-0.009}$&$0.642^{+0.004}_{-0.004}$&$0.766^{+0.017}_{-0.016}$\\
$$&$b$&$0.176^{+0.007}_{-0.006}$&$0.334^{+0.024}_{-0.022}$&$0.141^{+0.001}_{-0.001}$&$0.202^{+0.002}_{-0.003}$\\
\hline\hline
\end{tabular}}\label{tablePVA1}
\end{center}
\end{table}

\begin{table}[H]
\caption{Same as Table \ref{tablePPX1} except for the transitions $D_{(s)}\to K_{1A},K_{1B}$.}
\begin{center}
\scalebox{1.15}{
\begin{tabular}{c|c|cccc}
\hline\hline
Transition&&$A(0)$&$V_0(0)$&$V_1(0)$&$V_2(0)$\\
\hline
$D \to K_{1A}$&$F(0)$&$0.133^{+0.012}_{-0.012}$&$0.311^{+0.004}_{-0.005}$&$1.566^{+0.036}_{-0.040}$&$0.019^{+0.003}_{-0.003}$\\
\hline
$$&$F(q^2_{max})$&$0.138^{+0.000}_{-0.000}$&$0.330^{+0.000}_{-0.011}$&$1.641^{+0.046}_{-0.032}$&$0.020^{+0.000}_{-0.000}$\\
$$&$a$&$0.344^{+0.016}_{-0.011}$&$0.350^{+0.009}_{-0.008}$&$0.555^{+0.009}_{-0.011}$&$0.801^{+0.023}_{-0.025}$\\
$$&$b$&$0.164^{+0.008}_{-0.009}$&$0.115^{+0.005}_{-0.006}$&$0.112^{+0.001}_{-0.001}$&$0.197^{+0.008}_{-0.007}$\\
\hline
$D \to K_{1B}$&$F(0)$&$0.010^{+0.005}_{-0.003}$&$0.091^{+0.037}_{-0.025}$&$0.239^{+0.102}_{-0.066}$&$-0.036^{+0.010}_{-0.014}$\\
\hline
$$&$F(q^2_{max})$&$0.010^{+0.000}_{-0.000}$&$0.093^{+0.021}_{-0.042}$&$0.249^{+0.066}_{-0.023}$&$-0.042^{+0.011}_{-0.015}$\\
$$&$a$&$0.500^{+0.003}_{-0.004}$&$0.664^{+0.005}_{-0.001}$&$0.591^{+0.003}_{-0.004}$&$0.449^{+0.008}_{-0.012}$\\
$$&$b$&$0.100^{+0.001}_{-0.001}$&$0.116^{+0.001}_{-0.002}$&$0.105^{+0.000}_{-0.000}$&$0.096^{+0.002}_{-0.004}$\\
\hline
$D_s \to K_{1A}$&$F(0)$&$0.155^{+0.011}_{-0.012}$&$0.293^{+0.005}_{-0.006}$&$1.435^{+0.035}_{-0.041}$&$0.053^{+0.003}_{-0.003}$\\
\hline
$$&$F(q^2_{max})$&$0.173^{+0.022}_{-0.011}$&$0.295^{+0.003}_{-0.016}$&$1.513^{+0.036}_{-0.043}$&$0.052^{+0.000}_{-0.011}$\\
$$&$a$&$0.340^{+0.011}_{-0.017}$&$0.831^{+0.056}_{-0.083}$&$0.556^{+0.017}_{-0.014}$&$0.625^{+0.013}_{-0.014}$\\
$$&$b$&$0.264^{+0.027}_{-0.023}$&$1.201^{+0.183}_{-0.118}$&$0.134^{+0.001}_{-0.001}$&$0.211^{+0.013}_{-0.010}$\\
\hline
$D_s \to K_{1B}$&$F(0)$&$0.017^{+0.007}_{-0.004}$&$0.148^{+0.057}_{-0.039}$&$0.339^{+0.135}_{-0.091}$&$-0.055^{+0.014}_{-0.020}$\\
\hline
$$&$F(q^2_{max})$&$0.021^{+0.000}_{-0.000}$&$0.174^{+0.062}_{-0.047}$&$0.355^{+0.095}_{-0.028}$&$-0.064^{+0.021}_{-0.024}$\\
$$&$a$&$0.524^{+0.004}_{-0.005}$&$0.512^{+0.003}_{-0.003}$&$0.615^{+0.005}_{-0.008}$&$0.457^{+0.011}_{-0.017}$\\
$$&$b$&$0.115^{+0.001}_{-0.002}$&$0.338^{+0.003}_{-0.002}$&$0.119^{+0.001}_{-0.000}$&$0.110^{+0.003}_{-0.005}$\\
\hline\hline
\end{tabular}}\label{tablePVA2}
\end{center}
\end{table}

\begin{figure}[H]
\vspace{0.60cm}
  \centering
  \subfigure[]{\includegraphics[width=0.28\textwidth]{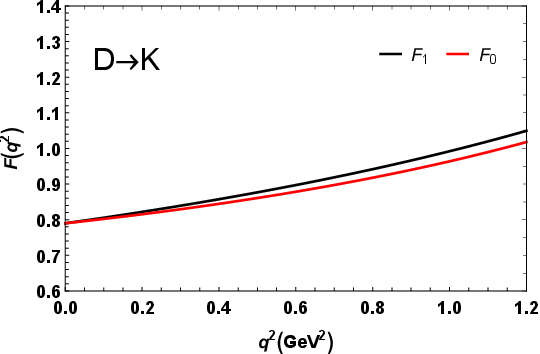}\quad}
  \subfigure[]{\includegraphics[width=0.28\textwidth]{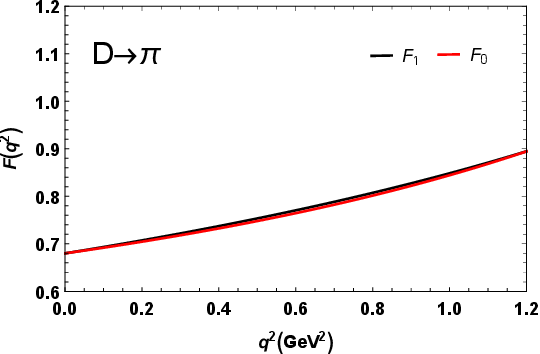}\quad}
  \subfigure[]{\includegraphics[width=0.28\textwidth]{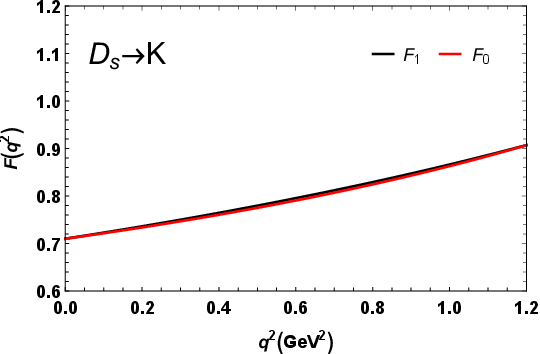}\quad}
  \subfigure[]{\includegraphics[width=0.28\textwidth]{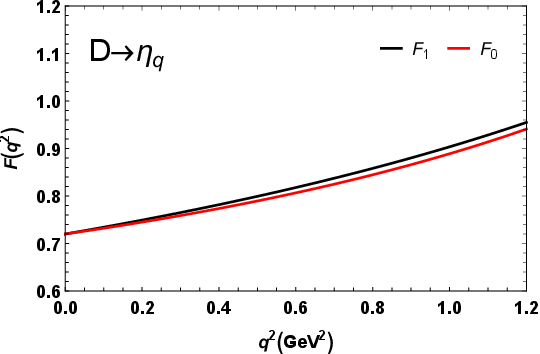}\quad}
  \subfigure[]{\includegraphics[width=0.28\textwidth]{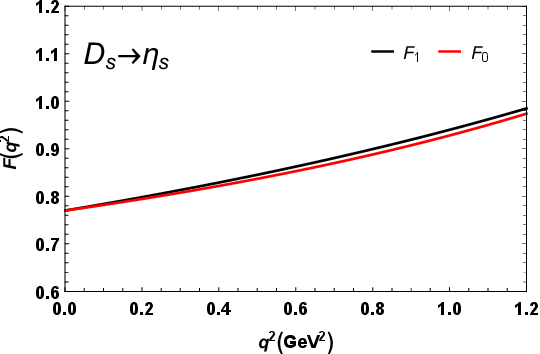}\quad}
\caption{Form factors $F_0(q^2)$ and $ F_{1}(q^2)$ of the transitions $D_{(s)} \rightarrow  K, \pi, \eta_q, \eta_s$ .}\label{figP2}
\end{figure}

\begin{figure}[H]
\vspace{0.60cm}
  \centering
  \subfigure[]{\includegraphics[width=0.28\textwidth]{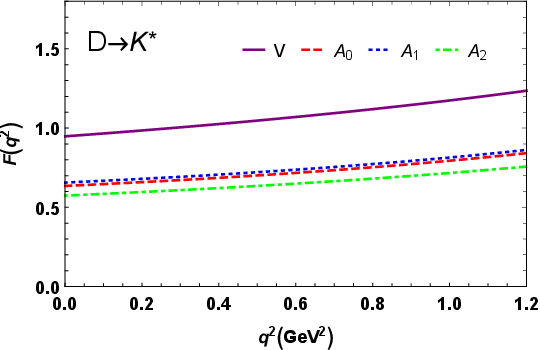}\quad}
  \subfigure[]{\includegraphics[width=0.28\textwidth]{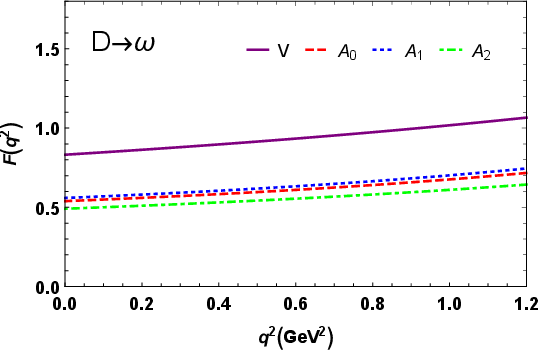}\quad}
  \subfigure[]{\includegraphics[width=0.28\textwidth]{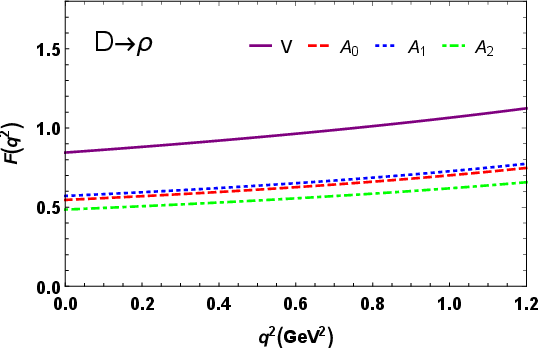}\quad}
  \subfigure[]{\includegraphics[width=0.28\textwidth]{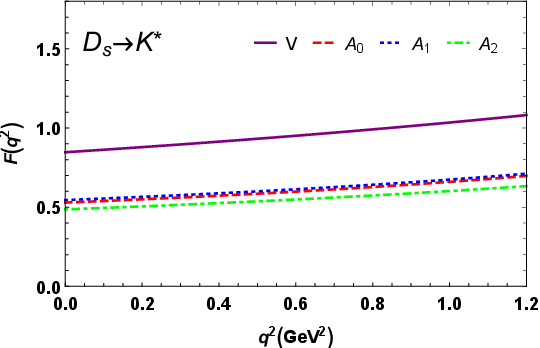}\quad}
  \subfigure[]{\includegraphics[width=0.28\textwidth]{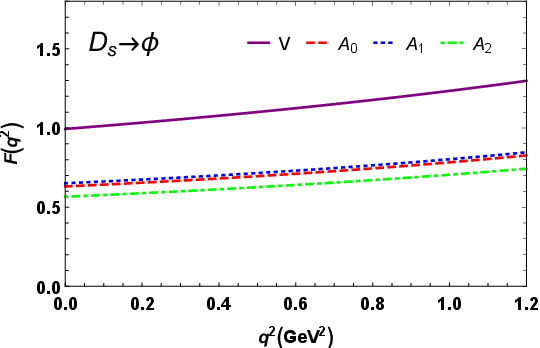}\quad}
\caption{Form factors $V(q^2)$, $A_{0}(q^2)$, $A_{1}(q^2)$ and $A_{2}(q^2)$ of the transitions $D_{(s)} \rightarrow  K^*, \omega, \rho, \phi$ .}\label{figV1}
\end{figure}

\begin{figure}[H]
\vspace{0.60cm}
  \centering
  \subfigure[]{\includegraphics[width=0.28\textwidth]{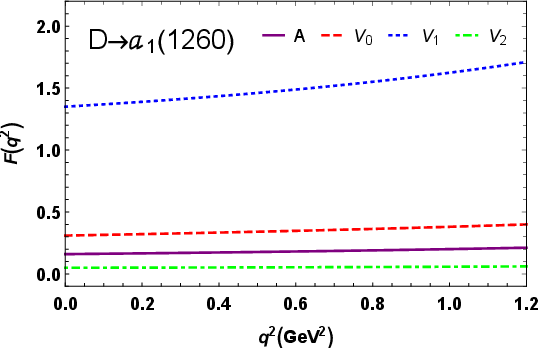}\quad}
  \subfigure[]{\includegraphics[width=0.28\textwidth]{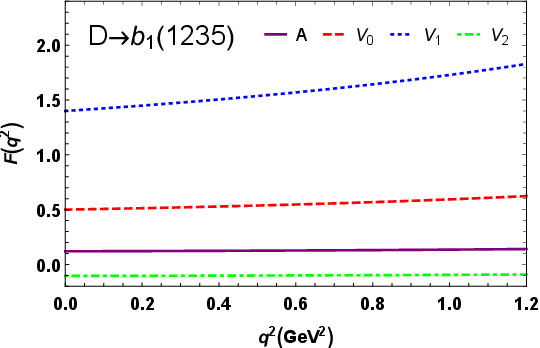}\quad}
  \subfigure[]{\includegraphics[width=0.28\textwidth]{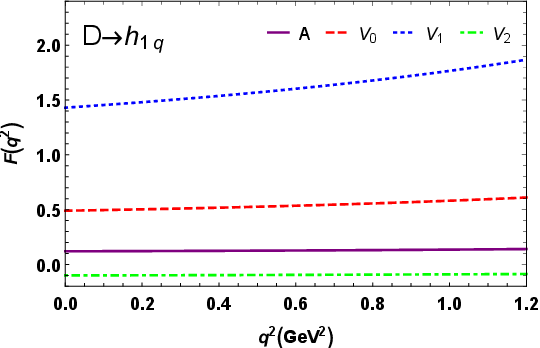}\quad}
  \subfigure[]{\includegraphics[width=0.28\textwidth]{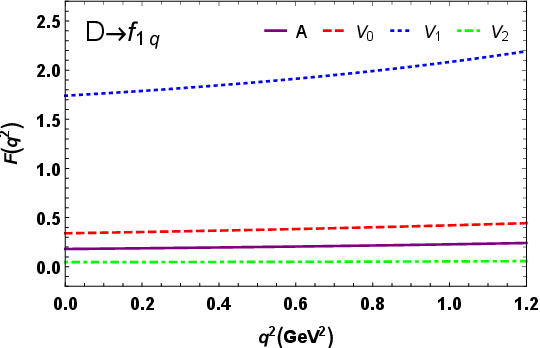}\quad}
  \subfigure[]{\includegraphics[width=0.28\textwidth]{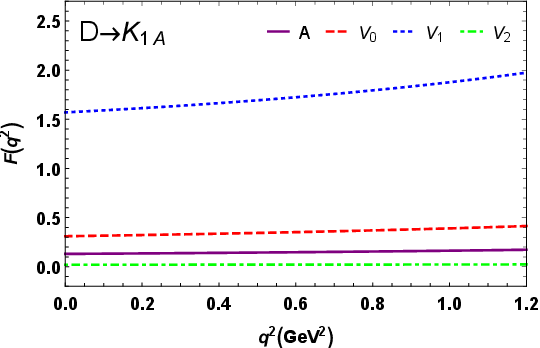}\quad}
  \subfigure[]{\includegraphics[width=0.28\textwidth]{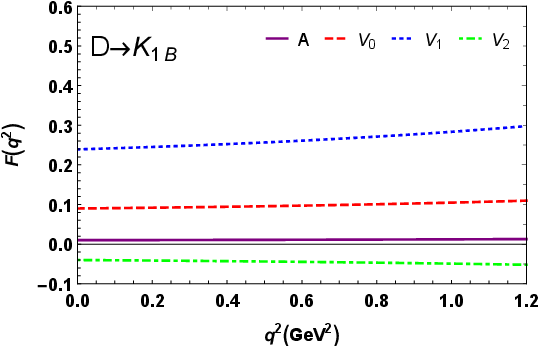}\quad}
  \subfigure[]{\includegraphics[width=0.28\textwidth]{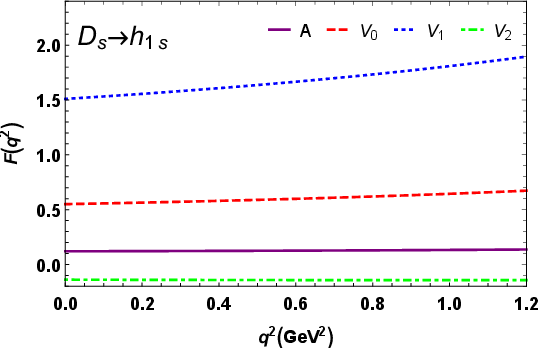}\quad}
  \subfigure[]{\includegraphics[width=0.28\textwidth]{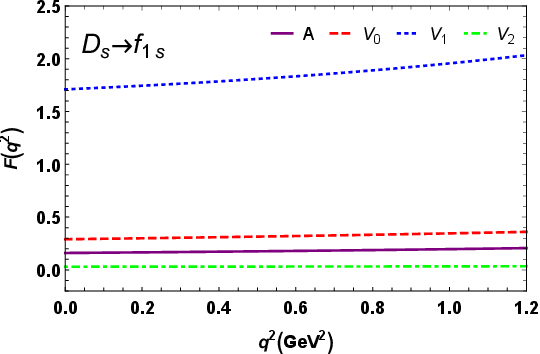}\quad}
  \subfigure[]{\includegraphics[width=0.28\textwidth]{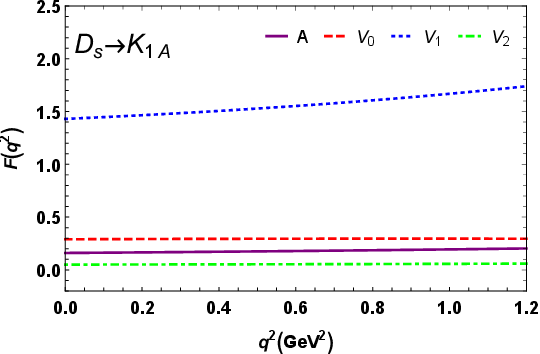}\quad}
  \subfigure[]{\includegraphics[width=0.28\textwidth]{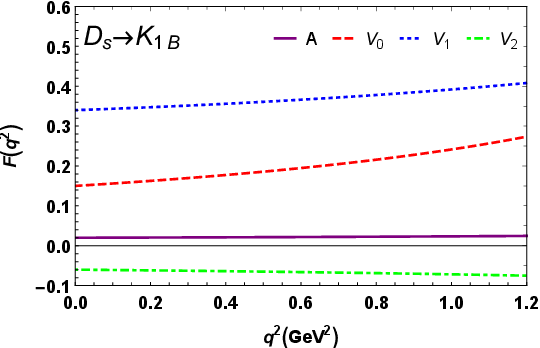}\quad}
\caption{Form factors $A(q^2)$, $V_{0}(q^2)$, $V_{1}(q^2)$ and $V_{2}(q^2)$ of the transitions $D_{(s)} \to a_1(1260),b_1(1235),h_{1q},h_{1s},f_{1q},f_{1s},K_{1A},K_{1B}$.}\label{figA3}
\end{figure}

\begin{figure}[H]
\vspace{0.60cm}
  \centering
  \subfigure[]{\includegraphics[width=0.28\textwidth]{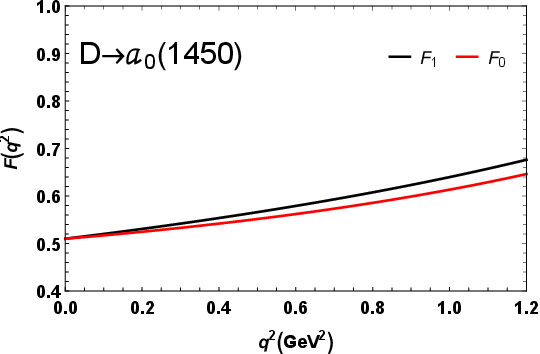}\quad}
  \subfigure[]{\includegraphics[width=0.28\textwidth]{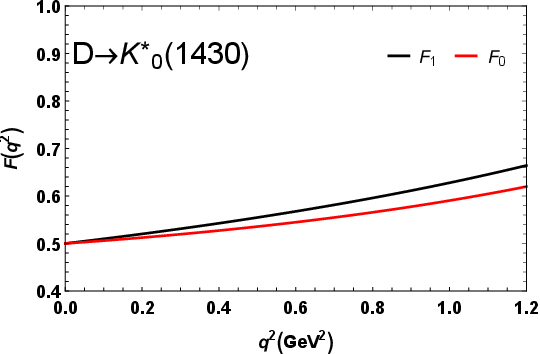}\quad}
  \subfigure[]{\includegraphics[width=0.28\textwidth]{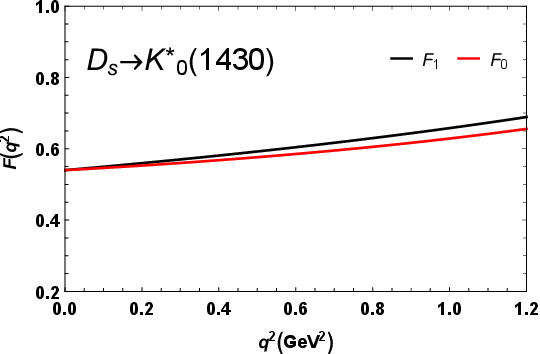}\quad}
  \subfigure[]{\includegraphics[width=0.28\textwidth]{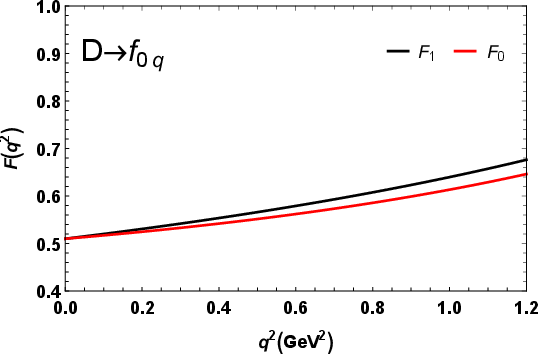}\quad}
  \subfigure[]{\includegraphics[width=0.28\textwidth]{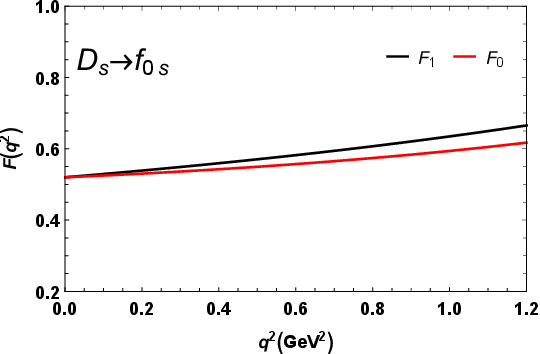}\quad}
\caption{Form factors $F_0(q^2)$ and $F_1(q^2)$ of the transitions $D_{(s)} \rightarrow a_0(1450),K^*_0(1430),f_{0q},f_{0s}$.}\label{figS4}
\end{figure}

\section{Some specific rules under the $p^-$ integration}
When integrating, it is important to include the zero-mode contribution for proper integration in the CLFQM. Specifically, we follow the rules outlined in Ref. \cite{Cheng0310}
\be
\hat{p}_{1 \mu}^{\prime} &\doteq & P_{\mu} A_{1}^{(1)}+q_{\mu} A_{2}^{(1)}, \\
\hat{p}_{1 \mu}^{\prime} \hat{p}_{1 \nu}^{\prime} &\doteq & g_{\mu \nu} A_{1}^{(2)}+P_{\mu} P_{\nu} A_{2}^{(2)}+\left(P_{\mu} q_{\nu}+q_{\mu} P_{\nu}\right) A_{3}^{(2)}+q_{\mu} q_{\nu} A_{4}^{(2)}, \\
\hat{p}_{1 \mu}^{\prime} \hat{p}_{1 \nu}^{\prime} \hat{p}_{1 \alpha}^{\prime} &\doteq & \left(g_{\mu \nu} P_{\alpha}+g_{\mu \alpha} P_{\nu}+g_{\nu \alpha} P_{\mu}\right) A_{1}^{(3)}+\left(g_{\mu \nu} q_{\alpha}+g_{\mu \alpha} q_{\nu}+g_{\nu \alpha} q_{\mu}\right) A_{2}^{(3)} \non
&& +P_{\mu} P_{\nu} P_{\alpha} A_{3}^{(3)}+\left(P_{\mu} P_{\nu} q_{\alpha}+P_{\mu} q_{\nu} P_{\alpha}+q_{\mu} P_{\nu} P_{\alpha}\right) A_{4}^{(3)} \non
&& +\left(q_{\mu} q_{\nu} P_{\alpha}+q_{\mu} P_{\nu} q_{\alpha}+P_{\mu} q_{\nu} q_{\alpha}\right) A_{5}^{(3)}+q_{\mu} q_{\nu} q_{\alpha} A_{6}^{(3)}, \\
\hat{N}_{2} &\rightarrow & Z_{2}, \quad x_{1} \hat{N}_{2} \rightarrow 0, \\
\hat{p}_{1 \mu}^{\prime} \hat{N}_{2} &\rightarrow & q_{\mu}\left[A_{2}^{(1)} Z_{2}+\frac{q \cdot P}{q^{2}} A_{1}^{(2)}\right], \\
\hat{p}_{1 \mu}^{\prime} \hat{p}_{1 \nu}^{\prime} \hat{N}_{2} &\rightarrow & g_{\mu \nu} A_{1}^{(2)} Z_{2}+q_{\mu} q_{\nu}\left[A_{4}^{(2)} Z_{2}+2 \frac{q \cdot P}{q^{2}} A_{2}^{(1)} A_{1}^{(2)}\right] \\
\hat{p}_{1 \mu}^{\prime} \hat{p}_{1 \nu}^{\prime} p_{1 \alpha}^{\prime} \hat{N}_{2} &\rightarrow & \left(g_{\mu \nu} q_{\alpha}+g_{\mu \alpha} q_{\nu}+g_{\nu \alpha} q_{\mu}\right)\left[A_{2}^{(3)} Z_{2}+\frac{q \cdot P}{3 q^{2}}\left(A_{1}^{(2)}\right)^{2}\right] \non
&& +q_{\mu} q_{\nu} q_{\alpha}\left\{A_{6}^{(3)} Z_{2}+3 \frac{q \cdot P}{q^{2}}\left[A_{2}^{(1)} A_{2}^{(3)}-\frac{1}{3 q^{2}}\left(A_{1}^{(2)}\right)^{2}\right]\right\},\\
\en

\be
A_{1}^{(1)}&=&\frac{x_{1}}{2}, \quad A_{2}^{(1)}=A_{1}^{(1)}-\frac{p_{\perp}^{\prime} \cdot q_{\perp}}{q^{2}},  \\
Z_{2}&=&\hat{N}_{1}^{\prime}+m_{1}^{\prime 2}-m_{2}^{2}+\left(1-2 x_{1}\right) M^{\prime 2}+\left(q^{2}+q \cdot P\right) \frac{p_{\perp}^{\prime} \cdot q_{\perp}}{q^{2}}, \\
A_{1}^{(2)}&=&-p_{\perp}^{\prime 2}-\frac{\left(p_{\perp}^{\prime} \cdot q_{\perp}\right)^{2}}{q^{2}}, \quad A_{2}^{(2)}=\left(A_{1}^{(1)}\right)^{2}, \quad A_{3}^{(2)}=A_{1}^{(1)} A_{2}^{(1)}, \\
A_{4}^{(2)}&=&\left(A_{2}^{(1)}\right)^{2}-\frac{1}{q^{2}} A_{1}^{(2)}, \quad A_{1}^{(3)}=A_{1}^{(1)} A_{1}^{(2)}, \quad A_{2}^{(3)}=A_{2}^{(1)} A_{1}^{(2)},\\
A_{3}^{(3)}&=&A_{1}^{(1)} A_{2}^{(2)}, \quad A_{4}^{(3)}=A_{2}^{(1)} A_{2}^{(2)}, \quad A_{5}^{(3)}=A_{1}^{(1)} A_{4}^{(2)} \\
A_{6}^{(3)}&=&A_{2}^{(1)} A_{4}^{(2)}-\frac{2}{q^{2}} A_{2}^{(1)} A_{1}^{(2)}
\en

The following formulas are the analytical expressions of the traces in the transtion amplitudes and the form factors.
For the transition $D_{(s)} \to P$, the trace $S_{ \mu}^{D_{(s)} P}$ and the form factors $F^{D_{(s)} P}_{1,0}(q^2)$ are listed as
\be
S_{ \mu}^{D_{(s)} P}&= & 2 p_{1 \mu}^{\prime}\left[M^{\prime 2}+M^{\prime \prime 2}-q^{2}-2 N_{2}-\left(m_{1}^{\prime}-m_{2}\right)^{2}-\left(m_{1}^{\prime \prime}-m_{2}\right)^{2}+\left(m_{1}^{\prime}-m_{1}^{\prime \prime}\right)^{2}\right] \non
&& +q_{\mu}\left[q^{2}-2 M^{\prime 2}+N_{1}^{\prime}-N_{1}^{\prime \prime}+2 N_{2}+2\left(m_{1}^{\prime}-m_{2}\right)^{2}-\left(m_{1}^{\prime}-m_{1}^{\prime \prime}\right)^{2}\right] \non
&& +P_{\mu}\left[q^{2}-N_{1}^{\prime}-N_{1}^{\prime \prime}-\left(m_{1}^{\prime}-m_{1}^{\prime \prime}\right)^{2}\right],\\
F^{D_{(s)} P}_1 \left(q^{2}\right)&=& \frac{N_{c}}{16 \pi^{3}} \int d x_{2} d^{2} p_{\perp}^{\prime} \frac{h_{D_{(s)}}^{\prime} h_{P}^{\prime \prime}}{x_{2} \hat{N}_{1}^{\prime} \hat{N}_{1}^{\prime \prime}}\left[x_{1}\left(M_{0}^{\prime 2}+M_{0}^{\prime \prime 2}\right)+x_{2} q^{2}-x_{2}\left(m_{1}^{\prime}-m_{1}^{\prime \prime}\right)^{2}\right. \non
&& \left.\quad-x_{1}\left(m_{1}^{\prime}-m_{2}\right)^{2}-x_{1}\left(m_{1}^{\prime \prime}-m_{2}\right)^{2}\right] \\
F^{D_{(s)} P}_0 \left(q^{2}\right)&=& F^{D_{(s)} P}_1(q^2)+\frac{q^2}{(q \cdot P)} \frac{N_{c}}{16 \pi^{3}} \int d x_{2} d^{2} p_{\perp}^{\prime} \frac{2 h_{D_{(s)}}^{\prime} h_{P}^{\prime \prime}}{x_{2} \hat{N}_{1}^{\prime} \hat{N}_{1}^{\prime \prime}}\left\{-x_{1} x_{2} M^{\prime 2}-p_{\perp}^{\prime 2}-m_{1}^{\prime} m_{2}\right. \non
&& \quad+\left(m_{1}^{\prime \prime}-m_{2}\right)\left(x_{2} m_{1}^{\prime}+x_{1} m_{2}\right)+2 \frac{q \cdot P}{q^{2}}\left(p_{\perp}^{\prime 2}+2 \frac{\left(p_{\perp}^{\prime} \cdot q_{\perp}\right)^{2}}{q^{2}}\right)+2 \frac{\left(p_{\perp}^{\prime} \cdot q_{\perp}\right)^{2}}{q^{2}} \non
&& \left. \quad-\frac{p_{\perp}^{\prime} \cdot q_{\perp}}{q^{2}}\left[M^{\prime \prime 2}-x_{2}\left(q^{2}+q \cdot P\right) -\left(x_{2}-x_{1}\right) M^{\prime 2}+2 x_{1} M_{0}^{\prime 2}\right.\right.\non
&& \left.\left. \quad-2\left(m_{1}^{\prime}-m_{2}\right)\left(m_{1}^{\prime}+m_{1}^{\prime \prime}\right)\right]\right\}.
\en
The trace $S_{ \mu}^{D_{(s)} S}$ and the form factors $F^{PS}_{0,1}(q^2)$ for the transtion $D_{(s)}\to S$ are related those of the transtion $D_{(s)} \to P$ by
\be
S_{ A\mu}^{D_{(s)} S}&=&-iS_{V \mu}^{D_{(s)} P}(m^{\prime\prime}\to-m^{\prime\prime})\\
F^{D_{(s)}S}_{0,1}&=&-F^{D_{(s)}P}_{0,1}\left(m_{1}^{\prime \prime} \rightarrow-m_{1}^{\prime \prime}, h_{P}^{\prime \prime} \rightarrow h_{S}^{\prime \prime}\right).
\en

For the transition $D_{(s)} \to V$, the trace $S_{\mu \nu}^{D_{(s)} V}(q^2)$ and the form factors $V^{D_{(s)} V}(q^{2}),$ $A_0^{D_{(s)} V}(q^{2}), A_1^{D_{(s)} V}(q^{2}), A_2^{D_{(s)} V}(q^{2})$ are given as
\begin{footnotesize}
\be
S_{\mu \nu}^{D_{(s)} V}&=&\left(S_{V}^{D_{(s)} V}-S_{A}^{D_{(s)} V}\right)_{\mu \nu}\non
&=&\operatorname{Tr}\left[\left(\gamma_{\nu}-\frac{1}{W_{V}^{\prime \prime}}\left(p_{1}^{\prime \prime}-p_{2}\right)_{\nu}\right)\left(p_{1}^{\prime \prime}
+m_{1}^{\prime \prime}\right)\left(\gamma_{\mu}-\gamma_{\mu} \gamma_{5}\right)\left(\not p_{1}^{\prime}+m_{1}^{\prime}\right) \gamma_{5}\left(-\not p_{2}
+m_{2}\right)\right] \non
&=&-2 i \epsilon_{\mu \nu \alpha \beta}\left\{p_{1}^{\prime \alpha} P^{\beta}\left(m_{1}^{\prime \prime}-m_{1}^{\prime}\right)
+p_{1}^{\prime \alpha} q^{\beta}\left(m_{1}^{\prime \prime}+m_{1}^{\prime}-2 m_{2}\right)+q^{\alpha} P^{\beta} m_{1}^{\prime}\right\} \non
&&+\frac{1}{W_{V}^{\prime \prime}}\left(4 p_{1 \nu}^{\prime}-3 q_{\nu}-P_{\nu}\right) i \epsilon_{\mu \alpha \beta \rho} p_{1}^{\prime \alpha} q^{\beta} P^{\rho}\non &&
+2 g_{\mu \nu}\left\{m_{2}\left(q^{2}-N_{1}^{\prime}-N_{1}^{\prime \prime}-m_{1}^{\prime 2}-m_{1}^{\prime \prime 2}\right)
-m_{1}^{\prime}\left(M^{\prime \prime 2}-N_{1}^{\prime \prime}-N_{2}-m_{1}^{\prime \prime 2}-m_{2}^{2}\right)\right.\non
&&\left.-m_{1}^{\prime \prime}\left(M^{\prime 2}-N_{1}^{\prime}-N_{2}-m_{1}^{\prime 2}-m_{2}^{2}\right)
-2 m_{1}^{\prime} m_{1}^{\prime \prime} m_{2}\right\} \non &&
+8 p_{1 \mu}^{\prime} p_{1 \nu}^{\prime}\left(m_{2}-m_{1}^{\prime}\right)-2\left(P_{\mu} q_{\nu}
+q_{\mu} P_{\nu}+2 q_{\mu} q_{\nu}\right) m_{1}^{\prime}+2 p_{1 \mu}^{\prime} P_{\nu}\left(m_{1}^{\prime}-m_{1}^{\prime \prime}\right)\non &&
+2 p_{1 \mu}^{\prime} q_{\nu}\left(3 m_{1}^{\prime}-m_{1}^{\prime \prime}-2 m_{2}\right)
+2 P_{\mu} p_{1 \nu}^{\prime}\left(m_{1}^{\prime}+m_{1}^{\prime \prime}\right)+2 q_{\mu} p_{1 \nu}^{\prime}\left(3 m_{1}^{\prime}+m_{1}^{\prime \prime}-2 m_{2}\right)\non &&
+\frac{1}{2 W_{V}^{\prime \prime}}\left(4 p_{1 \nu}^{\prime}-3 q_{\nu}-P_{\nu}\right)\left\{2 p_{1 \mu}^{\prime}\left[M^{\prime 2}
+M^{\prime \prime 2}-q^{2}-2 N_{2}+2\left(m_{1}^{\prime}-m_{2}\right)\left(m_{1}^{\prime \prime}+m_{2}\right)\right]\right.\non&&
+q_{\mu}\left[q^{2}-2 M^{\prime 2}+N_{1}^{\prime}-N_{1}^{\prime \prime}+2 N_{2}-\left(m_{1}+m_{1}^{\prime \prime}\right)^{2}+2\left(m_{1}^{\prime}-m_{2}\right)^{2}\right]\non&&
\left.+P_{\mu}\left[q^{2}-N_{1}^{\prime}-N_{1}^{\prime \prime}-\left(m_{1}^{\prime}+m_{1}^{\prime \prime}\right)^{2}\right]\right\}, \\
V^{D_{(s)} V}(q^{2})&=&\frac{N_{c}(M^{'}+M^{''})}{16 \pi^{3}} \int d x_{2} d^{2} p_{\perp}^{\prime} \frac{2 h_{D_{(s)}}^{\prime}
 h_{V}^{\prime \prime}}{x_{2} \hat{N}_{1}^{\prime} \hat{N}_{1}^{\prime \prime}}\left\{x_{2} m_{1}^{\prime}
 +x_{1} m_{2}+\left(m_{1}^{\prime}-m_{1}^{\prime \prime}\right) \frac{p_{\perp}^{\prime} \cdot q_{\perp}}{q^{2}}\right.\non &&\left.
 +\frac{2}{w_{D_{(s)}}^{\prime \prime}}\left[p_{\perp}^{\prime 2}+\frac{\left(p_{\perp}^{\prime} \cdot q_{\perp}\right)^{2}}{q^{2}}\right]\right\},\\
A_0^{D_{(s)} V}(q^{2})&=& \frac{M^{'}+M^{''}}{2M^{''}}A_1^{D_{(s)} V}(q^{2})-\frac{M^{'}-M^{''}}{2M^{''}}A_2^{D_{(s)} V}(q^{2})-\frac{q^2}{2M^{''}}\frac{N_{c}}{16 \pi^{3}} \int d x_{2} d^{2} p_{\perp}^{\prime} \frac{h_{D_{(s)}}^{\prime} h_{V}^{\prime \prime}}{x_{2} \hat{N}_{1}^{\prime}
\hat{N}_{1}^{\prime \prime}}\left\{2\left(2 x_{1}-3\right)\right.\non &&\left.\left(x_{2} m_{1}^{\prime}+x_{1} m_{2}\right)-8\left(m_{1}^{\prime}-m_{2}\right)
\times\left[\frac{p_{\perp}^{\prime 2}}{q^{2}}
+2 \frac{\left(p_{\perp}^{\prime} \cdot q_{\perp}\right)^{2}}{q^{4}}\right]-\left[\left(14-12 x_{1}\right) m_{1}^{\prime}\right.\right. \non &&\left.\left.-2 m_{1}^{\prime \prime}-\left(8-12 x_{1}\right) m_{2}\right] \frac{p_{\perp}^{\prime} \cdot q_{\perp}}{q^{2}}
+\frac{4}{w_{D_{(s)}}^{\prime \prime}}\left(\left[M^{\prime 2}+M^{\prime \prime 2}-q^{2}+2\left(m_{1}^{\prime}-m_{2}\right)\left(m_{1}^{\prime \prime}
+m_{2}\right)\right]\right.\right.\non &&\left.\left.\times\left(A_{3}^{(2)}+A_{4}^{(2)}-A_{2}^{(1)}\right)
+Z_{2}\left(3 A_{2}^{(1)}-2 A_{4}^{(2)}-1\right)+\frac{1}{2}\left[x_{1}\left(q^{2}+q \cdot P\right)
-2 M^{\prime 2}-2 p_{\perp}^{\prime} \cdot q_{\perp}\right.\right.\right.\non &&\left.\left.\left.-2 m_{1}^{\prime}\left(m_{1}^{\prime \prime}+m_{2}\right)
-2 m_{2}\left(m_{1}^{\prime}-m_{2}\right)\right]\left(A_{1}^{(1)}+A_{2}^{(1)}-1\right) q \cdot P\left[\frac{p_{\perp}^{\prime 2}}{q^{2}}
+\frac{\left(p_{\perp}^{\prime} \cdot q_{\perp}\right)^{2}}{q^{4}}\right]\right.\right.\non &&\left.\left.\times\left(4 A_{2}^{(1)}-3\right)\right)\right\},\\
A_1^{D_{(s)} V}(q^{2})&=& -\frac{1}{M^{'}+M^{''}}\frac{N_{c}}{16 \pi^{3}} \int d x_{2} d^{2} p_{\perp}^{\prime} \frac{h_{D_{(s)}}^{\prime} h_{V}^{\prime \prime}}{x_{2}
	\hat{N}_{1}^{\prime}
	\hat{N}_{1}^{\prime \prime}}\left\{2 x_{1}\left(m_{2}-m_{1}^{\prime}\right)\left(M_{0}^{\prime 2}+M_{0}^{\prime \prime 2}\right)
-4 x_{1} m_{1}^{\prime \prime} M_{0}^{\prime 2}\right.\non
&&\left.+2 x_{2} m_{1}^{\prime} q \cdot P+2 m_{2} q^{2}-2 x_{1} m_{2}\left(M^{\prime 2}+M^{\prime \prime 2}\right)+2\left(m_{1}^{\prime}-m_{2}\right)\left(m_{1}^{\prime}
+m_{1}^{\prime \prime}\right)^{2}+8\left(m_{1}^{\prime}-m_{2}\right) \right.\non &&
\left. \times\left[p_{\perp}^{\prime 2}+\frac{\left(p_{\perp}^{\prime}
	\cdot q_{\perp}\right)^{2}}{q^{2}}\right]+2\left(m_{1}^{\prime}+m_{1}^{\prime \prime}\right)\left(q^{2}+q \cdot P\right) \frac{p_{\perp}^{\prime} \cdot q_{\perp}}{q^{2}}
-4 \frac{q^{2} p_{\perp}^{\prime 2}+\left(p_{\perp}^{\prime} \cdot q_{\perp}\right)^{2}}{q^{2} w_{D_{(s)}}^{\prime \prime}}
\right.\non && \left.\times\left[2 x_{1}\left(M^{\prime 2}+M_{0}^{\prime 2}\right)-q^{2}-q \cdot P-2\left(q^{2}+q \cdot P\right) \frac{p_{\perp}^{\prime} \cdot q_{\perp}}{q^{2}}-2\left(m_{1}^{\prime}-m_{1}^{\prime \prime}\right)\left(m_{1}^{\prime}-m_{2}\right)\right]\right\},
\end{eqnarray}
\end{footnotesize}
\begin{footnotesize}
\begin{eqnarray}
A_2^{D_{(s)} V}(q^{2})&=& \frac{N_{c}(M^{'}+M^{''})}{16 \pi^{3}} \int d x_{2} d^{2} p_{\perp}^{\prime} \frac{2 h_{D_{(s)}}^{\prime} h_{V}^{\prime \prime}}{x_{2} \hat{N}_{1}^{\prime}
\hat{N}_{1}^{\prime \prime}}\left\{\left(x_{1}-x_{2}\right)\left(x_{2} m_{1}^{\prime}+x_{1} m_{2}\right)-\frac{p_{\perp}^{\prime} \cdot q_{\perp}}{q^{2}}\left[2 x_{1} m_{2}
+m_{1}^{\prime \prime} \right.\right.\non &&
\left.\left.+\left(x_{2}-x_{1}\right) m_{1}^{\prime}\right]-2 \frac{x_{2} q^{2}+p_{\perp}^{\prime} \cdot q_{\perp}}{x_{2} q^{2} w_{D_{(s)}}^{\prime \prime}}\left[p_{\perp}^{\prime} \cdot p_{\perp}^{\prime \prime}
+\left(x_{1} m_{2}+x_{2} m_{1}^{\prime}\right)\left(x_{1} m_{2}-x_{2} m_{1}^{\prime \prime}\right)\right]\right\}.
\end{eqnarray}
\end{footnotesize}

Similarly, the analytic expressions for the transition $D_{(s)} \to A$ can be obtained from those of the transtion $D_{(s)} \to V$ by the following replacements:
\begin{footnotesize}
\be
S^{D_{(s)}A}_V&=&iS^{D_{(s)}V}_A, S^{D_{(s)}A}_A=iS^{D_{(s)}V}_{V}(m^{\prime\prime}\to -m^{\prime\prime}, W^{\prime\prime}_V\to W^{\prime\prime}_A),\\
A^{D_{(s)} A}(q^{2})&=&V^{D_{(s)} V}(q^{2}), V^{D_{(s)} A}_{0}(q^{2})=A^{D_{(s)} V}_{0}(q^{2}),V^{D_{(s)} A}_{1}(q^{2})=A^{D_{(s)} V}_{1}(q^{2}),\nonumber \\
V^{D_{(s)} A}_{2}(q^{2})&=&A^{D_{(s)} V}_{2}(q^{2})\left(M^{\prime\prime}\to -M^{\prime\prime}, m_{1}^{\prime \prime} \to-m_{1}^{\prime \prime}, h_{V}^{\prime \prime} \to h^{\prime \prime}_{A}, w_{V}^{\prime \prime} \to w_{A}^{\prime \prime}\right),
\en
\end{footnotesize}
where we should be cautious that the replacement of $m^{\prime\prime}\to -m^{\prime\prime}$ can not be applied to $m^{\prime\prime}$ in $w^{\prime \prime}$ and $h^{\prime \prime}$, $A$ refers two types axial-vector mesons with $J^{PC}=1^{++}$ and $1^{+-}$ denoted as $^3A$ and $^1A$, respectively.



\begin{thebibliography}{99}
\bibitem{AJaus}
W. Jaus, Phys. Rev. D {\bf41}, 3394 (1990).

\bibitem{BJaus}
W. Jaus, Phys. Rev. D {\bf44}, 2851 (1991).

\bibitem{CJaus}
W. Jaus, Phys. Rev. D {\bf60}, 054026 (1999).

\bibitem{Cheng0310}
H. Y. Cheng, C. K. Chua and C. W. Hwang, Phys. Rev. D {\bf69}, 074025 (2004) [arXiv:hep-ph/0310359].

\bibitem{CLF1103}
R. C. Verma, J. Phys. G: Nucl. Part. Phys {\bf39}, 025005 (2012)[arXiv:1103.2973 [hep-ph]].

\bibitem{BESIII2411}
M. Ablikim \textit{et al.} (BESIII Collaboration), Phys. Rev. D {\bf111}, L091501 (2025) [arXiv:2411.07730 [hep-ex]].

\bibitem{BESIII2303}
M. Ablikim \textit{et al.} (BESIII Collaboration), Phys.Rev.Lett. {\bf132}, 14, 141901 (2024) [arXiv:2303.12927 [hep-ex]].

\bibitem{BESIII250302}
M. Ablikim \textit{et al.} (BESIII Collaboration), 	Phys. Rev. Lett. {\bf135}, 091801 (2025) [arXiv:2503.02196  [hep-ex]].

\bibitem{BESIII250203}
M. Ablikim \textit{et al.} (BESIII Collaboration), 	Phys. Rev. D {\bf111}, L071101 (2025) [arXiv:2502.03828 [hep-ex]].

\bibitem{Wirbel}
M. Wirbel, B. Stech and M. Bauer, Z. Phys. C {\bf29}, 637 (1985); M. Bauer, B. Stech, and M. Wirbel, ibid, 42, 671 (1989).

\bibitem{Sun2023}
Z. J. Sun, Z. Q. Zhang, Y. Y. Yang and H. Yang, Eur. Phys. J. C {\bf84}, 65 (2024) [arXiv:2311.04431 [hep-ph]].

\bibitem{Gaussian1988}
P. L. Chung, F. Coester, and W. N. Polyzou, Phys. Lett. B 205, 545 (1988).

\bibitem{PDG2024}
S. Navas \textit{et al.} [Particle Data Group], Phys. Rev. D {\bf110}, 030001 (2024)
\bibitem{yang}
K.~C.~Yang,
Nucl. Phys. B \textbf{776}, 187-257 (2007)
[arXiv:0705.0692 [hep-ph]].
\bibitem{Momeni2004}
S. Momeni, Eur. Phys. J. C {\bf80}, 553 (2020) [arXiv:2004.02522 [hep-ph]].

\bibitem{BESIII2110}
M. Ablikim \textit{et al.} (BESIII Collaboration), Phys. Rev. Lett. {\bf127}, 171801 (2021) [arXiv:2106.02218 [hep-ex]].

\bibitem{HFLAV}
Y. S. Amhis \textit{et al.} [Heavy Flavor Averaging Group (HFLAV)], Eur. Phys. J. C  {\bf81} 226 (2021) [arXiv:1909.12524 [hep-ex]].

\bibitem{BESIII1804}
M. Ablikim \textit{et al.} (BESIII Collaboration), 	Phys. Rev. D {\bf98}, 072005 (2018) [arXiv:1804.05536 [hep-ex]].

\bibitem{Zuo0604}
Y. L. Wu, M. Zhong, Y. B. Zuo, Int. J. Mod. Phys. {\bf21} 6125-6172, (2006) [arXiv:hep-ph/0604007].

\bibitem{Momeni1903}
S. Momeni and R. Khosravi, J. Phys. G {\bf46}, 105006 (2019) [arXiv:1903.00860 [hep-ph]].

\bibitem{LCSROffen}
N. Offen, F. A. Porkert, and A. Schafer, Phys. Rev. D {\bf88}, 034023 (2013) [arXiv:1307.2797 [hep-ph]].

\bibitem{LCSRDuplancic}
G. Duplancic and B. Melic, J. High Energy Phys. {\bf11} (2015) 138 [arXiv:1508.05287 [hep-ph]].

\bibitem{CCQM1904}
M. A. Ivanov, J. G. Korner, J. N. Pandya, P. Santorelli, N. R. Soni and C. T. Tran, Frontiers of Phys {\bf14}, 64401 (2019)[arXiv:1904.07740 [hep-ph]]

\bibitem{Kang1911}
R. N. Faustov, V. O. Galkin and X. W. Kang, Phys. Rev. D {\bf101}, 013004 (2020)[arXiv:1911.08209 [hep-ph]].

\bibitem{2019etap}
M. Ablikim \textit{et al.} [BESIII], Phys. Rev. Lett. \textbf{122}, 121801 (2019)[arXiv:1901.02133 [hep-ex]].

\bibitem{LQCD1406}
G. S. Bali, S. Collins, S. D¡§urr, and I. Kanamori, Phys. Rev. D {\bf91}, 014503 (2015) [arXiv:1406.5449 [hep-lat]].

\bibitem{BESIII150807}
M. Ablikim \textit{et al.} (BESIII Collaboration), Phys. Rev. D {\bf92}, 072012 (2015) [arXiv:1508.07560[hep-ex]].

\bibitem{BABAR0704}
B. Aubert \textit{et al.} (BABAR Collaboration), Phys. Rev. D {\bf76}, 052005 (2007) [arXiv:0704.0020[hep-ex]].

\bibitem{AblikimBES}
M. Ablikim et al. (BESIII Collaboration), Phys. Rev. Lett. {\bf122}, 121801 (2019) 50 [arXiv:1901.02133].

\bibitem{MS0008}
D. Melikhov and B. Stech, Phys. Rev. D {\bf62}, 014006 (2000)[arXiv:hep-ph/0001113]

\bibitem{QSR9305}
P. Ball, Phys. Rev. D {\bf48}, 3190-3203 (1993)[arXiv:hep-ph/9305267].

\bibitem{BSW}
M. Wirbel, S. Stech and M. Bauer, Z. Phys. C {\bf29}, 637 (1985); M. Bauer, B. Stech, and M. Wirbel, Z. Phys. C {\bf34}, 103 (1987).

\bibitem{BESIII181111}
M. Ablikim \textit{et al.} (BESIII Collaboration), Phys. Rev. D {\bf99}, 011103 (2019) [arXiv:1811.11349 [hep-ex]].

\bibitem{BESIII180906}
M. Ablikim \textit{et al.} (BESIII Collaboration), Phys. Rev. Lett. {\bf122}, 062001 (2019) [arXiv:1809.06496 [hep-ex]].

\bibitem{BESIII181102}
M. Ablikim \textit{et al.} (BESIII Collaboration), Phys. Rev. Lett. {\bf122}, 061801 (2019) [arXiv:1811.02911 [hep-ex]].

\bibitem{Cheng1503}
H. Y. Cheng, C. K. Chua and K. F. Liu, Phys. Rev. D {\bf92}, no. 9, 094006 (2015)[arXiv:1503.06827 [hep-ph]];

\bibitem{lee}
W. Lee and D. Weingarten, Phys. Rev. D {\bf61}, 014015 (2000) [arXiv:hep-lat/9910008].

\bibitem{Huang2102}
Q. Huang, Y. J. Sun, D. Gao, G. H. Zhao, B. Wang and W. Hong, (2021) [arXiv:2102.12241 [hep-ph]].

\bibitem{Hong2409}
W. Hong, D. Gao and Y. J. Sun, Nucl. Phys. A {\bf1064} 123219 (2025) [arXiv:2409.15776 [hep-ph]].

\bibitem{Soni2020}
N. R. Soni, A. N. Gadaria, J. J. Patel, J. N. Pandya, Phys. Rev. D {\bf102}, 016013 (2020) [arXiv:2001.10195 [hep-ph]].

\bibitem{yangmaozhi2017}
X. D. Cheng, H. B. Li, B. Wei, Y. G. Xu and M. Z. Yang,	Phys. Rev. D {\bf96}, 033002 (2017), [arXiv:1706.01019 [hep-ph]].

\bibitem{wangwei1701}
Y. J. Shi, W. Wang and S. Zhao,	Eur. Phys. J. C {\bf77}, 452 (2017) [arXiv:1701.07571 [hep-ph]].

\bibitem{Colangelo}
P. Colangelo, F. D. Fazio and W. Wang, Phys. Rev. D {\bf81}, 074001 (2010) [arXiv:1002.2880 [hep-ph]].

\bibitem{kehongwei2009}
H. W. Ke, X. Q. Li and Z. T. Wei, Phys. Rev. D {\bf80}, 074030 (2009) [arXiv:0907.5465 [hep-ph]].

\bibitem{Bennich2009}
B. El-Bennich, O. Leitner, J. P. Dedonder and B. Loiseau, Phys. Rev. D {\bf79}, 076004 (2009), [arXiv:0810.5771 [hep-ph]].

\bibitem{Yang2409}
Y. L. Yang, H. J. Tian, Y. X. Wang, H. B. Fu, T. Zhong and S. Q. Wang,  Phys. Rev. D {\bf110}, 116030 (2024) [arXiv:2409.01512 [hep-ph]].

\bibitem{Huang2211}
D. Huang, T. Zhong, H. B. Fu, Z. H. Wu, X. G. Wu, H. Tong, 	Eur. Phys. J. C {\bf83}, 680 (2023) [arXiv:2211.06211 [hep-ph]].

\bibitem{Yang0509}
M. Z. Yang, Phys.Rev.D {\bf73}, 034027 (2006) [arXiv:hep-ph/0509103].

\bibitem{Kang1707}
H. Y. Cheng and X. W. Kang, Eur. Phys. J. C {\bf77}, 587 (2017) [arXiv:1707.02851 [hep-ph]].

\bibitem{axialmix}
H. Y. Cheng, Phys. Lett. B \textbf{707}, 116-120 (2012) [arXiv:1110.2249 [hep-ph]].

\bibitem{Zuo2016}
Y. B. Zuo \textit{et al.} Int. J. Mod. Phys. A  {\bf31}, (2016) [arXiv:1608.03651 [hep-ph]].

\bibitem{wuxinggang2107}
D. D. Hu, H. B. Fu, T. Zhong, Z. H. Wu and X. G. Wu, Eur. Phys. J. C {\bf82}, 603 (2022) [arXiv:2107.02758 [hep-ph]].

\bibitem{Khosravi0812}
R. Khosravi, K. Azizi and N. Ghahramany, Phys. Rev. D {\bf79}, 036004 (2009)[arXiv:0812.1352 [hep-ph]].

\bibitem{LCSR1508}
G. Duplancic and B. Melic, J. High Energy Phys. {\bf11}, (2015) 138 [arXiv:1508.05287[hep-ex]].

\bibitem{BESIII2506}
M. Ablikim \textit{et al.} (BESIII Collaboration), [arXiv:2506.02521 [hep-ex]].

\bibitem{BESIII2410}
M. Ablikim \textit{et al.} (BESIII Collaboration), Phys. Rev. Lett. {\bf134}, 111801 (2025) [arXiv:2410.08603 [hep-ex]].

\bibitem{CLEO1011}
J. Yelton \textit{et al.} (CLEO Collaboration), Phys. Rev. D {\bf84}, 032001 (2011) [arXiv:1011.1195[hep-ex]].

\bibitem{BESIII2306}
M. Ablikim \textit{et al.} (BESIII Collaboration), 	Physical Review D {\bf108}, 092003 (2023) [arXiv:2306.05194 [hep-ex]].

\bibitem{BESIII2307}
M. Ablikim \textit{et al.} (BESIII Collaboration), Physical Review Letters {\bf132}, 091802 (2024) [arXiv:2307.12852 [hep-ex]].

\bibitem{CLEO2015}
J. Hietala, D. Cronin-Hennessy, T. Pedlar, I. Shipsey, [CLEO Collaboration], Phys. Rev. D {\bf92}, (2015) 012009. [arXiv:1505.04205[hep-ex]].


\bibitem{FajferP}
S. Fajfer and J. F. Kamenik, Phys. Rev. D {\bf71}, 014020 (2005) [arXiv:hep-ph/0412140].

\bibitem{BESIII1802}
M. Ablikim \textit{et al.} (BESIII Collaboration), Phys. Rev. Lett. {\bf121}, 171803 (2018) [arXiv:1802.05492[hep-ex]].

\bibitem{BESIII2408}
M. Ablikim \textit{et al.} (BESIII Collaboration), [arXiv:2408.09087 [hep-ex]].

\bibitem{wuxinggang2405}
H. J. Tian, Y. L. Yang, D. D. Hu, H. B. Fu, T. Zhong and X. G. Wu,  Phys. Lett. B. {\bf857}, 138975 (2024) [arXiv:2405.07154 [hep-ph]].

\bibitem{CQM}
D. Melikhov and B. Stech, Phys. Rev. D {\bf62}, 014006 (2000) [arXiv:hep-ph/0001113].

\bibitem{CLEO1505}
J. Hietala, D. Cronin-Hennessy, T. Pedlar, and I. Shipsey, Phys. Rev. D {\bf92}, 012009 (2015)[arXiv:1505.04205[hep-ex]].

\bibitem{BESIII2406}
M. Ablikim \textit{et al.} (BESIII Collaboration), Phys. Rev. D 2024 [arXiv:2406.19190 [hep-ex]].

\bibitem{Sekihara}
T. Sekihara and E. Oset, Phys. Rev. D {\bf92}, 054038 (2015) [arXiv:1507.02026 [hep-ph]].

\bibitem{BESIII230703}
M. Ablikim \textit{et al.} (BESIII Collaboration), JHEP {\bf12}, 072(2023) [arXiv:2307.03024[hep-ex]].

\bibitem{BABAR0807}
B. Aubert \textit{et al.} (BABAR Collaboration), Phys. Rev. D {\bf78}, 051101 (2008) [arXiv:hep-ex/0807.1599].

\bibitem{BESIII240310}
M. Ablikim \textit{et al.} (BESIII Collaboration), Phys. Rev. Lett. {\bf134}, 011803 (2025) [arXiv:2403.10877 [hep-ex]].

\bibitem{CLEO0506}
T. E. Coan \textit{et al.} (CLEO Collaboration), Phys. Rev. Lett. {\bf95}, 181802 (2005)[arXiv:hep-ex/0506052].

\bibitem{CLEO1004}
R. A. Briere \textit{et al.} (CLEO Collaboration), Phys. Rev. D {\bf81}, 112001 (2010) [arXiv:1004.1954[hep-ex]].

\bibitem{BESIII2409}
M. Ablikim \textit{et al.} (BESIII Collaboration), [arXiv:2409.04276 [hep-ex]].

\bibitem{BESIII2106}
M. Ablikim \textit{et al.} (BESIII Collaboration), Phys. Rev. D {\bf104},  L091103 (2021) [arXiv:2106.02292 [hep-ex]].

\bibitem{BESIII1809}
M. Ablikim \textit{et al.} (BESIII Collaboration), Phys. Rev. Lett. {\bf122}, 062001 (2019) [arXiv:1809.06496 [hep-ex]].

\bibitem{CLEO}
S. Dobbs \textit{et al.} (CLEO Collaboration), Phys. Rev. Lett. {\bf110}, 131802 (2013)[arXiv:1112.2884[hep-ex]].

\bibitem{BESIII2002}
M. Ablikim \textit{et al.} (BESIII Collaboration), Phys. Rev. D {\bf101}, 072005 (2020) [arXiv:2002.10578 [hep-ex]].

\bibitem{wuxinggang2211}
Z. H. Wu, H. B. Fu, T. Zhong, D. Huang, D. D. Hu and X. G. Wu, Nuclear Physics A {\bf1036}, 122671(2023), [arXiv:2211.05390 [hep-ph]].

\bibitem{wangrumin2023}
R. M. Wang, Y. X. Liu, C. Hua, J. H. Sheng and Y. G. Xu,  [arXiv:2301.00079 [hep-ph]]

\bibitem{BESIII1803}
M. Ablikim \textit{et al.} (BESIII Collaboration), Phys. Rev. Lett. {\bf121}, 081802 (2018) [arXiv:1803.02166 [hep-ex]].

\bibitem{CLEO0907}
K. M. Ecklund \textit{et al.} (CLEO Collaboration), Phys. Rev. D {\bf80}, 052009 (2009) [arXiv:0907.3201 [hep-ex]].

\bibitem{zhaozhenxing2015}
W. Wang and Z. X. Zhao, Eur.Phys.J. C {\bf76}, 59 (2016) [arXiv:1511.06998 [hep-ph]].

\bibitem{Bian2105}
L. Z. Bian, L. Sun and W. Wang, Phys. Rev. D {\bf104}, 053003 (2021) [arXiv:2105.06207 [hep-ph]].

\bibitem{BESIII240319}
M. Ablikim \textit{et al.} (BESIII Collaboration), JHEP \textbf{09}, 089 (2024) [arXiv:2403.19091 [hep-ex]].

\bibitem{BESIII2309}
M. Ablikim \textit{et al.} (BESIII Collaboration), Phys. Rev. D {\bf108} 11, 112002 (2023) [arXiv:2309.04090 [hep-ex]].

\bibitem{wangwei}
W.~Wang, Y.~L.~Shen and C.~D.~Lu,
Phys. Rev. D \textbf{79}, 054012 (2009)
[arXiv:0811.3748 [hep-ph]].

\end{thebibliography}
\end{document}